\pdfoutput=1
\documentclass[preprint2]{aastex}
\usepackage[caption=false]{subfig}
\usepackage{rotating}
\usepackage{placeins}
\usepackage{color}

\newcommand{\gtabouteq}{\,\hbox{\raise 0.5 ex \hbox{$>$}\kern-.77em 
                    \lower 0.5 ex \hbox{$\sim$}$\,$}}       
\newcommand{\ltabouteq}{\,\hbox{\raise 0.5 ex \hbox{$<$}\kern-.77em 
                     \lower 0.5 ex \hbox{$\sim$}$\,$}}

\slugcomment{To be submitted to Astronomical Journal}

\shorttitle{CHANG-ES XX: Data Release 3}
\shortauthors{Irwin et al.}

\begin{document}

%% LaTeX will automatically break titles if they run longer than
%% one line. However, you may use \\ to force a line break if
%% you desire.

\title{CHANG-ES XX.  High Resolution Radio Continuum Images of Edge-on Galaxies and their AGNs -- Data Release 3}

%% Use \author, \affil, and the \and command to format
%% author and affiliation information.
%% Note that \email has replaced the old \authoremail command
%% from AASTeX v4.0. You can use \email to mark an email address
%% anywhere in the paper, not just in the front matter.
%% As in the title, use \\ to force line breaks.

\author{Judith Irwin\altaffilmark{1},  Theresa Wiegert\altaffilmark{1}, Alison Merritt\altaffilmark{1}, Marek We{\.z}gowiec\altaffilmark{2,9},  Lucas Hunt\altaffilmark{3}, Alex Woodfinden\altaffilmark{1}, Yelena Stein\altaffilmark{4,9},   Ancor Damas-Segovia\altaffilmark{5}, Jiangtao Li\altaffilmark{6}, Q. Daniel Wang\altaffilmark{7},   Megan Johnson\altaffilmark{3}, Marita Krause\altaffilmark{8},
  Ralf-J{\"u}rgen Dettmar\altaffilmark{9},
  Jisung Im\altaffilmark{1},
  Philip Schmidt\altaffilmark{8},
  Arpad Miskolczi\altaffilmark{9},
  Timothy T. Braun\altaffilmark{10},
  D. J. Saikia\altaffilmark{11},
  Jayanne English\altaffilmark{12},
  Mark L. A. Richardson\altaffilmark{1,13}
}

\altaffiltext{1}{Dept. of Physics, Engineering Physics \& Astronomy, 
  Queen's University, Kingston, ON, Canada, K7L 3N6, {\tt irwinja@queensu.ca, theresa.wiegert@gmail.com, merritt.j.alison@gmail.com, 17aw14@queensu.ca, j4im@uwaterloo.ca, Mark.Richardson@queensu.ca.}}
\altaffiltext{2}{Obserwatorium Astronomiczne Uniwersytetu Jagiello{\'n}skiego, ul. Orla 171, 30-244 Krak{\'o}w, Poland
  {\tt markmet@oa.uj.edu.pl}.}
\altaffiltext{3}{US Naval Observatory, Astrometry Dept., 3450 Massachusetts Ave. NW, Washington 20392 D.C.
  {\tt lrhunt87@gmail.com, meg7770@yahoo.com}.}
\altaffiltext{4}{Centre de Donn{\'e}es astronomiques de Strasbourg, Observatoire de Strasbourg 11, rue de l'Universit{\'e} - 67000 Strasbourg, France
  {\tt yelena.stein@astro.unistras.fr}.}
\altaffiltext{5}{Departamento de Astronom{\'i}a Extragal{\'a}ctica, Instituto de Astrof{\'i}sica de Andalucia, Glorieta de la Astronom{\it i}a sn, 18008, Granada, Spain 
  {\tt adamas@iaa.es}.}
\altaffiltext{6}{Department of Astronomy, University of Michigan, 409 West Hall, 1085 S. University,
Ann Arbor, MI, USA,  48109 {\tt jiangtal@umich.edu}.}
\altaffiltext{7}{Dept. of Astronomy, University of Massachusetts, 710 North
Pleasant St., Amherst, MA, 01003, USA, {\tt wqd@astro.umass.edu}}
\altaffiltext{8}{Max-Planck-Institut f{\"u}r Radioastronomie,  Auf dem H{\"u}gel 69, 53121, Bonn, Germany,
{\tt mkrause@mpifr-bonn.mpg.de, pschmidt@mpifr-bonn.mpg.de}} 
\altaffiltext{9}{Faculty of Physics \& Astronomy, Astronomical Institute, Ruhr-University Bochum, 44780 Bochum, Germany
{\tt miskolczi@astro.rub.de, dettmar@astro.ruhr-uni-bochum.de}} 
\altaffiltext{10}{Dept. of Physics and Astronomy, University of New Mexico, 1919 Lomas Boulevard NE, Albuquerque, NM 87131, USA, 0000-0003-2048-4228
{\tt ttbraun@unm.edu}} 
\altaffiltext{11}{Inter-University Centre for Astronomy and Astrophysics, Post Bag 4, Ganeshkhind, Pune, Maharashtra, 411007, India
  {\tt djsaikia21jan@gmail.com}}
\altaffiltext{12}{Department of Physics \& Astronomy, University of Manitoba, MB, R3T 2N2, Canada {\tt jayanne\_english@umanitoba.ca}}
\altaffiltext{13}{McDonald Institute, 64 Bader Lane, Queen's University, Kingston, Ontario, Canada, K7L 3N6
  {\tt mark.richardson@mcdonaldinstitute.ca}}
%\and

%\author{Last name}
%\affil{Affiliation}

%% Notice that each of these authors has alternate affiliations, which
%% are identified by the \altaffilmark after each name.  Specify alternate
%% affiliation information with \altaffiltext, with one command per each
%% affiliation.

%\altaffiltext{1}{Visiting Astronomer, Cerro Tololo Inter-American Observatory.
%CTIO is operated by AURA, Inc.\ under contract to the National Science
%Foundation.}
%\altaffiltext{2}{Society of Fellows, Harvard University.}
%\altaffiltext{3}{present address: Center for Astrophysics,
%    60 Garden Street, Cambridge, MA 02138}
%\altaffiltext{4}{Visiting Programmer, Space Telescope Science Institute}
%\altaffiltext{5}{Patron, Alonso's Bar and Grill}

%% Mark off your abstract in the ``abstract'' environment. In the manuscript
%% style, abstract will output a Received/Accepted line after the
%% title and affiliation information. No date will appear since the author
%% does not have this information. The dates will be filled in by the
%% editorial office after submission.

\begin{abstract}

  The CHANG-ES galaxy sample consists of 35 nearby edge-on galaxies that have been observed using the VLA at 1.6 GHz and 6.0 GHz.  Here we present the 3rd data release of our sample, namely the B-configuration 1.6 GHz sample. In addition, we make available the {\it band-to-band} spectral index maps between 1.6 GHz and 6.0 GHz, the latter taken in the matching resolution C-configuration.  The images can be downloaded from https://www.queensu.ca/changes.  These are our highest resolution images ($\approx$ 3 arcsec) and we examine the possible presence of low luminosity active galactic nuclei in the sample as well as some in-disk structure.  New features can be seen in the spectral index maps that are masked in the total intensity emission, including hidden spiral arms in NGC~3448 and two previously unknown radio lobes on either side of the nucleus of NGC~3628. Our AGN detection rate, using only radio criteria, is 55\% which we take as a lower limit because some weaker embedded AGNs are likely present which could be revealed at higher resolution. Archival XMM-Newton data were used to search for further fingerprints of the AGNs in the studied sample.  In galaxy disks, discrete regions of flat spectral index are seen, likely due to a thermal emission fraction that is higher than the global average. 
\end{abstract}

%% Keywords should appear after the \end{abstract} command. The uncommented
%% example has been keyed in ApJ style. See the instructions to authors
%% for the journal to which you are submitting your paper to determine
%% what keyword punctuation is appropriate.


%% Authors who wish to have the most important objects in their paper
%% linked in the electronic edition to a data center may do so by tagging
%% their objects with \objectname{} or \object{}.  Each macro takes the
%% object name as its required argument. The optional, square-bracket 
%% argument should be used in cases where the data center identification
%% differs from what is to be printed in the paper.  The text appearing 
%% in curly braces is what will appear in print in the published paper. 
%% If the object name is recognized by the data centers, it will be linked
%% in the electronic edition to the object data available at the data centers  
%%
%% Note that for sources with brackets in their names, e.g. [WEG2004] 14h-090,
%% the brackets must be escaped with backslashes when used in the first
%% square-bracket argument, for instance, \object[\[WEG2004\] 14h-090]{90}).
%%  Otherwise, LaTeX will issue an error. 

\section{Introduction}
% Right now it goes from very specific (survey details) to the most broad. 
%change around a bit:

%1. Why are galaxy halos interesting
%2. what is the changes project and what are the goals
%3. how does this paper contribute
%4. brief survey and sample overview and refer people to paper 1 for details. 

This paper presents the third data release from the Continuum Halos in Nearby Galaxies -- an EVLA Survey (CHANG-ES) program. The Expanded Very Large Array (EVLA) is now known as the Karl G. Jansky Very Large Array (hereafter, the VLA).  CHANG-ES  is a large program that has observed 35 nearby edge-on galaxies in B, C, and D configurations of the VLA at two radio frequencies: 6 GHz (hereafter C-band) and 1.6 GHz (hereafter L-band).  Both frequencies were observed in C and D configurations and only L-band was observed in B configuration.

The scientific goals of the project are outlined in \cite{irwI} and cover a range of topics, including investigating the physical conditions and origin of halos, characterizing cosmic-ray transport and wind speed, measuring Faraday rotation and mapping the magnetic field,  
probing the in-disk and extraplanar far-infrared/radio-continuum relation, and exploring the prevalence of active galactic nuclei (AGNs) in nearby spiral galaxies.

Our first data release \citep{wie15} provided D configuration images, including total intensity images at two different uv-plane weightings, linear polarization (polarization angle and polarized intensity) images, and {\it in-band} spectral index maps\footnote{In-band spectral index maps are formed from the slope of the intensity across a single spectral band.} with their error maps.  The data release details are outlined in \cite{wie15} and images in FITS (Flexible Image Transport System)  format can be downloaded from {\tt https://www.queensu.ca/changes}. The second data release \citep{var19} provides FITS files of the H$\alpha$ images that were used in the thermal/non-thermal separation as described in \cite{var18}. The fourth CHANG-ES data release (in prep.)  will include images from our C configuration data at both frequencies \citep{wal19}. 

This third data release presents the B configuration L-band images (hereafter BL). In addition, we include {\it band-to-band} spectral index maps between matching resolution C configuration C-band data (hereafter CC) and the BL data.  

Since these data sets represent the highest resolution observations from the CHANG-ES complement ($\approx$ 3 arcsec), they are ideally suited to investigating compact, well-defined regions in the disk such as star forming regions, or active galactic nuclei (AGN).  In this paper, we focus mainly on AGNs, providing an estimate of the AGN frequency in our sample; we also discuss an example of radio emission from a compact region in a galaxy disk. % Of particular interest are the `dwarf' AGNs, i.e. low-luminosity AGNS (LLAGNs) whose radio luminosities, $L_{H\alpha}, fall below $10^{40}$ erg s$^{-1}$ \citep{ho97}

The galaxies and their positions with adopted distances are given in Table~\ref{tab:observationsB}.

{\renewcommand{\arraystretch}{1.1}
\begin{deluxetable}{ccccccccc}
\tabletypesize{\scriptsize}
%\rotate
\tablecaption{L-band B-Configuration Observations \label{tab:observationsB}}
\tablewidth{0pt}
\tablehead{
  \colhead{Galaxy} & \colhead{RA} & \colhead{DEC} & \colhead{Distance$^a$}& \colhead{Date$^b$} & \colhead{SB ID} & \colhead{Prim. Cal.$^c$} & \colhead{Zero Pol. Cal.$^c$} & \colhead{Sec. Cal.} \\
  \colhead{} &  \colhead{(h m s)} & \colhead{($^\circ$ $^\prime$ $^{\prime\prime}$)}& (Mpc)& \colhead{} &  \colhead{} & \colhead{}  & \colhead{} & \colhead{}
}
\startdata
N 660	& 01 43 02.40	& +13 38 42.2   & 12.3  & 24-Jun-12  & 9833270  & 3C48  & 3C84  & J0204+1514  \\
N 891	& 02 22 33.41	& +42 20 56.9   & 9.1   & 24-Jun-12  & 9833270  & 3C48  & 3C84  & J0230+4032  \\
N 2613	& 08 33 22.84	& -22 58 25.2   & 23.4  & 21-Mar-11  & 3699677  & 3C286 & OQ208 & J0846-2610  \\
N 2683	& 08 52 41.35	& +33 25 18.5   & 6.27  & 16-Jun-12  & 10459911 & 3C286 & OQ208 & J0837+2454  \\
N 2820	& 09 21 45.58	& +64 15 28.6   & 26.5  & 24-Jun-12  & 9872780  & 3C286 & OQ208 & J0921+6215  \\
N 2992	& 09 45 42.00	& -14 19 35.0   & 34    & 21-Mar-11  & 3699677  & 3C286 & OQ208 & J0943-0819  \\
N 3003	& 09 48 36.05	& +33 25 17.4   & 25.4  & 16-Jun-12  & 10459911 & 3C286 & OQ208 & J0956+2515  \\
N 3044	& 09 53 40.88	& +01 34 46.7   & 20.3  & 21-Mar-11  & 3699677  & 3C286 & OQ208 & J1007-0207  \\
N 3079	& 10 01 57.80	& +55 40 47.3   & 20.6  & 24-Jun-12  & 9878453  & 3C286 & OQ208 & J1035+5628  \\
N 3432	& 10 52 31.13	& +36 37 07.6   & 9.42  & 24-Jun-12  & 9872780  & 3C286 & OQ208 & J1130+3815  \\
N 3448	& 10 54 39.24	& +54 18 18.8   & 24.5  & 23-Jun-12  & 9878453  & 3C286 & OQ208 & J1035+5628  \\
N 3556	& 11 11 30.97	& +55 40 26.8   & 14.09 & 23-Jun-12  & 9878453  & 3C286 & OQ208 & J1035+5628  \\
N 3628	& 11 20 17.01	& +13 35 22.9   & 8.5   & 30-Jun-12  & 3746030  & 3C286 & OQ208 & J1120+1420  \\
N 3735	& 11 35 57.30	& +70 32 08.1   & 42    & 09-Jun-12  & 9872780  & 3C286 & OQ208 & J1313+6735  \\
N 3877	& 11 46 07.80	& +47 29 41.2   & 17.7  & 17-Jun-12  & 3752357  & 3C286 & OQ208 & J1219+4829  \\
	&       	&               &       & 04-Jul-12  &          &       &       &             \\
N 4013	& 11 58 31.38	& +43 56 47.7   & 16    & 11-Aug-12  & 9888001  & 3C286 & OQ208 & J1146+3958  \\
N 4096	& 12 06 01.13	& +47 28 42.4   & 10.32 & 11-Aug-12  & 9888001  & 3C286 & OQ208 & J1331+3030  \\
N 4157	& 12 11 04.37	& +50 29 04.8   & 15.6  & 17-Jun-12  & 3752357  & 3C286 & OQ208 & J1219+4829  \\
        &               &               &       & 04-Jul-12  &          &       &       &             \\
N 4192	& 12 13 48.29	& +14 54 01.2   & 13.55V& 24-Jun-12  & 3746030  & 3C286 & OQ208 & J1254+1141  \\
N 4217	& 12 15 50.90	& +47 05 30.4   & 20.6  & 11-Aug-12  & 9888001  & 3C286 & OQ208 & J1331+3030  \\
N 4244	& 12 17 29.66	& +37 48 25.6   & 4.4   & 09-Jun-12  & 9889954  & 3C286 & OQ208 & J1227+3635  \\
N 4302	& 12 21 42.48	& +14 35 53.9   & 19.41V& 29-Jul-12  & 9892023  & 3C286 & OQ208 & J1254+1141  \\
N 4388	& 12 25 46.75	& +12 39 43.5   & 16.6V & 29-Jul-12  & 9892023  & 3C286 & OQ208 & J1254+1141  \\
N 4438	& 12 27 45.59	& +13 00 31.8   & 10.39V& 29-Jul-12  & 9892023  & 3C286 & OQ208 & J1254+1141  \\
N 4565	& 12 36 20.78	& +25 59 15.6   & 11.9  & 02-Jun-12  & 9893234  & 3C286 & OQ208 & J1221+2813  \\
N 4594	& 12 39 59.43	& -11 37 23.0   & 12.7  & 17-Mar-11  & 3693390  & 3C286 & OQ208 & J1248-1959  \\
N 4631	& 12 42 08.01	& +32 32 29.4   & 7.4   & 03-Jun-12  & 9893234  & 3C286 & OQ208 & J1221+2813  \\
N 4666	& 12 45 08.59	& -00 27 42.8   & 27.5  & 10-Jun-12  & 9893236  & 3C286 & OQ208 & J1246-0730  \\
N 4845	& 12 58 01.19	& +01 34 33.0   & 16.98V& 10-Jun-12  & 9893236  & 3C286 & OQ208 & J1246-0730  \\
N 5084	& 13 20 16.92	& -21 49 39.3   & 23.4  & 17-Mar-11  & 3693390  & 3C286 & OQ208 & J1248-1959  \\
N 5297	& 13 46 23.68	& +43 52 20.5   & 40.4  & 09-Jun-12  & 9889954  & 3C286 & OQ208 & J1327+4326  \\
N 5775	& 14 53 57.60	& +03 32 40.0   & 28.9  & 05-Apr-11  & 3681627  & 3C286 & OQ208 & J1445+0958  \\
N 5792	& 14 58 22.71	& -01 05 27.9   & 31.7  & 05-Apr-11  & 3681627  & 3C286 & OQ208 & J1505+0326  \\
N 5907	& 15 15 53.77	& +56 19 43.6   & 16.8  & 08-Mar-11  & 3070258  & 3C286 & OQ208 & J1438+6211  \\
U10288	& 16 14 24.80	& -00 12 27.1   & 34.1  & 05-Apr-11  & 3681627  & 3C286 & OQ208 & J1557-0001  \\
\enddata
%% Text for table notes should follow after the \enddata but before
%% the \end{deluxetable}. Make sure there is at least one \tablenotemark
%% in the table for each \tablenotetext.
\tablecomments{Observations of the galaxies indicating the pointing center, distance, date, scheduling block (SB) identification number (ID), and primary, zero polarization leakage, and secondary calibrators.  Blanks mean that the value is the same as the previous row. }
\tablenotetext{a}{Distances are from \cite{wie15}. `V' designates a Virgo Cluster galaxy. }
\tablenotetext{b}{`Date' refers to the UT start date.}
\tablenotetext{c}{Alternate names for primary and zero polarization calibrators:  3C286=J1331+3030, 3C84=J0319+4130, OQ208 = QSO B1404+2841 or J1407+2827.}
\end{deluxetable}
} %end the arraystretch

In Sect.~\ref{sec:obs_red}, we describe the data acquisition and reductions for the radio and some supplementary X-ray data.  Sect.~\ref{sec:images_products} outlines the images and data products. Sect.~\ref{sec:images} presents some general results from the images. Sect.~\ref{sec:agns_general} lists our criteria for determining the presence of AGNs and compares these results to what has been previously known. Sect.~\ref{sec:disks} discusses the galaxy disks and possible contamination from background sources and Sect.~\ref{sec:conclusions} presents a summary and conclusions.

\section{Observations and Data Reductions}

\subsection{CHANG-ES VLA Data}
\label{sec:obs_red}

Table~\ref{tab:observationsB} provides observational details for our sample.  The 
observational set-up has previously been described in \cite{wie15} and earlier CHANG-ES papers.
The B-configuration L-band setup was consistent with L-band observations in the other  VLA configurations.  The frequency was centered at 1.575 GHz, with a bandwidth of 512 MHz.  The band was split into two sections, the first from 1.247 to 1.503 GHz and the second from 1.647 to 1.903 GHz, with the central gap set to avoid known interference. For all galaxies except NGC~4438, the central frequency was therefore $\sim$ 1.58 GHz\footnote{Very small variations in the central frequency of the resulting images may occur because of the flagging of radio-frequency-interference (RFI); the FITS header reveals the exact value.}. For NGC~4438, only the upper half of the frequency band was useable, hence its central frequency was 1.77 GHz.

The total 512 MHz bandwidth was divided into 32 spectral windows (spws), each with 64 channels for a total of 2048 spectral channels.  The high spectral resolution allowed us to identify and flag interference and was also adopted so that rotation measure (RM) analysis could be carried out at a later date, should the data warrant such an analysis. The spectral resolution also facilitated multi-frequency synthesis\citep[e.g.][]{sau99} and permitted the fitting of an in-band spectral index.

Approximately 2 hours of observing time were allocated for each galaxy.  Every galaxy was observed in a `scheduling block' (SB) that included other galaxies so that the total time on each galaxy could be spread out over a wide uv plane range, typically before and after transit.  The primary phase and amplitude calibrators (hereafter, the primary calibrators) are sources of known flux density for each baseline and frequency, and the secondary phase and amplitude calibrators (hereafter, the secondary calibrators) are sources that are closer than 10 degrees from the galaxy in the sky. The secondary calibrator was observed before and after each scan on the target galaxy and the primary calibrator was observed once per SB.  A zero polarization calibrator was also observed once during a SB
%in order to calibrate the polarization leakage terms
for the linear polarization calibration (see below). A list of calibrators is given in Table~\ref{tab:observationsB}.

The VLA uses right (R) and left (L) handed circular feeds.  Total intensity images (Stokes $I$) are formed from correlation of the parallel hands (RR and LL) and linear polarization images are formed from the correlation of the cross hands (RL and LR). A summary of the relation between the observed correlations and Stokes parameters is given in \cite{irwXI}. Each correlation was calibrated separately.

Data reductions were carried out using the {\it Common Astronomy Software Applications} package, CASA\footnote{See {\tt casa.nrao.edu}.}. \cite{irwII} and \cite{irwIII} describe the CHANG-ES data reduction process in detail. Briefly, calibration involved first Hanning smoothing the data in frequency and correcting for antenna-based delays.  The flux density scale was then set using the primary calibrator and the bandpass response was also corrected using the primary calibrator.  Complex gains (amplitudes and phases) were then determined as a function of time using the secondary calibrator.  At each step, the data were flagged for interference, as needed,  and the process was iterative such that the calibration was redone after each flagging episode.  In forming each new correction table, previously-determined tables were applied on the fly.

The polarization calibration required, in addition, determining the absolute position angle of the linearly polarized flux based on the known angle of the primary calibrator, determining any residual antenna-based delays for the cross-hands as well as solving for polarization leakage terms using the zero-polarization calibrator, according to standard practice\footnote{See {\tt https://evlaguides.nrao.edu/index.php?title=Category:Polarimetry}.}.

Imaging parameters are provided in Table~\ref{tab:imagingparameters}. 

{\renewcommand{\arraystretch}{1.1}
\begin{deluxetable}{lcccccccc}
\tabletypesize{\scriptsize}
%\rotate
\tablecaption{L-band Imaging parameters \label{tab:imagingparameters}}
\tablewidth{0pt}
\tablehead{
\colhead{Galaxy} &  \colhead{weighting\tablenotemark{a}} & \colhead{Beam size\tablenotemark{b}} &  \colhead{rms$_I$\tablenotemark{c}} & \colhead{I$_{max}$\tablenotemark{d}} & \colhead{rms$_{Q,U}$\tablenotemark{e}} & \colhead{$P_{max}$\tablenotemark{f}} &\colhead{$P$ comments\tablenotemark{g}} \\
                 &                      &(arcsec, arcsec, deg) & ($\mu$Jy/beam)& (mJy/beam)& ($\mu$Jy/beam) & ($\mu$Jy/beam) & (\%) 
}
\startdata 
N 660  & rob 0      & 3.39, 3.27, 44.4    & 24 (1.5$\times$)  &  245  & 18.2    & 134    & $P_{max}/I<0.5\%$\\
       & uvtap 16   & 6.14, 5.79, 76.1    & 28 (1.5$\times$)  &  299  & 21.0    & 132    & $P_{max}/I<0.5\%$ \\
N 891  & rob 0      & 3.15, 2.90, 54.2    & 16.0      & 4.91  & 18.0    & 89.3   & \\
       & uvtap 17   & 5.69, 5.33, 53.2    & 17.0      & 9.91  & 18.0    & 87.1   & \\
N 2613 & rob 0      & 5.18, 3.02, -179.7  & 19.6      & 0.39  & 18.5    & 85.3   & \\
       & uvtap 16   & 7.03, 6.05, -3.2    & 19.4      & 0.58  & 16.7    & 67.8   & \\
N 2683 & rob 0      & 3.06, 2.98, 57.8    & 14.5      & 1.06  & 14.6    & 65.7   & \\
       & uvtap 16   & 5.86, 5.77, 52.5    & 15.8      & 1.43  & 15.4    & 66.4   & \\
N 2820 & rob 0      & 3.23, 3.17, 52.8    & 16.3      & 0.87  & 16.8    & 73.4   & \\
       & uvtap 16   & 5.95, 5.92, 27.4    & 18.5      & 2.16  & 16.9    & 60.8   & \\ 
N 2992 & rob 0      &  4.87, 3.57, 16.4   & 16.5 (1.2$\times$)   & 84.3  & 16.7    & 78.8   &\\
       & uvtap 16   & 6.53, 6.33, -70.3   & 16.5 (1.3$\times$)   &  118  & 15.9    & 61.9   &\\
N 3003 & rob 0      &  3.11, 3.00, 70.1   & 14.0      & 0.77  & 15.1    & 67.2   &\\
       & uvtap 16   &  5.89, 5.76, 59.2   & 15.0      & 1.48  & 15.5    & 52.2   &\\
N 3044 & rob 0      & 3.67, 3.39, 68.8    & 15.0      & 3.36  & 14.7    & 71.8   &\\
       & uvtap 16   & 6.54, 5.54, 86.9    & 16.0      & 6.26  & 14.7    & 67.1   &\\
N 3079 & rob 0      & 3.14, 3.00, 58.4    & 18.0 (3$\times$)  &  123  & 16.4 (2$\times$) & 436   & $P_{max}/I=3.8\%$\\
       & uvtap 16   & 6.01, 5.80, 47.5    & 25.0 (5$\times$)  &  172  & 17.3 (3$\times$)  & 607   &$P_{max}/I=1.8\%$\\
N 3432 & rob 0      & 3.20, 3.12, 82.8    & 21.0      &  0.46 & 21.9    & 96.8   &\\
       & uvtap 16   & 5.95, 5.75, -23.8   & 24.0      &  0.84 & 22.5    & 97.8   &\\
N 3448 & rob 0      & 3.16, 2.98, 63.9    & 17.0      &  2.92 & 16.3    & 67.1   &\\
       & uvtap 17.5 & 5.59, 5.32, 52.9    & 18.0      &  5.94 & 17.0    & 60.8   &\\
N 3556 & rob 0      & 3.12, 2.98, 58.2    & 16.0      &  2.05 & 16.0    & 73.9   &\\
       & uvtap 17.5 & 5.56, 5.36, 49.5    & 16.5      &  3.47 & 16.5    & 75.8   &\\
N 3628 & rob 0      & 3.21, 3.13,  3.7    & 14.5 (1.6$\times$)  &  79.9 & 12.5    & 59.1   &\\
       & uvtap 16   & 6.01, 5.70, 86.2    & 20.0 (2$\times$)  &   134 & 13.9    & 68.2   &\\ 
N 3735 & rob 0      & 3.24, 3.11, 33.8    & 16.0      &  2.08 & 16.3    & 77.1   & $P_{max}\,<\,5\sigma_{Q,U}$\\
       & uvtap 16   & 6.01, 5.92, 27.5    & 16.5      &  4.08 & 15.7    & 88.3   & $P_{max}\,>\,5\sigma_{Q,U}$ \\
N 3877 & rob 0      & 3.01, 2.87, 22.4    & 11.5      &  1.26 & 11.4    & 50.2   &\\
       & uvtap 16   & 5.99, 5.86, 36.0    & 11.5      &  2.10 & 11.3    & 44.3   &\\
N 4013 & rob 0      & 3.01, 2.90, -84.2   & 14.0      &  4.14 & 13.7    & 64.4   &\\
       & uvtap 16   & 5.83, 5.79, 81.9    & 16.0      &  6.93 & 14.5    & 58.0   &\\
N 4096 & rob 0      & 3.06, 2.94, -84.9  & 14.5      &  0.24 & 14.8    & 70.6   &\\
       & uvtap 16   & 5.86 x 5.79, 70.6   & 14.5      &  0.55 & 15.8    & 76.7   &\\
N 4157 & rob 0      & 3.02, 2.84, 29.5    & 11.7 (2$\times$)  &  0.57 & 12.0    & 52.6   &\\
       & uvtap 16   & 5.98, 5.86, 37.3    & 12.0 (3$\times$)  &  1.75 & 11.5    & 52.2   &\\
N 4192 & rob 0      & 3.21, 3.07, -7.5    & 14.5 (1.2$\times$)  &  3.52 & 14.1    & 59.5   &\\
       & uvtap 16   & 6.00, 5.68, 83.6    & 22.0 (1.2$\times$)  &  6.25 & 14.2    & 65.5   &\\
N 4217 & rob 0      & 3.07, 2.94, -85.5   & 14.5      &  2.21 & 14.0    & 58.1   &\\
       & uvtap 16   & 5.86, 5.80, 76.7    & 15.1      &  3.94 & 15.0    & 57.0   & \\
N 4244 & rob 0      & 3.09, 3.00, 45.0    & 14.4      &  0.79 & 22.5    & 121.4  & noise peak\\
       & uvtap 16   & 5.86, 5.82, 40.0    & 15.2      &  1.27 & 22.5    & 120.3  & noise peak\\
N 4302$^j$ & rob 0  & 3.50, 3.13, -8.4    & 13.5      &  1.13 & 12.6    & 52.7   &\\
N 4388$^j$ & rob 0  & 3.57, 3.22, -2.1    & 16.0 (4$\times$)  &  25.2 & 12.3    & 59.0   &\\
N 4438$^{j,k}$& rob 0& 3.32, 2.91, -6.3    & 25.0 (1.6$\times$) &  35.7 & 26.0 (1.6$\times$)   & 306    & $P_{max}/I=1.1\%$ \\
N 4565 & rob 0      & 3.31, 3.01, 45.5    & 15.0      &  1.61 & 14.0    & 68.0  &\\
       & uvtap 17   & 5.87, 5.36, 50.9    & 15.0      &  1.81 & 14.0    & 59.1  &\\
N 4594 & rob 0      & 4.36, 3.25, -14.0   & 17.5      &  70.2 & 15.6    & 60.3  &\\
       & uvtap 16   & 6.09, 5.85, 87.7    & 17.5      &  71.0 & 14.4    & 52.7  &\\
N 4631 & rob 0      & 3.40, 3.05, 63.4    & 16.0      &  7.56 & 16.3    & 81.5  &\\
       & uvtap 16   & 6.08, 5.85, 57.1    & 20.0      &  11.2 & 15.6    & 68.9  &\\
N 4666$^j$ & rob 0  & 3.80, 3.48, 39.6    & 15.0 (1.5$\times$)  &  4.97 & 14.0    & 66.1  &\\
N 4845 & rob 0      & 3.51, 3.33, 22.7    & 18.0 (1.4$\times$)  &  209  & 15.0   &382 &   $P_{max}/I<0.5\%$      \\
N 5084 & rob 0      & 5.63, 2.95, -11.8   & 17.0 (1.15$\times$)  &  28.2 & 17.1    & 68.9  &\\
       & uvtap 16   & 6.87, 5.86, -8.9    & 16.0 (1.25$\times$)  &  30.4 & 14.8    & 67.8  &\\
N 5297 & rob 0      & 3.13, 2.99, 52.8    & 13.6      &  0.15 & 22.5    & 87.5  &\\
       & uvtap 16   & 5.92, 5.82, 54.0    & 15.8      &  0.29 & 22.9    & 75.6  &\\
N 5775 & rob 0      & 3.65, 3.44, 64.3    & 14.0      &  2.14 & 13.9    & 62.6  &\\
       & uvtap 16   & 6.47, 5.70, 84.7    & 18.0      &  4.94 & 14.2    & 67.9  &\\
N 5792 & rob 0      & 3.89, 3.42, 48.6    & 15.0 (1.25$\times$) &  4.89 & 14.3    & 48.9  &\\
       & uvtap 16   & 6.55, 5.74, 78.6    & 15.0 (1.25$\times$) &  8.95 & 14.5    & 47.0  &\\
N 5907 & rob 0      & 3.35, 2.79, -4.6    & 13.5     &  20.8$^l$ & 12.5& 53.8$^l$     &\\
       & uvtap 16   & 5.94, 5.72, 2.1     & 15.7     &  24.6$^l$ & 10.8& 48.6$^l$ &\\
U 10288& rob 0      & 3.80, 3.58, 66.2    & 14.0 (1.2$\times$) &  0.55$^m$ & 12.7& 52.0$^m$\\
\enddata
%% Text for table notes should follow after the \enddata but before
%% the \end{deluxetable}. Make sure there is at least one \tablenotemark
%% in the table for each \tablenotetext.
\tablecomments{}
\tablenotetext{a}{Weighting applied in the uv plane: rob 0 = Robust zero, uvtap = Robust zero plus an outer Gaussian uvtaper, where the number specifies the size in klambda units (see Sect.~\ref{sec:obs_red}).}
\tablenotetext{b}{Synthesized beam parameters of the total intensity images: Major axis, Minor axis, Position Angle. Values for Stokes Q and U are the same or negligibly different.}
\tablenotetext{c}{rms map noise of the total intensity images prior to primary beam correction. Values in parentheses indicate the factor by which the rms should be multiplied when considering values close to the source (Sect.~\ref{sec:noise}).}
\tablenotetext{d}{Peak specific intensity of the total intensity images, as measured on the galaxy itself, after primary beam correction (Sect.~\ref{sec:total_intensity_images}).} %as opposed to some background source or on a companion galaxy
\tablenotetext{e}{rms map noise of Stokes Q and U images prior to primary beam correction. Values in parentheses indicate the factor by which the rms should be multiplied when considering values to the source.}
\tablenotetext{f}{Peak linearly polarized specific intensity on the galaxy itself after primary beam correction (Sect.~\ref{sec:pol_images}). }
\tablenotetext{g}{Comments on polarization for cases in which $P_{max}$ is $>5\,\sigma_{Q,U}$ (Sect.~\ref{sec:pol_images}).}
\tablenotetext{j}{The uvtapered map was not included since it was of poor quality.}
\tablenotetext{k}{N~4438 was made using only half the total bandwidth (spw 16 to 31 only) for a central frequency of 1.77 GHz.}
\tablenotetext{l}{The peak is at the location of the double-lobed radio source that is superimposed at the far SE of the disk. The peak polarization measurement excluded this region.} 
\tablenotetext{m}{The peak value belongs to the southern radio lobe of the background quasar which shines through the galaxy, see \cite{irwIII}.}
\tablenotetext{n}{Peak is unresolved and at different locations in the rob 0 and uvtap 16 maps.}
\end{deluxetable}
}  %end arraystretch

Maps of Stokes $I$, $Q$, and $U$ were made as follows \cite[for information on Stokes $V$, see][]{irwXI}.

The multi-scale/multi-frequency synthesis (ms-mfs) algorithm \citep{rau11} was applied, including 
w-projection \citep{cor08} and the Cotton-Schwab clean \citep{sch84}.  A very wide field image  was initally made to identify background sources whose sidelobes could cause cleaning problems. In most cases, depending on the field, at least one full primary beam was finally imaged and sometimes much more. % thus the synthesized beam, which was typically $3$ arcsec across at L-band, was well sampled. %Pixels sizes were $0.5$ arcsec square for almost all galaxies. In a few cases (5 galaxies) they ranged up to 1 arcsec;  thus the synthesized beam, which was typically $3$ arcsec across at L-band, was well sampled.
%cells exceptions N891 =0.5 for rob0 but 1 for uvtap, 2613: 0.75 for both, 2820: rob0=1 and uvtap=0.5, 4192: 1arcsec for both, U10288: 0.75 for both
Two maps were made: one using Briggs robust 0 uv weighting as implemented in CASA \citep{bri95}  (hereafter referred to as `rob 0' maps) and one using robust 0 weighting with an additional outer Gaussian uv taper (hereafter `uvtap' maps or `uvtap' followed by the taper size).  The uv taper size was adopted so that the uvtap synthesized beam was roughly twice the diameter of the non-tapered map.  Almost all images used a cell size of 0.5 arcsec, hence the synthesized beam was well-sampled.  In a few cases, image results and/or efficiency were improved by using 0.75 or 1 arcsec cells.

We attempted self-calibration for most of the galaxies according to the prescription given in \cite{wie15}.  For many galaxies, however, the emission was too weak for effective self-calibration.  In the end, fifteen galaxies were effectively self-calibrated, either phase-only or amplitude and phase together. 

%The rms map noise varies with the field and will be discussed in Sect.~\ref{sec:total_intensity_images}.

During imaging/cleaning, a spectral index is fit across the band, producing maps of the {\it in-band} spectral index and associated error maps.  A fit of the form, $I_\nu\,\propto\,\nu^\alpha$ is performed, hence a straight-line in log space is fit across the 512 MHz L-band bandwidth. A $5\sigma$ cutoff was applied when making the in-band spectral index maps, where  $\sigma$ is the rms noise of the total intensity image.  
Since B configuration emission is generally weak, however, we are not releasing the in-band spectral index maps and instead are releasing  {\it band-to-band}, BL to CC spectral index maps (Sect.~\ref{sec:spectral_indices}). Users who wish to obtain the in-band spectral index maps for brighter sources should email the first author of this paper.  See also Sect. 3.4.3 of \citet{wie15} for more detailed discussion of in-band spectral index errors.

For polarization imaging, we form the linearly polarized intensity, $P_{lin}$, and polarization angle, $\chi$,  images from the Stokes $Q$ and $U$ maps according to,
\begin{eqnarray}
  P_{lin}&=&\sqrt{Q^2+U^2 - \sigma_{Q,U}^2}\label{eqn:lin_pol}\\
  \chi&=& \left(1/2\right)arctan\left(U/Q\right)
\end{eqnarray}
where $\sigma_{Q,U}$ is the rms noise of maps $Q$ and $U$; no significant difference was found between the $Q$ and $U$ noise. The latter term in Eqn.~\ref{eqn:lin_pol} makes a zeroth-order correction for the fact that $P$ images are positively biased \citep[e.g.][]{sim85, eve01,vai06} and is the only correction that is currently implemented in CASA.  For maps of $\chi$, we formed two different maps, one using a $3\sigma_{Q,U}$ cut off and one with a $5\sigma_{Q,U}$ cut off, for each uv weighting.

Finally, we corrected all maps, including in-band spectral index maps, for the new primary beam (PB) following \cite{per16}, whose full-width at half-maximum (FWHM) at our central frequency is 25.8 arcmin.

In this paper, we show total intensity images that are uncorrected for the PB (uniform noise) but make measurements from the images that have been corrected for the PB (corrected flux). Both are downloadable from our website.

\subsection{XMM Data}
\label{sec:xmm_obs}

To supplement our data and provide additional information about possible AGNs in the sample, we used XMM-Newton archive data and obtained 
spectra for 19 of our galaxies{\footnote {Twenty-four galaxies were observed but good spectra could not be extracted for five of them.}}. All data were processed with the SAS 15.0.0 package \citep{gabriel04}.
For each observation the event lists for two EPIC-MOS cameras \citep{turner01} and the EPIC-pn camera \citep{strueder01} were filtered for periods of intense background radiation and therefore prepared 
for the spectral analysis. The background spectra were obtained using blank sky event lists \citep[see][]{carter07}, filtered using the same procedures as for the source event lists.
For each spectrum, response matrices and effective area files were produced, and for a good sampling of the PSF of the instrument, a region size of 12\arcsec\, in radius was used.

The resulting spectra were merged using the SAS task $epicspeccombine$ into a final background subtracted source spectrum and then fitted using XSPEC~12 \citep{arnaud96}. 
Due to a limited resolution of the XMM-Newton it was impossible to extract the spectra in close vicinity to a central source.  Therefore, 
for the spectral analysis of the emission from the galactic core regions we used a model consisting of a gaseous component represented by a {\it mekal} model \citep{mewe85,kaastra92}, 
and a power-law model (absorbed, if needed) to account for the emission of the hot gas and the central source(s), respectively. Finally, in two galaxies, 
NGC\,4388 and NGC\,4666, where the iron Fe-K{$\alpha$} line was clearly visible in the extracted spectrum, the power-law component was accompanied by a simple Gaussian to fit this line.

As we are interested in supplementary evidence for AGNs in the CHANG-ES sample, we focus only on those galaxies that show evidence for an AGN from the XMM data of which there are eight galaxies.  Observational details for these galaxies are presented in Table~\ref{xdat}, the fitted models are in Table~\ref{models}, and the derived photon indices and luminosities of the central sources in given in Table~\ref{lumins}.  Note that all fluxes and luminosities derived from X-ray spectra refer to the energy range, 0.3 - 12 keV. A more thorough analysis of XMM data will be presented in a separate paper.  Other examples of XMM analyses for the galaxies NGC~4666 and NGC~4013 can be found in \citet{ste19} and \citet{ste19b}.

{\renewcommand{\arraystretch}{1.1}
\begin{deluxetable}{lccccc}
  \tablecaption{XMM-Newton X-ray observations of the selected galaxies \label{xdat}}
  \tablewidth{0pt}
\tablehead{
\colhead{Galaxy}&\colhead{ObsID}& \colhead{Obs. date}	&\colhead{column density} & \colhead{Total/clean$^b$}   & \colhead{Total/clean} \\
	&	&		&\colhead{$N_{\rm H}\,^a$}	& \colhead{MOS time$\,^c$ (ks)} & \colhead{pn time (ks)}
}
\startdata
N 660  & 0093641001	& 2001-01-07	& 4.64		                & 22/22	        & 7/7	        \\
       & 0671430101	& 2011-07-18	&				& 37/3	        & n/a	        \\
%NGC891	& 0670950101	& 2011-08-26	& 6.80				& 256/232	& 126/107	\\
N 2613 & 0149160101	& 2003-04-23	& 6.01				& 19/8	        & 12/1          \\
       & 0149160201	& 2003-05-20	&				& 59/54	        & 28/22         \\
N 2683 & 0671430201	& 2011-05-05	& 2.51				& 42/22	        & 22/6          \\
N 2992 & 0147920301	& 2003-05-19	& 4.87				& 55/47	        & 26/22         \\
N 3079 & 0110930201	& 2001-04-13	& 0.89				& 49/21	        & 20/5          \\
       & 0147760101	& 2003-10-14	& 				& 82/66	        & 39/14         \\
N 3628 & 0110980101	& 2000-11-27	& 1.97				& 109/98        & 50/30         \\
%NGC4013 & 0306060201	& 2005-11-13	& 1.26				& 156/124	& 56/53		\\
       & 0306060301	& 2005-11-15	& 				& 34/34		& 15/14		\\
%NGC4157 & 0203170101    & 2004-05-16	& 2.02				& 102/74	& 48/34		\\
%NGC4302 & 0306060101	& 2005-12-05	& 2.53				& 192/176	& 93/76		\\
N 4388 & 0110930701	& 2002-12-12	& 2.58				& 23/23	        & 8/7           \\
%NGC4438 & 0210270101	& 2004-12-19	& 2.31				& 52/52		& 25/25		\\
%	& 0741160301	& 2015-01-16	&				& 98/94		& 45/38		\\
%NGC4565 & 0112550301	& 2001-07-01	& 1.17				& 28/28		& 10/10		\\
N 4594 & 0084030101	& 2001-12-28	& 3.70				& 85/85	        & n/a	        \\
%NGC4631 & 0110900201	& 2002-06-28	& 1.30				& 108/96	& 50/42		\\
N 4666 & 0110980201	& 2002-06-27	& 1.73				& 116/116       & 54/54         \\
N 4845 & 0658400601	& 2011-01-22	& 1.44				& 42/41         & 19/12         \\
%NGC5775 & 0150350101	& 2003-07-28	& 3.54				& 82/60		& 38/21		\\
%NGC5907 & 0145190201    & 2003-02-20	& 1.21				& 88/53		& 41/14		\\
% 	& 0145190101	& 2003-02-28	& 				& 100/43	& 47/12		\\
\enddata
\tablecomments{See Sect.~\ref{sec:xmm_obs}.}
\tablenotetext{a}{Column density in (10$^{20}$ cm$^{-2}$) weighted average value after LAB Survey of Galactic HI \citet{kalberla05}.}
\tablenotetext{b}{Filtered for high background radiation.}
\tablenotetext{c}{MOS1+MOS2.}
\end{deluxetable}
}  %end arraystretch

{\renewcommand{\arraystretch}{1.1}
%\begin{table*}
%\centering       
\begin{deluxetable}{lcc}
  \tablecaption{Model type and reduced $\chi_{\rm red}^2$.}
\tablewidth{0pt}
\tablehead{ 
\colhead{Galaxy}  & \colhead{model} 			& \colhead{$\chi_{\rm red}^2$} \\
}
\startdata
NGC660 	& wabs(mekal+wabs*powerlaw) 			& 1.19 	\\
%NGC891	& wabs(mekal+mekal+wabs*powerlaw)		& 0.95	\\
NGC2613	& wabs(mekal+powerlaw+wabs*powerlaw) 		& 1.13  \\
NGC2683	& wabs(powerlaw+wabs*powerlaw) 			& 1.57  \\
NGC2992	& wabs(powerlaw+wabs*powerlaw)			& 1.21	\\
NGC3079	& wabs(mekal+mekal+powerlaw)			& 1.40	\\	
NGC3628	& wabs(mekal+wabs*powerlaw)			& 1.16	\\	
%NGC4013 & wabs(mekal+powerlaw)				& 1.21	\\
%NGC4157 & wabs*powerlaw				& 1.08  \\
%NGC4302 & wabs(mekal+powerlaw)				& 1.24  \\
NGC4388	& wabs(mekal+powerlaw+wabs(powerlaw+gauss))	& 1.31	\\
%NGC4438 & wabs(mekal+mekal+wabs*powerlaw)		& 0.76  \\
%NGC4565 & wabs*wabs*powerlaw				& 0.90  \\
NGC4594	& wabs*wabs*powerlaw				& 0.94	\\
%NGC4631 & wabs(mekal+powerlaw)				& 0.88  \\
NGC4666	& wabs(mekal+mekal+wabs(powerlaw+gauss))	& 1.35	\\
NGC4845	& wabs*wabs*powerlaw				& 1.23	\\
%NGC5775 & wabs(mekal+powerlaw)				& 1.04  \\
%NGC5907 & wabs*powerlaw				& 1.29  \\
\enddata
\tablecomments{Models: {\it wabs} - a photoelectric absorption using Wisconsin \citep{mor83}
cross-sections; {\it mekal} - emission from the
hot gas, based on the calculation of \cite{mewe85} and \cite{kaastra92}; {\it powerlaw} - a simple photon
power law model.}
\label{models}
\end{deluxetable}
} %end arraystretch

{\renewcommand{\arraystretch}{1.1}
%\begin{table*}
%\centering       
\begin{deluxetable}{lcc}
  \tablecaption{\label{lumins} Photon indices and luminosities of the central sources in the studied galaxies.}
\tablewidth{0pt}
\tablehead{ 
  \colhead{Region}  & \colhead{Photon} 	& Luminosity \\
                    & \colhead{Index$^a$}   & (10$^{40}$\,erg/s)
}
\startdata
NGC660  & 2.45$^{+0.81}_{-0.63}$& 0.47$^{+1.09}_{-0.20}$ \\
%NGC891	& 1.65$\pm$0.11		& 0.10$^{+0.03}_{-0.02}$ \\	
NGC2613 & 1.89$^{+1.71}_{-0.86}$& 1.84$^{+347}_{-1.73}$  \\
NGC2683 & 1.69$^{+1.05}_{-0.92}$& 0.92$^{+27.5}_{-0.83}$ \\
NGC2992 & 1.66$\pm$0.05		& 764$^{+57.6}_{-50.8}$  \\
NGC3079 & 1.46$^{+0.10}_{-0.12}$& 0.88$^{+0.26}_{-0.19}$ \\
NGC3628 & 1.46$^{+0.07}_{-0.06}$& 0.73$^{+0.13}_{-0.11}$ \\
%NGC4013 & 1.18$^{+0.22}_{-0.23}$& 0.07$^{+0.06}_{-0.03}$ \\
%NGC4157 & 1.15$^{+0.13}_{-0.12}$& 0.12$^{+0.04}_{-0.03}$ \\
%NGC4302 & 1.62$^{+0.37}_{-0.47}$& 0.04$^{+0.05}_{-0.02}$ \\
NGC4388 & 1.41$^{+0.12}_{-0.11}$& 230$^{+111}_{-72.8}$   \\
%NGC4438 & 2.36$^{+0.30}_{-0.34}$& 0.22$^{+0.12}_{-0.08}$ \\
%NGC4565 & 1.94$^{+0.30}_{-0.34}$& 0.44$^{+0.16}_{-0.11}$ \\
NGC4594 & 1.76$\pm$0.05		& 3.26$^{+0.36}_{-0.31}$ \\
%NGC4631 & 1.22$^{+0.24}_{-0.26}$& 0.02$\pm$0.0		 \\	
NGC4666 & 1.77$^{+0.17}_{-0.16}$& 2.55$^{+1.23}_{-0.99}$ \\
NGC4845 & 2.09$\pm$0.03		& 539$^{+39.2}_{-35.9}$  \\
%NGC5775 & 0.90$^{+0.26}_{-0.25}$& 0.44$^{+0.45}_{-0.23}$ \\
%NGC5907 & 1.14$\pm$0.16	& 0.16$^{+0.05}_{-0.04}$ \\
\enddata
\tablenotetext{a}{Energy spectral index of photon counts (0.3 - 12 keV).}

\end{deluxetable}
} %end arraystretch

\section{Images and Data products}
\label{sec:images_products}

For the BL data release, we have regridded all images to the same pixel size (0.5 arcsec square) and trimmed to the same field size: 2000 x 2000 pixels, or 16.7 arcmin on a side.  %The maximum distance from the field center for any background source is therefore 23.6 arcmin.
Panels and a description of these data products are given in Appendix~\ref{app:B} and a sample image is in Fig.~\ref{fig:sample_panel}. An explanation is provided below.

\subsection{Noise Measurements}
\label{sec:noise}

The rms noise values, measured prior to PB-correction, are given in Table~\ref{tab:imagingparameters}.  The rms values show some variation with position on the image, depending on the locations and strengths of the sources in the field. For consistency between maps and also consistent with the approach taken in \cite{wie15}, we quote rms values in Table~\ref{tab:imagingparameters} that are far from the center. Because most sources are quite weak in B-configuration, residual sidelobes from cleaning were usually not excessive and the quoted rms noise variations over the map were typically $\ltabouteq\,10\%$.

Exceptions are also present, though. For example, sources whose peak brightness exceeded $70$ mJy/beam contained some residual sidelobes that remained after cleaning.   
Sources which were near other bright background sources in the sky, or sources in more crowded fields in general had rms noise values that were larger closer to the target galaxy than in regions more distant from the center.   Sources that fell into these categories have their rms values followed by an approximate factor by which the rms increases near the source itself.

Finally, the sources sometimes are found in a `negative bowl' which is expected when broad scale flux is missing ($>$ 2 arcmin in L-band at B configuration), as is the case for many of our galaxies.

{ Aside from the above-mentioned variations, the rms noise values on maps have been listed with an accuracy that reflects measurements on the maps themselves.  Stokes Q and U maps, for example, show rms noise values that are very `clean' with little relative variation from location to location on any given map.  In such cases, it is possible to measure the rms to a fraction of a $\mu$Jy/beam.  The reader is reminded, however, that the absolute flux calibration scale at the VLA for the primary calibrator, 3C~286, is $\approx$ 1\% \citep{per13}. Individuals who download the images may wish to make their own measurements according to their scientific goals.}

\subsection{Total Intensity Images}
\label{sec:total_intensity_images}

The total intensity images (Stokes $I$) are shown in panels in the {first} row of Appendix~\ref{app:B} and Fig.~\ref{fig:sample_panel}. The corresponding  rms values and peak brightnesses ($I_{max}$)  are given in Table~\ref{tab:imagingparameters}.  %For clarity and to represent uniform noise, we show images prior to PB-correction, but make measurements on PB-corrected maps.
Note that the peak value on a map could occur in an off-galaxy background source, so measurements of $I_{max}$ were restricted to the region of the galaxy, itself.  

%\begin{figure}[!tbp]
\begin{figure*}[htp]
  \centering
  \includegraphics[width=\textwidth]{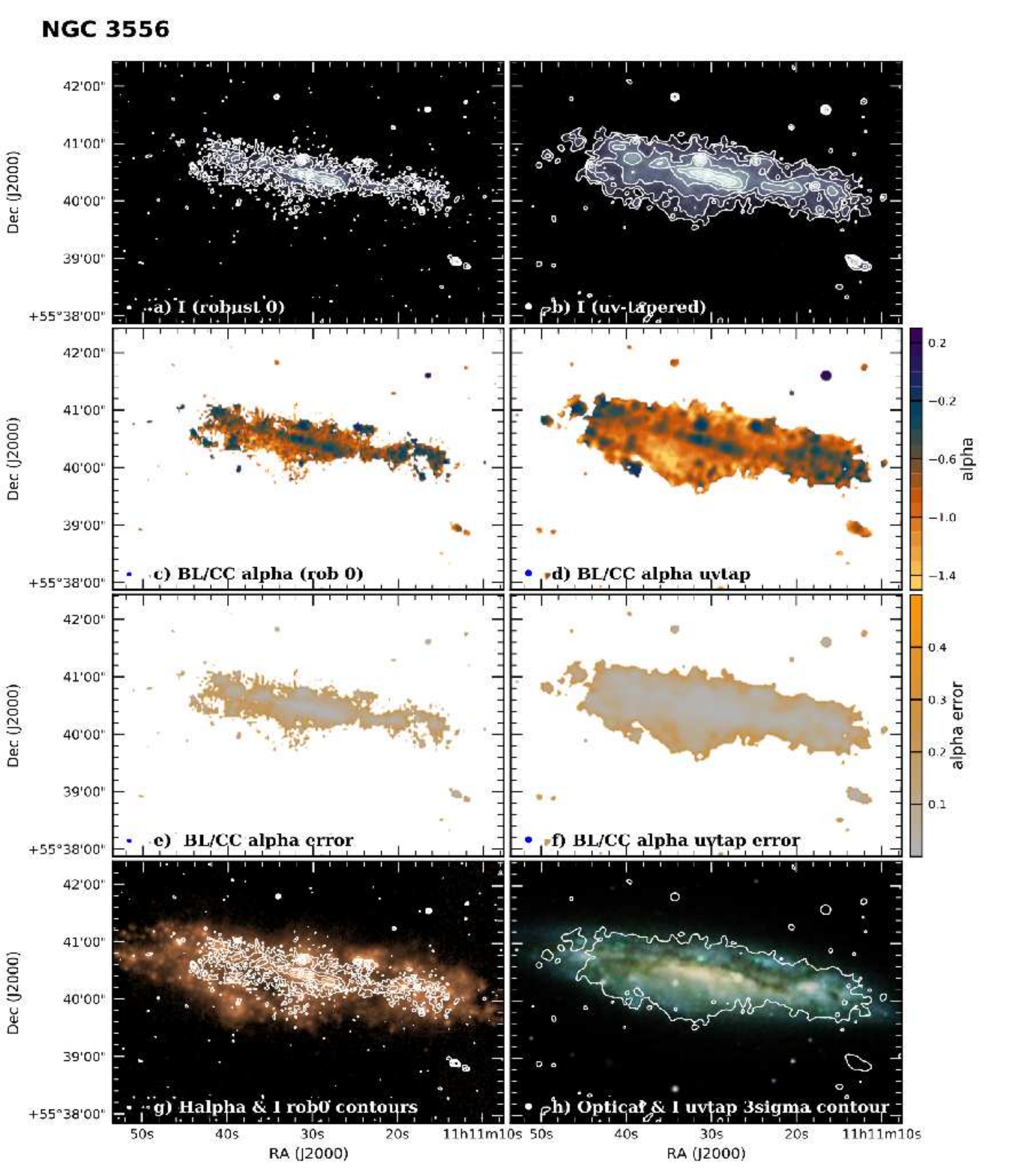}
  \caption{Sample panel from Appendix~~\ref{app:B}.}
\label{fig:sample_panel}
\end{figure*}

At the high resolution achieved for B configuration, though, and given the fact that steep-spectrum background sources will usually be brighter at L-band than at C-band, it is entirely possible that a discrete source seen in the region of the disk of a galaxy could in fact be a background source. We explore this possibility further in Sect.~\ref{sec:background_sources}.

In several cases, uvtapered images were not included because of poorer image quality; generally this was because of residual sidelobes either from the source itself or from an interfering background source.

\subsection{Linear Polarization Images}
\label{sec:pol_images}

In almost all cases, the linearly polarized images are nothing more than noise maps. As with the total intensity images, we measured the peak polarization, $P_{max}$, from the PB-corrected maps in the region of the galaxy itself.  A scan down this column of Table~\ref{tab:imagingparameters} shows that almost every galaxy displays {\it no} emission above 5$\sigma_{Q,U}$.  Those galaxies that show higher values of $P_{max}$ have comments in the last column as discussed next.

If there is emission above 5$\sigma_{Q,U}$ but the peak values for rob 0 and uvtapered images are at different locations, then they are labelled as `noise peaks', such as is the case for NGC~4244. For this galaxy, these peaks are unresolved points and have the appearance of random noise.  The total intensity values at these peaks also show only noise. 

If there is emission above 5$\sigma_{Q,U}$ in one uv weighting but below 5$\sigma_{Q,U}$ in the other weighting, then the emission is also not considered to be real.  Such is the case for NGC~3735.

If higher emission is seen at both weightings, we then calculate the ratio, $P_{max}/I$ (\%) at the location of  $P_{max}$.  Since the polarization calibration is not considered to be reliable below about 0.5\% \citep{irwV}, signals that fall below this limit are also not considered to be real.  This is the case for NGC~660 and NGC~4845, for example.

\begin{figure}[!tbp]
  \centering
  \begin{minipage}[b]{0.55\textwidth}
    \includegraphics[width=\textwidth,trim=0 12cm 0 0,clip]{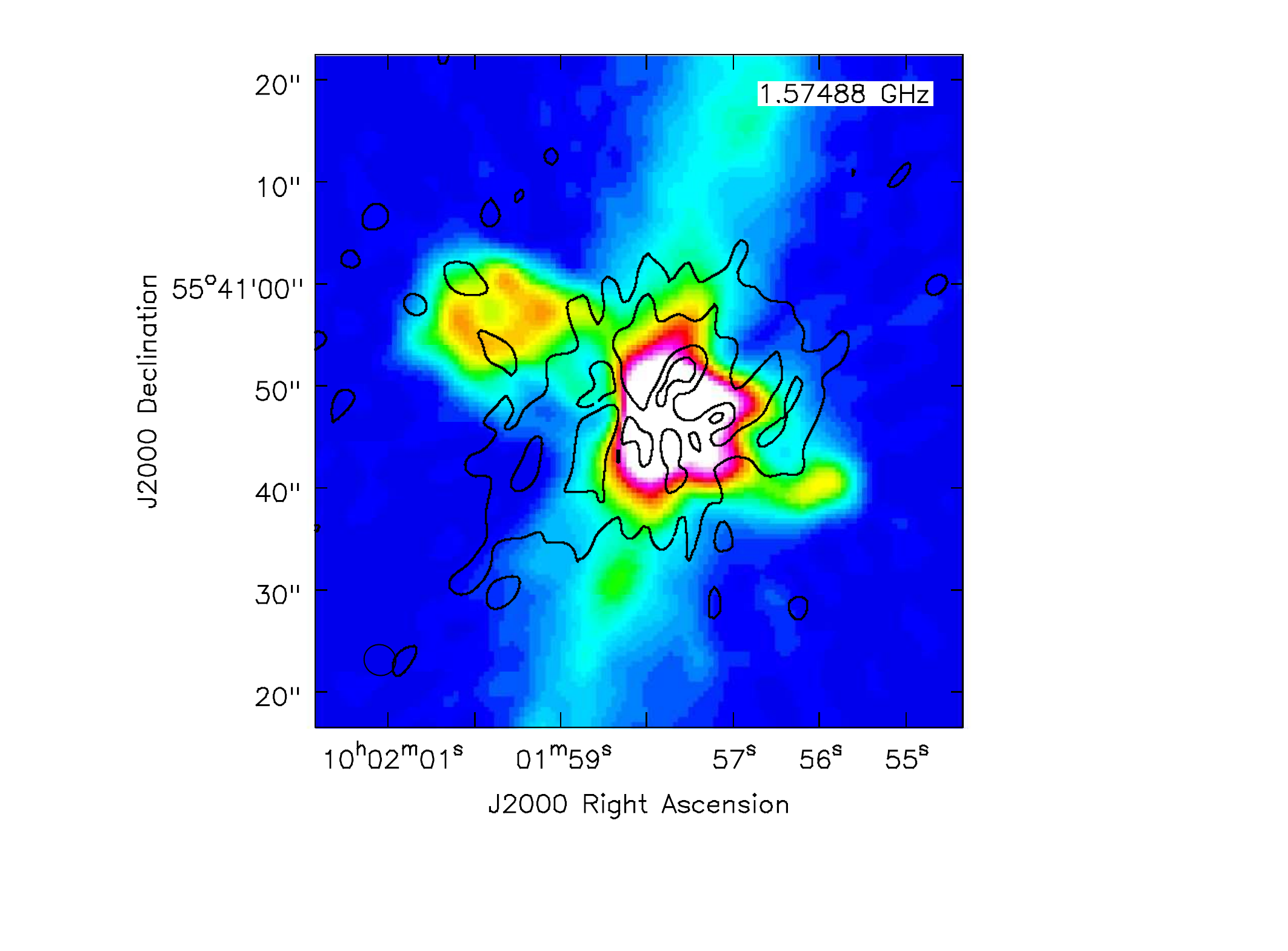}
  \end{minipage}
%\vspace{-2truein}
  \hfill
  \begin{minipage}[b]{0.55\textwidth}
    \centering
    \includegraphics[width=\textwidth,trim=0 12cm 0 0,clip]{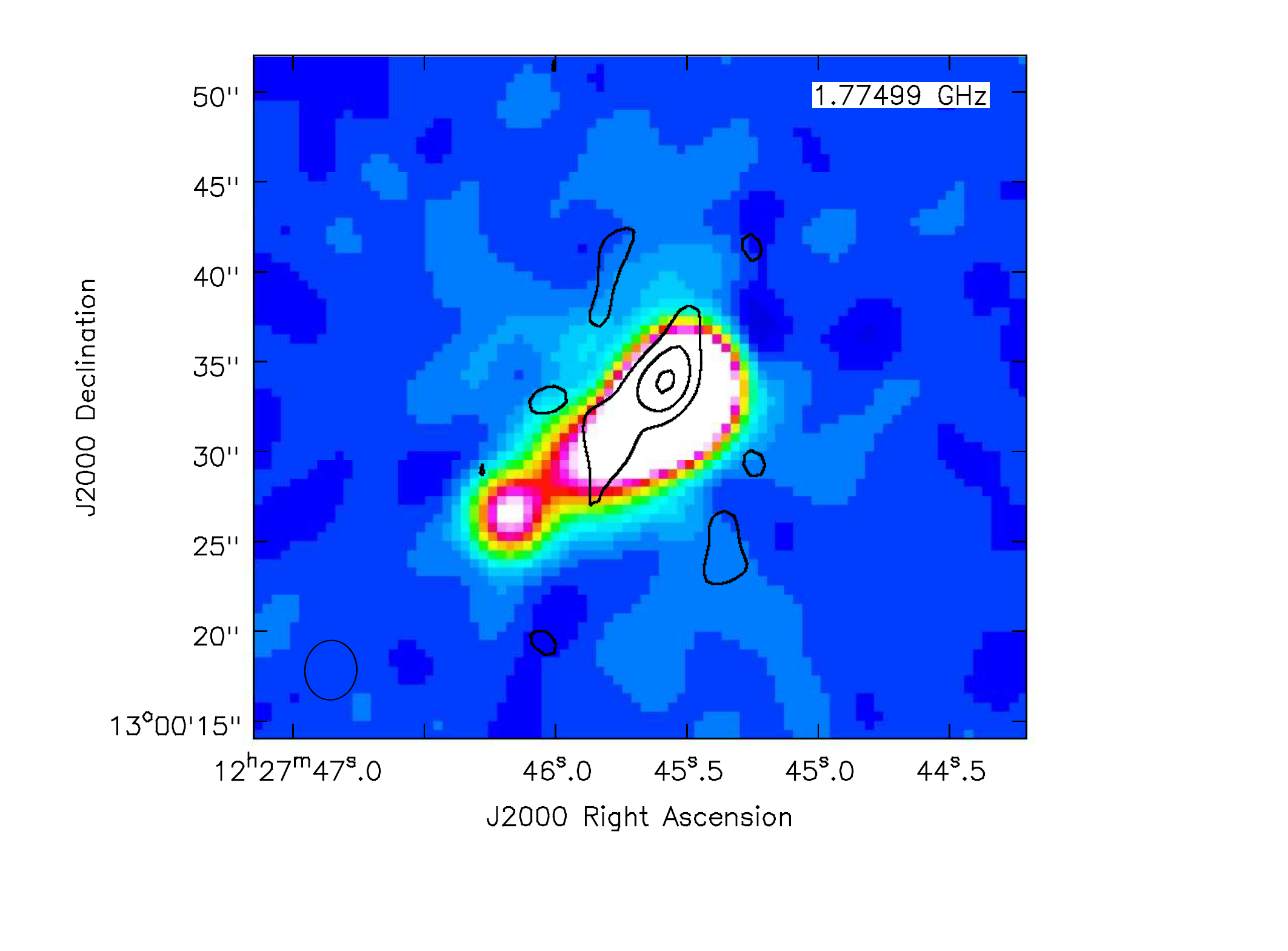}
  \end{minipage}
%\vspace{-2truein} 
  \caption{L-band rob 0 linear polarization (contours) over the total intensity image (colour). The synthesized beam is shown as a black circle at lower left and frequency is given at upper right. 
   {\bf (a)} NGC~3079. Contours are at
3, 7, and 15 $\times\,\sigma_{Q,U}$ ($\sigma_{Q,U} = 24.6~\mu$Jy beam$^{-1}$, Table~\ref{tab:imagingparameters}).   The major axis extends NW to SE. {\bf (b)} NGC~4438.  Contours are at
3, 5, and 7 $\times\,\sigma_{Q,U}$ ($\sigma_{Q,U} = 41.6~\mu$Jy beam$^{-1}$, Table~\ref{tab:imagingparameters}).  }
\label{fig:lin_pol}
\end{figure}

We are finally left with only two galaxies that could have real linearly polarized BL emission, these being NGC~3079 and NGC~4438, both of which are displayed in Fig.~\ref{fig:lin_pol}. These galaxies are known to have strong nuclear activity.

In summary, given how little believable linear polarization is seen in the BL data, we do not release any polarization images. By contrast, the matching resolution CC data show quite significant polarization for most galaxies \citep{wal19}, so it is clear that the lower frequency L-band emission suffers from Faraday de-polarization.  It is possible that some L-band linear polarization could be reclaimed via a rotation measure synthesis analysis \citep[e.g.][]{dam16}.  %We also know that some galaxies with AGNs have had linearly polarized flux in the AGN converted to circular polarization \citep{irwXI}.
Further such analysis is, however, beyond the scope of the current paper.

%NGC~3735 shows a very weak linearly polarized emission peak greater than 5$\sigma_{Q,U}$ in the uvtapered image at location, RA = 11 35 58.64, DEC = 70 32 10.7 (black contours).  An elongated feature is also seen in the rob 0 image (red contours), though only at a 4$\sigma_{Q,U}$ level. This peak is seen independently in the C configuration C-band image as well (not shown).

%{
%Duric et al. 1983 found no polarization at L-band for N3079 -- note the strong gradient in total intensity for argument against reality..
%uvtap of N3079:  within a single beam centered at the P peak (for uvtap) the I goes from 106.4 mJy/beam to 14.9 mJy/beam so a gradient of %a factor of 7.1
%for rob0: from 48.2 mJy/beam to 6.07 so factor of 7.9
%For example, I measure the flux density within a beam and take those ratios, then I get rob0 = 2.8856e-4/1.13e-2 *100 = 2.55\%
%and for uvtap = 
%3.50e-3/2.06e-2*100=16.99\%
%But suppose I move the circle by half a beam.  then the flux density ratio is
%uvtap=3.11e-4/1.09e-2*100 = 2.8\%  I think you just have to say that the apparent signal is along a strong gradient and is unlikely to be %real.}

\subsection{BL to CC (Band-to-Band) Spectral Index Maps}
\label{sec:spectral_indices}

Spectral indices, $\alpha$, are defined according to
\begin{equation}\label{eqn:alpha_defn}
I_\nu\,\propto\,\nu^\alpha
  \end{equation}

Because of the weakness of most of the sources in B-configuration, {\it in-band} spectral indices were noisier than desired for most galaxies.  An example of a BL in-band spectral index map for one of our stronger sources, however, can be seen in \cite{irwV}.

We therefore made {\it band-to-band} spectral index maps from the BL central frequency of $\nu_L\,=\,1.58$ GHz)  to the similar resolution CC central frequency of $\nu_C\,=\,6.00$ GHz (the latter data have a 2 GHz bandwidth).  The resulting spectral index maps ($\alpha$ maps) have a higher signal-to-noise (S/N), but no variation in $\alpha$  between the bands can be recovered, should such variations be present. 

Although the BL and CC data sets were very close in resolution and pixel size, some minor processing was required.  The non-PB-corrected images were first regridded so that both had the same (smaller) pixel size.  Then minor  smoothing ($<$ 2 arcsec) was carried out so that the two images had exactly the same (larger) spatial resolution.  New rms values ($\sigma$ in the following discussion) were then measured for the BL and CC data sets.  These rms values  and adjusted beam sizes are listed in Table~\ref{tab:smoothed_data}.

{\renewcommand{\arraystretch}{1.1}
\begin{deluxetable}{lcccccccccc}
\tabletypesize{\scriptsize}
%\rotate
\tablecaption{Smoothed BL and CC map parameters as input for Band-to-Band spectral index maps \label{tab:smoothed_data}}
\tablewidth{0pt}
\tablehead{
  \colhead{Galaxy}   & \multicolumn{5}{c}{Robust 0 maps }& \multicolumn{5}{c}{Uvtapered maps}\\
  &\colhead{BL rms}   & \colhead{CC rms} & \colhead{bmaj}&\colhead{bmin}&\colhead{bpa}&
  \colhead{BL rms}   & \colhead{CC rms} & \colhead{bmaj}&\colhead{bmin}&\colhead{bpa}\\
      &  ($\mu$Jy/beam) & ($\mu$Jy/beam) & (arcsec) & (arcsec) & (deg) & ($\mu$Jy/beam) & ($\mu$Jy/beam) & (arcsec) & (arcsec) & (deg) \\
}
\startdata 
      N 660   & 20.6 & 3.85 &  4.0 &  3.5 &   -5   & 28.6 & 5.06 &  6.8 &  6.2  &  10   \\
      N 891   & 16.2 & 2.81 &  3.8 &  3.3 &   70   & 19.5 & 3.61 &  6.2 &  6.0 &   60   \\
      N 2613  & 18.1 & 32.3 &  7.5 &  5.5 &  -100  & 19.3 & 3.56 &  7.6 &  6.6 &   -7   \\
      N 2683  & 14.1 & 4.64 &  3.4 & 3.3  &  13   & 17.7 & 5.62 &  6.5 &  6.3 &   12  \\
      N 2820  & 15.6 & 2.58 & 3.6 &  3.5 &   21   & 17.5 & 2.97 & 6.3 & 6.3 &   56   \\
      N 2992  & 16.3 & 3.16 & 5.3 &  3.9 &   4    & 17.0 & 3.73 & 6.9 & 6.6 &  -40   \\
      N 3003  & 13.9 & 15.2 &  3.8 &  3.4 &    0   & 15.8 & 2.87 &  6.3 &  6.25&    0   \\
      N 3044  & 15.5 & 3.57 &  4.1 &  3.9 &   15   & 18.1 & 4.34 &  7.5 &  6.6 &    3   \\
      N 3079  & 18.7 & 2.73 &  3.9 &  3.8 &  -15   & 30.0 & 3.73 &  6.8 &  6.4 &  -12   \\
      N 3432  & 20.7 & 2.97 & 3.4 & 3.4 &    2    & 22.6 & 4.45 & 6.5 & 6.3 & -5   \\
      N 3448  & 15.5 & 2.57 &  3.7 &  3.5 &   62   & 17.5 & 3.06 & 6.1 & 6.0 &   1   \\
      N 3556  & 15.8 & 2.62 &  3.5 &  3.3 &   49   & 16.8 & 3.03 & 6.1 & 6.0 &   0.5   \\
      N 3628  & 14.6 & 4.28 &  4.2 &  3.8 &  -37   & 21.9 & 5.80 &  6.8 &  6.4 &   45   \\
      N 3735  & 15.7 & 2.49 & 3.6 &  3.5 &   29    & 18.8 & 3.64 & 6.5 & 6.4 &   -16   \\
      N 3877  & 10.3 & 2.79 &  3.6 &  3.3 &  -27   & 11.6 & 3.44 &  6.5 &  6.4 &   67   \\
      N 4013  & 13.7 & 2.39 &  3.6 &  3.3 &  -85   & 18.1 & 2.97 &  6.4 &  6.2 &   11   \\
      N 4096  & 14.3 & 2.83 &  3.4 &  3.3 &  -80   & 16.1 & 3.07 &  6.4 &  6.3 &   80   \\
      N 4157  & 11.1 & 2.88 &  3.3 &  3.1 &  -27   & 11.8 & 3.44 &  6.4 &  6.3 &  -20   \\
      N 4192  & 14.3 & 3.01 &  3.5 &  3.3 &  -37   & 19.7 & 3.75 &  6.4 &  6.1 &   84   \\
      N 4217  & 14.0 & 2.65 &  3.4 &  3.2 &  -82   & 15.6 & 3.06 &  6.2 &  6.1 &    0   \\
      N 4244  & 13.7 & 2.41 &  3.4 &  3.3 &  -18   & 15.3 & 2.56 &  6.3 &  6.2 &   20   \\
      N 4302  & 13.2 & 2.96 & 3.9 & 3.7 &  -38    &  &  &  &  &      \\
      N 4388  & 17.6 & 2.93 & 4.0 & 3.7 &    -6   & & &      &      &        \\
      N 4438  & 378 & 29.1 & 3.6 & 3.4 &   -4    & & &      &      &        \\
      N 4565  & 14.6 & 2.62 &  3.6 &  3.3 &   -9   & 14.7 & 3.88 &  6.3 &  6.0 &   61 \\
      N 4594  & 17.0 & 2.69 & 4.8 &  3.7 &  -9    & 21.3 & 63.8 & 6.5 & 6.4 &  34   \\
      N 4631  & 16.0 & 2.77 &  3.7 &  3.5 &   -7   &  &  & & &    \\
      N 4666  & 16.9 & 3.38 &  4.1 &  3.8 &   7   &  &  &  &  &    \\
      N 4845  & 19.2 & 3.64 &  3.8 &  3.6 &   6   & & &      &      &        \\
      N 5084  & 17.0 & 2.34 &  5.9 &  3.3 &   -8   & 18.2 & 2.78 &  7.8 &  6.8 &   39   \\
      N 5297  & 14.1 & 2.60 &  3.4 &  3.3 &  -10   & 16.7 & 2.83 &  6.3 &  6.2 &   -2   \\
      N 5775  & 13.4 & 2.80 &  4.0 &  3.8 &   29   &  &  &  &  &     \\
      N 5792  & 14.7 & 2.89 &  4.2 &  3.8 &   21   & 16.1 & 3.83 &  7.5 &  6.7 &   -3   \\
      N 5907  & 12.2 & 2.53 &  3.7 &  3.2 &   40   & 11.8 & 3.58 & 6.3 & 6.2 &  -37   \\
      U 10288 & 13.5 & 2.88 &  4.2 &  3.9 &   28   &  &  &  &  &    \\ \hline
      \enddata
      \tablecomments{The matching beams for BL and CC are designated `bmaj', `bmin', and `bpa' for the FWHM of the major axis, the FWHM of the minor axis, and the position angle, respectively.}
    \end{deluxetable}

This process was then repeated for the PB-corrected images from which the spectral index maps were actually made.  A 3$\sigma$ cutoff was then applied to each of the BL and CC images prior to making band-to-band spectral index maps.  This cutoff was adopted because tests showed that a higher 5$\sigma$ cutoff removed some real features.  Moreover, since noise peaks higher than 3$\sigma$ are usually in different locations at the two frequencies, the 3$\sigma$ cutoff produced the best results.

It should be pointed out that
 there are still artifacts in a few spectral index maps.  This is because PB-corrected images have increasing noise with distance from the map center and some residual sidelobes remained for sources with strong emission.  Examples are NGC~660 and NGC~3079 in Appendix~\ref{app:B}. 

The final band-to-band spectral index maps shown in the Second row of Appendix~\ref{app:B} apply from 1.6 to 6.0 GHz.

\subsubsection{Uncertainties in Spectral Index Maps}
\label{sec:alpha_errors}

Spectral index error maps were formed using
\begin{equation}
  \label{eqn:error_maps}
  \sigma_\alpha\,=\,\frac{1}{ln\left(\frac{\nu_L}{\nu_C}\right)}
  \sqrt{\left(   \frac{\sigma_L}{I_L}   \right)^2\,+\,
  \left(\frac{\sigma_C}{I_C}\right)^2}
  \end{equation}
where the subscripts refer to the band and $I$ represents the specific intensity at a given location.
{Note that, since the rms values are measured from the maps, they include thermal noise as well as any residual sidelobes that extend throughout the map.  Uncertainties due to the deconvolution itself, as is the case for any imaging process, are typically not included.}

Although the $\alpha$ maps have correctly taken the PB response at the two frequencies into account, the $\alpha$-error maps, as given in Eqn.~\ref{eqn:error_maps}, have not.  If higher accuracy is desired for $\sigma_\alpha$, then the increasing noise with distance from the map center can be accounted for by multiplying the error maps by a position-dependent factor, $f(r)$, as given in Eqn.~\ref{eqn:errorfactor} of Appendix~\ref{app:A}.

To estimate the maximum correction factor that is required in our BL-CC $\alpha$-error maps, we examine our largest angular-size spectral index map, which is the uvtapered $\alpha$ map of NGC~5907 in Appendix~\ref{app:B}.  At the farthest NW edge on the galaxy a distance at $r\,=\,4.5$ arcmin from the center, $\alpha\,\approx\,-0.3$ in which case $f(r)\,=\,1.6$.  The given error at this point from Eqn.~\ref{eqn:error_maps} is $\sigma_\alpha\,\approx\,0.3$ so the corrected error is  ${\sigma_\alpha}_{corr}\,=\,0.46$.  As can be seen from the error maps in the panels (frames e and f) and the sample calculation in Appendix~\ref{app:A}, typical corrections to the error maps are much lower than this.

\subsubsection{Thermal contribution to $\alpha_{BL-CC}$}
\label{sec:alpha_thermal}

The band-to-band spectral index maps have not been corrected for possible contributions from thermal emission.  \cite{var18} have done extensive work on estimating the spatially resolved thermal/non-thermal fraction in CHANG-ES galaxies at $\approx$ 15 arcsec resolution.
To estimate the thermal contribution, these authors have used H$\alpha$ maps which require a significant correction for extinction using additional 22 $\mu$m images.  We do not pursue such a correction for our BL or CC images since appropriate infrared images are not available at equivalently high spatial resolution.
At the centers of the galaxies (of particular interest for understanding AGN fractions), the adopted electron temperature ($10^4$ K) could also introduce some uncertanties in the thermal fractions.  

A reasonable global estimate of an L-band thermal fraction is 8\% and a C-band thermal fraction is 20\% \citep{var18}.  Adopting these values, the
non-thermal spectral index, $\alpha_{BL-CC}\,(NT)$ would be steeper than the observed spectral index, 
$\alpha_{BL-CC}$, by only 0.1. % i.e. $\alpha_{BL-CC}\,(NT)\,=\,\alpha_{BL-CC}\,-\,0.1$.

In the following discussion, then, we interpret our B-configuration L-band data to be globally dominated by non-thermal emission and this conclusion is confirmed by surveying the spectral index values shown in Appendix~\ref{app:B} (see colour scales) which are, on average, all steeper than $\alpha_{TH}\,=\,-0.1$ (the value expected for thermal emission alone).
%A scan through Table~\ref{tab:mean_alphas}, for example, shows that every galaxy has spectral indices that are consistent with a dominant non-thermal component.
The galaxy with the flattest spectral index averaged globally, NGC~4594 ($\overline{\alpha_{BL-CC}}=-0.20 ~\pm~0.12$), whose value is numerically consistent with thermal emission, is straightforwardly explained by non-thermal emission from a central compact AGN (Sect.~\ref{sec:fraction_AGNS}).

Since the B-configuration data have such high resolution, however, there could be discrete regions in the disk, specificaly HII region complexes, for which the thermal fraction departs from this general result.  We provide one example in Sect.~\ref{sec:flat_disks}.

\section{The Images and High Resolution Structures}
\label{sec:images}

%\subsection{CHANG-ES BL Images}

The {first} row of the Panels in Appendix~\ref{app:B} show a wide variety of emission in these high resolution images.
In total intensity, we see compact cores of varying strength (e.g. NGC~2613, NGC~4845 and others), numerous compact regions in the disk -- likely HII region complexes (e.g. NGC~3432, NGC~4244 and others) and vertical diffuse emission away from the plane (e.g. NGC~3044, NGC~4666, and again others).

Particularly interesting is the complex structure that can be seen in many spectral index maps (second row) that may be masked in the total intensity images. The fact that more structure/information can be available from spectral index maps than from total intensity maps has been noted in the past \citep[e.g.][]{lee01} but several remarkable examples are seen in the CHANG-ES sample.

An example is NGC~3448 (Fig.~\ref{fig:alpha_shift} top).  In the spectral index map (colours) we see evidence for a spiral arm curving from  RA $\approx$ 10 54 38, DEC $\approx$ 54 18 23 to RA $\approx$ 10 54 39, DEC $\approx$ 54 18 33. The arm is delineated by three spectral index peaks that are somewhat flatter than their surroundings ($\approx -0.53$) and typical of multiple young supernova remnants (SNRs). 
Yet no such structure is visible in the {L-band total intensity image alone.  Only three contours are shown in Fig.~\ref{fig:alpha_shift} but an inspection of the image as a whole (see the N~3448 image in Appendix~\ref{app:B}) shows no such feature. The C-band image structure \citep{wal19} of course will differ from the L-band image, hence leading to the structure in spectral index, but such a feature is again not obvious at all at C-band.  It is really the spectral index map that clearly reveals this structure, otherwise masked in total intensity. }  A possible arm may also be present on the opposite side of the nucleus centered at RA $\approx$ 10 54 41, DEC $\approx$ 54 18 20.

\begin{figure}[!tbp]
  \centering
  \begin{minipage}[b]{0.5\textwidth}
    \includegraphics[width=\textwidth,trim=0 12cm 0 0,clip]{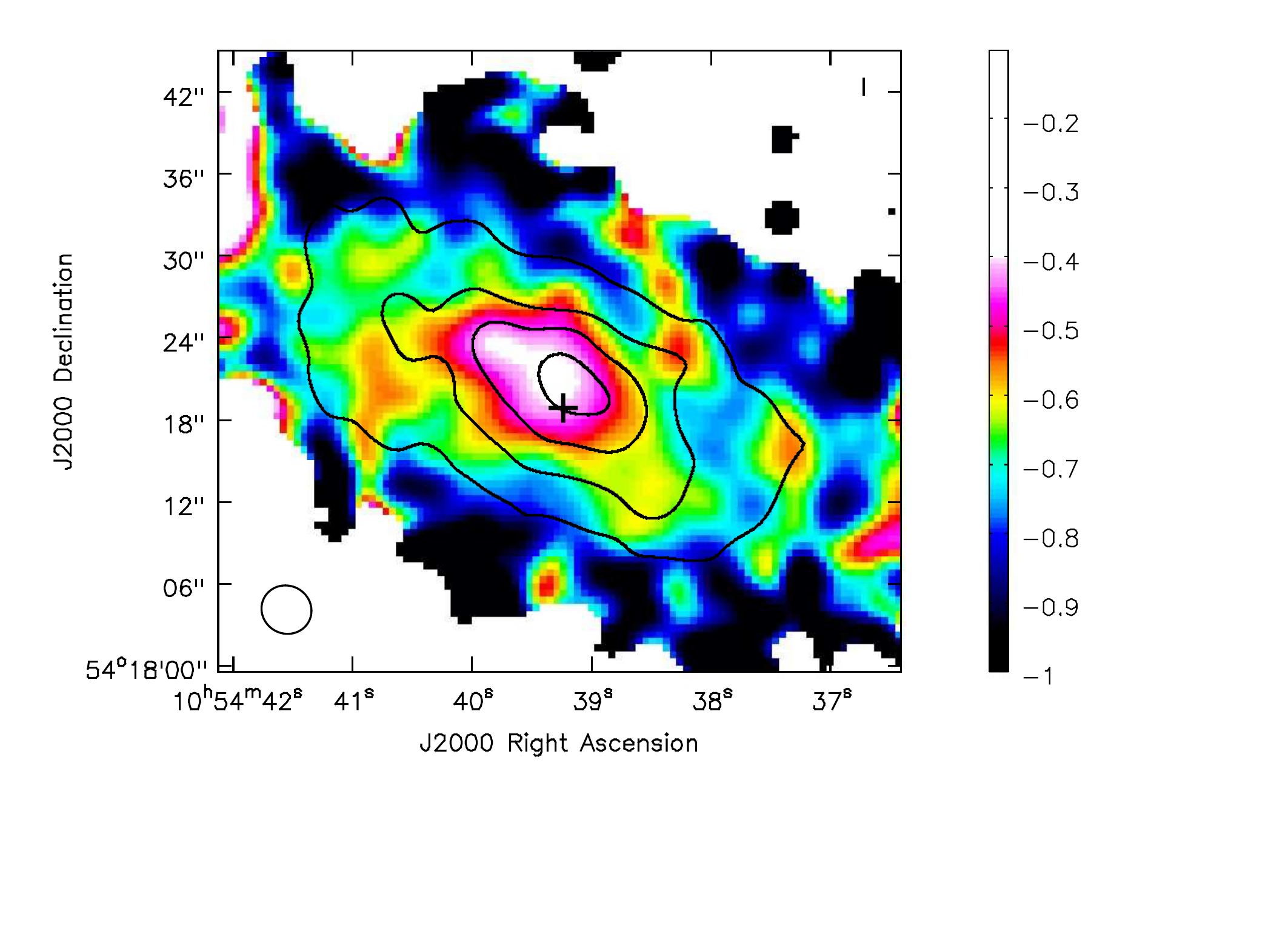}
%    \caption{Flower1.}
  \end{minipage}
  \centering
%\vspace{-2truein}
  \hfill
  \begin{minipage}[b]{0.5\textwidth}
    \includegraphics[width=\textwidth,trim=0 15cm 0 0,clip]{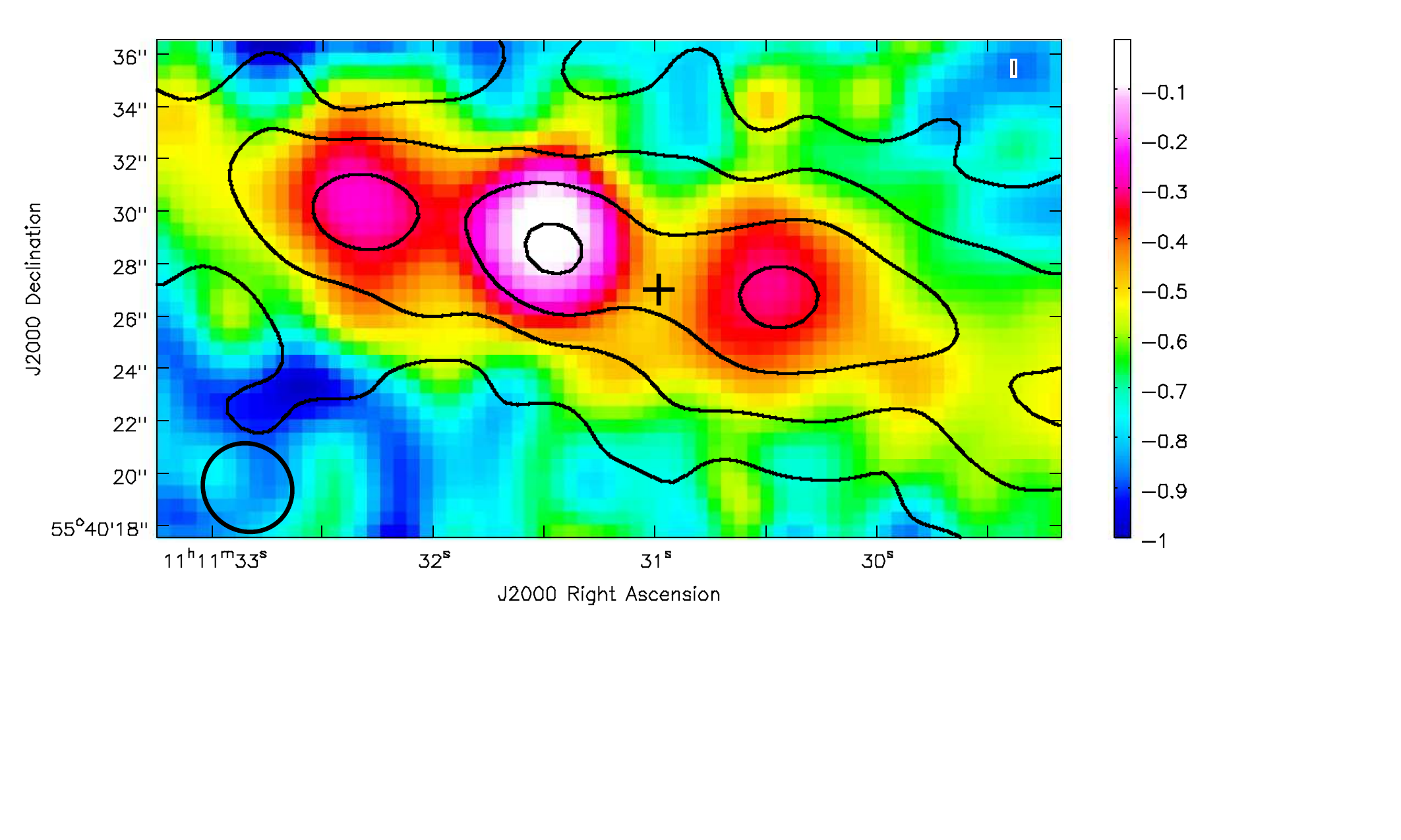}
 %   \caption{Flower two.}
  \end{minipage}
%\vspace{-2.5truein}
  \caption{{L-band} total intensity contours for the rob 0 weighting images over spectral index image with colour scale shown at right. The beam of the spectral index image is shown at lower left. 
   {\bf (a)} NGC~3448. Contours are at 170 (10$\sigma_I$), 400, 800, and 2000 $\mu$Jy beam$^{-1}$. The NED galaxy center is marked with a cross.
 {\bf (b)} NGC~3556. Contours are at 160 (10$\sigma_I$), 300, 600, and 1500 $\mu$Jy beam$^{-1}$. The NED galaxy center is marked with a cross. }
\label{fig:alpha_shift}
\end{figure}

Another example is NGC~3556 (Fig.~\ref{fig:alpha_shift} bottom) which shows three distinct peaks in the spectral index map that are present, but not so strongly obvious, in the total intensity image. The peaks are suggestive of radio lobes. For example, if
the white, flattest spectral index peak actually represents the galaxy center (offset from the NED center that is marked with a cross), then it is possible that the two other peaks on either side of it represent young radio lobes a distance $\approx$ 530 pc from the center.  
 However, since these features are in the disk in projection, it is possible that they are instead associated with other in-disk features, such as edge-on spiral arms or rings.

We provide another remarkable example in Sect.~\ref{sec:jets}.

\section{AGNs in the CHANG-ES Sample}
\label{sec:agns_general}

The fact that radio emission does not suffer from extinction is a strong incentive for using our high resolution BL images for determining whether or not an AGN is present. For edge-on galaxies, such incentive is even stronger.

In the past, it has been difficult to disentangle AGNs from nuclear starbursts in spiral galaxies.  The search for weak or hidden AGNs has been of particular interest, especially given the possibility that lower mass AGNs in galaxy nuclei could be on the same fundamental plane (i.e. the plane of black-hole accretion relating radio emission, X-ray emission, and mass of an accreting black hole) as supermassive black holes and stellar mass black holes \citep[e.g.][]{gul14}.

It has long been recognized that excessive luminosities within galaxy nuclei can indicate the presence of an AGN (e.g. Sect.~\ref{sec:radio_lums}.  However, 
`dwarf' AGNs, or `Low Luminosity AGNs' (LLAGNs), i.e. those with H$\alpha$ luminosities lower than $10^{40}$ erg s$^{-1}$ \citep{ho97} or with X-ray luminosities below  $10^{42}$ erg s$^{-1}$ \citep{ho09} may also be present yet hidden amongst other camouflaging emission.

In the CHANG-ES sample, there is already previous evidence for the presence of many AGNs or LLAGNs, for example,
 a Seyfert (Sy) categorization, or the presence of a low ionization nuclear emission region (LINER) for LLAGNs \citep{mao08}.  We identify these via their NED (NASA/IPAC Extragalactic Database) type which we list in Table~\ref{tab:AGNs}. A study by \cite{nag05} also identified $\approx$ 200 LLAGNs within which 8 CHANG-ES galaxies appear, although they do not reveal any new LLAGNs other than those already identified in Table~\ref{tab:AGNs}.

\subsection{Radio Criteria for Determining the Presence of an AGN}
\label{sec:criteria}

Our goal is to determine criteria for an AGN from the radio emission alone and then compare our results to what has been previously known.   In addition, we supplement our data with X-ray data for a subset of the galaxies (Sect.~\ref{sec:xmm_obs}).

Radio criteria that provide evidence of an AGN are:  1) the presence of an unresolved point-like core, 2) a luminosity that is too high to be accounted for from a collection of SNe alone, 3) the presence of radio jets or lobes, 4) the presence of a relatively flat or positive spectral index at the nucleus,
5) a brightness temperature, $T_B$ that is too high to be accounted for by thermal gas, 6) variability, and 7) the presence of circular polarization.  We will consider each of these in turn.

Our brightest source is  NGC~660 which has a peak specific intensity of 245 mJy beam$^{-1}$ at the highest resolution.  This corresponds to a brightness temperature of $T_B\,\approx\,10^4$ K, which does not put meaningful constraints on this value; in other words, the beam size is too large for any embedded AGN to distinguish itself, should it have a high brightness temperature (Point \#5).
%T_b=c^2/(2nu^2k) *I_nu = 1.32e18 *I_nu  for nu=1.57e9
%I_nu=245.3e-3*1e-23 erg/s/cm^2/Hz/beam
%one beam = 50.245 pixels (regions output from viewer) where pixel size is 0.5 arcsec square.  This should already have taken into account the ln(2) weighting.
%so 50.245/(0.5^2) = 12.56125 arcsec squared for the beam. which is then 2.952e-10 sr.
%so finally we have T_B = 1.32e18*245.3e-3*1e-23/(2.952e-10) = 1.10E4 K

Variability (Point \#6) is a clear indicator of an AGN.  Even a weak AGN, such as the LLAGN at the Galactic center, is variable on timescales of weeks to months \citep{con11}, and CHANG-ES data sets are separated by months (e.g. Table~\ref{tab:observationsB}). We do find variability of a central AGN by careful comparison and fitting of the central source brightness over all CHANG-ES data sets for the galaxy, NGC~4845 \citep{irwV}.  However, this source was rather extraordinary, displaying a Tidal Disruption Event (TDE). For the remaining galaxies,
a comparison was made between the flux densities of the B-configuration L-band images and the C-configuration L-band images (the former smoothed to the resolution of the latter) but no convincing variability was found beyond minor differences that could be attributed to differences in uv coverage.
%As a straightforward check, we smoothed all B-configuration L-band images to the C-configuration L-band resolution and compared the flux densities of the core -- didn't find differences beyond the kind of differences that were seen in some other position on the galaxy, the latter due to difference uv coverage.  
Similar analysis to NGC~4845, which will be feasible only for the stronger sources, is beyond the scope of this paper. 

Circular polarization {in the cores of the galaxies} (Point \#7) has been examined thoroughly for the BL data in \cite{irwXI}.  {This paper outlines the uncertainties involved in this process and adopts conservative criteria for concluding that circular polarization exists. The conclusion is that five} galaxies show evidence for circular polarization and these are indicated in Table~\ref{tab:AGNs}.
 
This leaves us with Points 1 through 4, which will be explored below.

\subsubsection{Point-like Cores}
\label{sec:point-like}

Table~\ref{tab:pointcores} lists the galaxies that have point-like cores.  We determine this by fitting a Gaussian to the centers of each galaxy and listing only those galaxies for which the FWHM of the source, after deconvolving from the synthesized beam, is less than the synthesized beam FWHM in {\it both} the major and minor axis directions. {This criterion picks out strong candidates for a LLAGN but allows for minor contributions from other emission components that may also be centrally concentrated.}

Thirteen galaxies fall into this category.
Of these galaxies, to our knowledge, five show a radio point core for the first time from the CHANG-ES sample. These galaxies are: NGC~2613, NGC~4845 \citep{irwV}, NGC~4666 \citep{ste19}, NGC~5084, and NGC~5297. Moreover, for the galaxy, NGC~2613, the radio point core provides the first evidence for a possible AGN in any band \citep[see also][]{des17}.

{\renewcommand{\arraystretch}{1.1}
\begin{deluxetable}{lcccccccc}
%\tabletypesize{\scriptsize}
%\rotate
\tablecaption{Galaxies with Point-Cores \label{tab:pointcores}}
\tablewidth{0pt}
\tablehead{
  \colhead{Galaxy} & \colhead{RA$^a$} & \colhead{DEC$^a$} & \colhead{S$_\nu$$^b$}& \colhead{Avg. FWHM$^c$} & \colhead{$L_\nu$$^d$} & \colhead{No.$_{SNR}$$^e$} & \colhead{} & \colhead{} \\
  \colhead{} &  \colhead{(h m s)} & \colhead{($^\circ$ $^\prime$ $^{\prime\prime}$)}& (mJy)& \colhead{(pc)} &  \colhead{$10^{19}$ W Hz$^{-1}$} & \colhead{}  & \colhead{} & \colhead{}
}
\startdata
N 660$^*$	& 01 43 02.320	& +13 38 44.88   & 296$\pm$ 5     & 176  & 536   &1786  &  &   \\
N 2613	& 08 33 22.772	& -22 58 24.81   & 0.37$\pm$0.02  & 398  & 2.42  & 8& &  \\
N 2683	& 08 52 41.31	& +33 25 18.79   & 1.13$\pm$0.03  & 81   & 0.531 & 2 &  &   \\
N 3079	& 10 01 57.793	& +55 40 47.27   & 152$\pm$5      & 272  & 772   & 2572 &  &   \\
N 3735	& 11 35 57.204	& +70 32 07.77   & 1.9$\pm$0.2    & 572  & 40.1  & 134 &  &  \\
N 4388$^*$	& 12 25 46.747	& +12 39 41.69   & 31.5$\pm$0.9   & 242  & 104   & 346 &  &   \\
N 4438$^*$	& 12 27 45.548	& +13 00 32.98   & 57$\pm$3       & 139  & 73.6  & 245 &  &  \\
N 4565	& 12 36 20.771	& +25 59 15.68   & 1.57$\pm$0.01  & 161  & 2.66  & 9 & &   \\
N 4594	& 12 39 59.433	& -11 37 23.02   & 70.50$\pm$0.06 & 205  & 136   & 453 \\
N 4666$^*$	& 12 45 08.632	& -00 27 42.99   & 4.2$\pm$0.2    & 430  & 38.0  & 127 &  &   \\
N 4845$^*$	& 12 58 01.196	& +01 34 32.42   & 210.8$\pm$0.6  & 250  & 727   & 2424 \\
N 5084$^*$	& 13 20 16.831	& -21 49 38.29   & 30.4$\pm$0.3   & 410  & 199   & 664 &  &   \\
N 5297	& 13 46 23.673	& +43 52 20.22   & 0.21$\pm$ 0.02 & 530  & 4.10  & 14 &  &   \\
\enddata
%% Text for table notes should follow after the \enddata but before
%% the \end{deluxetable}. Make sure there is at least one \tablenotemark
%% in the table for each \tablenotetext.
\tablecomments{Measured by doing Gaussian fitting of the cores of the rob 0 weighted maps. The fitted region was $\approx$ twice the size of the beam FWHM. Asterisks denote galaxies whose positions differ from the NED centers.}
\tablenotetext{a}{Fitted center of the core.  Uncertainties are $\approx$ 0.2 arcsec in both coordinates. }
\tablenotetext{b}{Flux density of the fitted Gaussian component.}
\tablenotetext{c}{Average linear size of the major and minor axes of the total intensity maps.}
\tablenotetext{d}{Spectral power corresponding to the flux density of the Gaussian component.}
\tablenotetext{e}{Number of average M~82 SNRs ($\bar L_\nu\,=\,3\,\times\,10^{18}$ W Hz$^{-1}$) corresponding to the measured spectral power, rounded to an integer.}
\end{deluxetable}
} %end the arraystretch

The fitted positions of these cores are also listed in Table~\ref{tab:pointcores}.  We have compared these positions to the suggested position of the source centers in NED. Six of the thirteen sources have positional differences greater than the error with four of those sources having positional offsets greater than three times the error. These six galaxies are marked with an astrisk in Table~\ref{tab:pointcores}.  We argue that these radio core positions likely designate the true centers of these galaxies (except N~4438, see Sect.~\ref{sec:jets}).

We stress that this criterion alone is not a definitive argument for an AGN.  For example, 
it is possible that a starburst could be so compact in the nuclear region of the galaxy (sizes typically several hundred pc, Table~\ref{tab:pointcores}) that a number of SNe might also appear point-like.  Alternatively, a weak
AGN could still  be present but masked (and missed) if other broader scale emission simply swamps that of a LLAGN.  Higher resolution data, e.g. Very Long Baseline Interferometry (VLBI) detections, should further reveal AGN activity \citep[e.g.][and references therein]{nag05}.

%JUDITH see Saikia et al. 2018 A\&A 616, 152 for 15 GHz survey of llagns.
%plus baldi et al. 2018
%also Panessa \& Giroletti 2013  N4565

\subsubsection{Radio Luminosities}
\label{sec:radio_lums}

It is well known that an AGN could (but need not) have a high luminosity in comparison to a collection of SNRs from a star forming (SF) region in a given area. 

The best known and well-studied extragalactic SF region is in the `starburst galaxy', M~82, which shows $\approx$  40 supernova remnants (SNRs) at 5 GHz \citep{mux94} and 20 SNRs at 408 MHz \citep{wil97} in a region
$\approx$ 600 pc in diameter.  The average spectral luminosity of the nuclear SNRs in M~82 at 20 cm is $L_\nu\,=\,3\,\times\,10^{18}$ W Hz$^{-1}$ \citep[data from][]{all98}, with a standard deviation of the same order of magnitude.  The star formation rate (SFR) of M~82 is 3.6 M$_\odot$ yr$^{-1}$ \citep{gri03} which exceeds all but two of the CHANG-ES sample \citep{wie15}.

%\citep{wil97, kro92}. 
%of diameter, $\approx\,59$ arcsec %2.88e-4 radians 52 arcsec is 2.5e-4 radians=807 pc
%5ghz sample says 600 pc they both use the same distance

%all98 measured spectra at l-band for 26 SNRs
%wil97 find 20 at 408Mhz -- fewer than higher frequency.
%muxlow et al. 1994 find 40 sources at 5 GHz. mux94
%``Surprisingly, many SNRs at L-band show a low-frequency turnover due to thermal absorption Muxlow pedlar and sanders 1995, iaa-iac-university of pixa workshop

Although such a comparison is not absolutely definitive, it is still useful to ask how many M~82-mean-SNRs could be accounted for by the measured luminosities in the cores of the galaxies of Table~\ref{tab:pointcores}.  The last column gives this result and the fifth column gives the size of the region to which it applies.
All galaxies except NGC~2613, NGC~2683, NGC4565, and NGC~5297 have luminosities that exceed what is seen in the starburst galaxy, M~82.  This suggests that the remaining nine galaxies likely harbour relatively active AGNs, with NGC~3079 being the most powerful.

For the galaxies that do not have point-like cores (Table~\ref{tab:nopointcores}) or at least have a deconvolved Gaussian size that is wider than the synthesized beam in at least one direction, we repeat this exercise.  From this list, NGC~2992, by far `outshines' all other galaxies. Here, almost 9000 M~82-mean-SNRs would be required to reproduce the observed luminosity in a region that is about the same as the M~82 nuclear starburst.  From this result, NGC~2992 clearly has an AGN and, in fact, has been found to have polarized radio jets that have been explored in detail in \cite{irwVIII}.

{\renewcommand{\arraystretch}{1.1}
\begin{deluxetable}{lcccccc}
%\tabletypesize{\scriptsize}
%\rotate
\tablecaption{Galaxies without Point-Cores \label{tab:nopointcores}}
\tablewidth{0pt}
\tablehead{
  \colhead{Galaxy} & \colhead{RA$^a$} & \colhead{DEC$^a$} & \colhead{S$_\nu$$^b$}& \colhead{Avg. FWHM$^c$} & \colhead{$L_\nu$$^d$} & \colhead{No.$_{SNR}$$^e$} \\
  \colhead{} &  \colhead{(h m s)} & \colhead{($^\circ$ $^\prime$ $^{\prime\prime}$)}& (mJy)& \colhead{(pc)} &  \colhead{($10^{19}$ W Hz$^{-1}$)} & \colhead{}  
}
\startdata
N 891	& 02 22 33.22	& +42 20 57.6   & 10.2$\pm$ 0.6     &  132 & 10.0  & 33   \\
N 2820	& 09 21 45.97	& +64 15 28.0   & 3.1$\pm$ 0.2      &  733 & 26.9  & 90  \\
N 2992  & 09 45 41.95   & -14 19 35.8   & 195$\pm$5         &  650 & 2700  & 8990 \\
N 3003  & 09 48 35.68   & +33 25 17.9   & 1.08$\pm$0.04     &  300 & 8.34  & 28    \\
N 3044  & 09 53 40.87   & +01 34 46.7   & 4.1$\pm$ 0.2      &  263 & 20.2  & 67    \\
N 3432  & 10 52 30.94   & +36 37 08.2   & 0.80$\pm$0.09     & 239  & 0.85  & 3\\
N 3448  & 10 54 39.17   & +54 18 20.5   & 6.8$\pm$0.3       & 441  & 48.8  &163\\
N 3556$^f$  & ---           & ---           & ---           &---   &---&---\\
N 3628  & 11 20 16.99   & +13 35 20.2   & 175$\pm$4         & 115  & 150   &501\\
N 3877  & 11 46 07.71   & +47 29 40.0   & 2.4$\pm$0.1     & 204  & 9.03  &30\\
N 4013  & 11 58 31.38   & +43 56 51.0   & 7.7$\pm$0.4       & 192  & 23.6  &79\\
N 4096  & 12 06 01.23   & +47 28 41.5   & 0.16$\pm$0.04     & 111  & 0.20  &1\\
N 4157$^f$  & ---       & ---           & ---               &---   &---    &---\\
N 4192  & 12 13 48.28   & +14 54 02.3   & 4.8$\pm$0.7       & 155  &10.5   &35\\
N 4217  & 12 15 50.95   & +47 05 29.2   & 4.1$\pm$0.3       & 271  &20.8   &69\\
N 4244$^g$  & ---       & ---           & ---               &---   &---    &---\\
N 4302  & 12 21 42.31   & +14 35 52.4   & 1.70$\pm$0.04     &212   &7.64   &25\\
N 4631  & 12 42 07.87   & +32 32 34.9   & 7.3$\pm$0.3       &151   &4.78   &16\\
N 5775$^h$  & 14 53 57.50   & +03 32 41.3   & 7.7$\pm$0.5       &1078  &76.9   &256\\
N 5792$^f$&---          & ---           & ---               & ---  &---    &---\\
N 5907$^f$ & ---        & ---           & ---               &---   &---    &---\\
U 10288 & 16 14 24.83   & -00 12 27.7    & 0.19$\pm$0.03    &399   &2.60   &9\\
\enddata
%% Text for table notes should follow after the \enddata but before
%% the \end{deluxetable}. Make sure there is at least one \tablenotemark
%% in the table for each \tablenotetext.
\tablecomments{Measured by doing Gaussian fitting of the highest total intensity peak at or closest to the cores of the rob 0 weighted maps. The fitted region was $\approx$ twice the size of the beam FWHM.}
\tablenotetext{a}{Fitted center of the core.  Uncertainties are $\approx$ 0.2 arcsec in both coordinates. }
\tablenotetext{b}{Flux density of the fitted Gaussian component.}
\tablenotetext{c}{Average linear size of the major and minor axes of the Gaussian fit, after deconcolving from the synthesized beam.}
\tablenotetext{d}{Spectral power corresponding to the flux density of the Gaussian component.}
\tablenotetext{e}{Number of average M~82 SNRs ($\overline{L_\nu}\,=\,3\,\times\,10^{18}$ W Hz$^{-1}$).}% corresponding to the measured spectral power, rounded to an integer.}
\tablenotetext{f}{No distinct peak at the core.}
\tablenotetext{g}{Emission is too weak to measure.}
\tablenotetext{h}{Peak is blended with extended emission to the NW.}
\end{deluxetable}
} %end the arraystretch

A second galaxy that stands out is NGC~3628, requiring $\approx$ 500 SNe within 115 pc to explain its luminosity. This galaxy also has circularly polarized emission \citep{irwXI} and radio lobes (next section).

%It is worth noting that these high resolution CHANG-ES observations are sensitive enough to detect a luminosity equivalent to a single mean M~82 supernova (e.g. N~4096, Table~\ref{tab:nopointcores})! 

\subsubsection{Jets/Lobes}
\label{sec:jets}

Determining whether a jet or lobe is present from radio data is somewhat more subjective than the previous criteria since it relies on morphology; such structures may be embedded in other emission, especially since these features might not (yet) have emerged from the disk.   Here, if we see bipolar structure, we interpret this as a jet or lobe.  Such a result does not preclude the possibility of additional winds from starbursts, but a bipolar feature seen in non-thermal radio emission is most likely explained by an AGN.  In the following, we refer to such structures as lobes, rather than jets, since our spatial resolution is insufficient to detect narrow jets that would be connected to the nucleus.

Several galaxies show clear radio lobes that have previously been known.  Examples are
NGC~4388 \citep[e.g.][among others]{hum91,dam16} and NGC~3079 \citep[][and others]{hum84} and these are also seen in the panels of Appendix~\ref{app:B}. Fig. 8 of \cite{irwVIII} also reveals how the lobes in NGC~3079 are more distinctly seen in linear polarization compared to total intensity.  % These jets are seen, not only in total intensity emission, but also in CP \citep{irwXI}.

In Fig.~\ref{fig:jets} (top), we show a CHANG-ES example of a previously known radio lobe, i.e. the inner region of NGC~4438 with total intensity contours revealing the NW-SE outflow direction.  The spectral index, moreover, also clearly reveals the same structure.  The mean spectral index within the 1 mJy beam$^{-1}$ total intensity contour (second contour) is $\alpha_{BL-CC}$ = -0.757$\pm$0.005. This is an example of a steep spectrum that is typically observed for radio lobes.  The position of the core from higher spatial resolution radio images suggest that its likely location is near the NED center marked in the figure \citep[J2000: 12h27m45.67s,  +13d00m31.54s,][]{hot07}.  The core is presumably a flatter spectrum source whose emission is too weak to perturb the spectral index of the lobe, as seen in the image.
% suggested core coordinates from hummel&saikia 1991 converted to J2000 is: 12h27m45.51781s   +13d00m33.3340s

\begin{figure}[!tbp]
  \centering
  \begin{minipage}[b]{0.4\textwidth}
    \includegraphics[width=\textwidth,trim=0 12cm 0 0,clip]{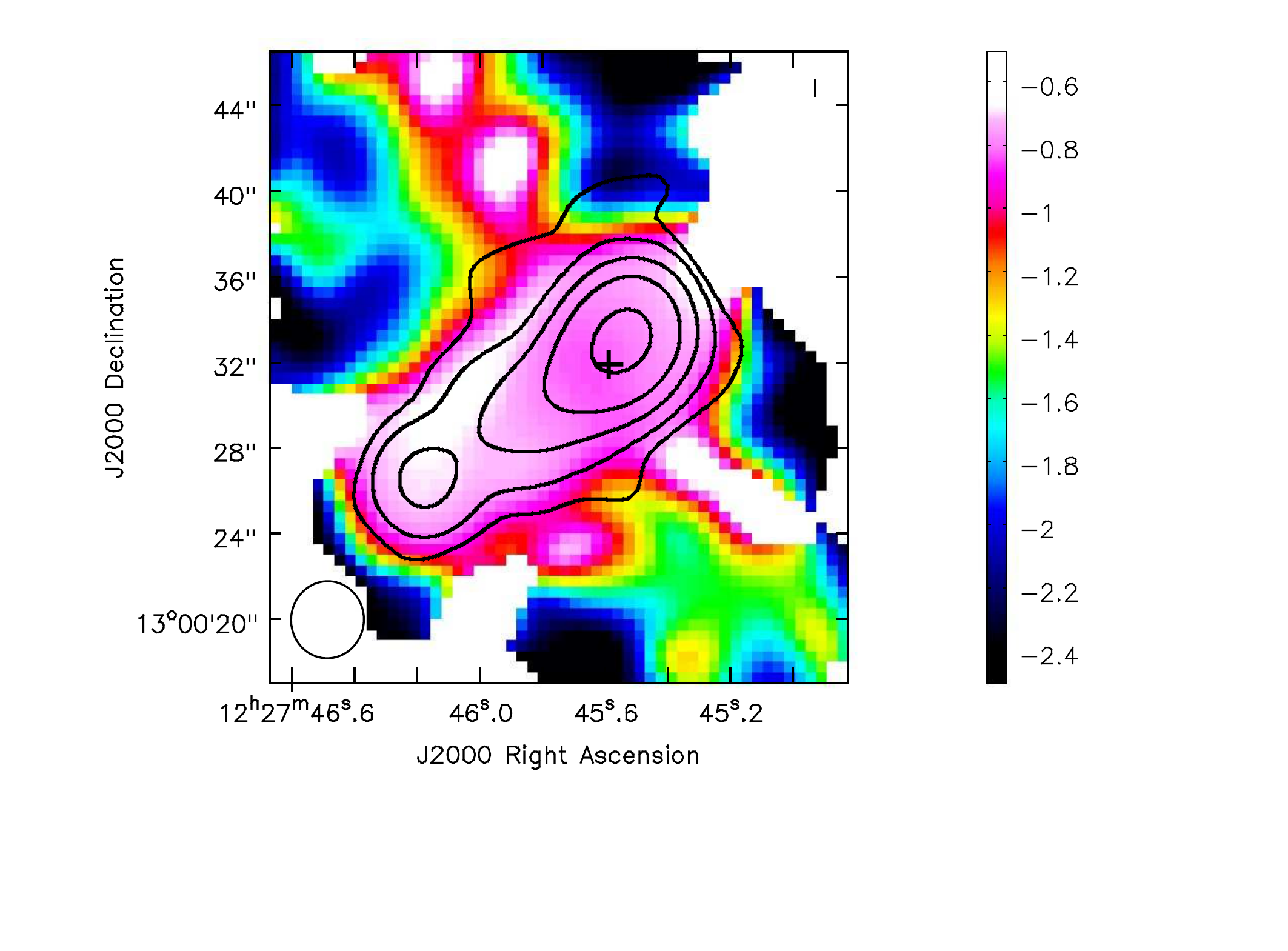}
  \end{minipage}
%\vspace{-1.5truein}
  \centering
  \hfill
  \begin{minipage}[b]{0.4\textwidth}
    \includegraphics[width=\textwidth,trim=0 15cm 0 0,clip]{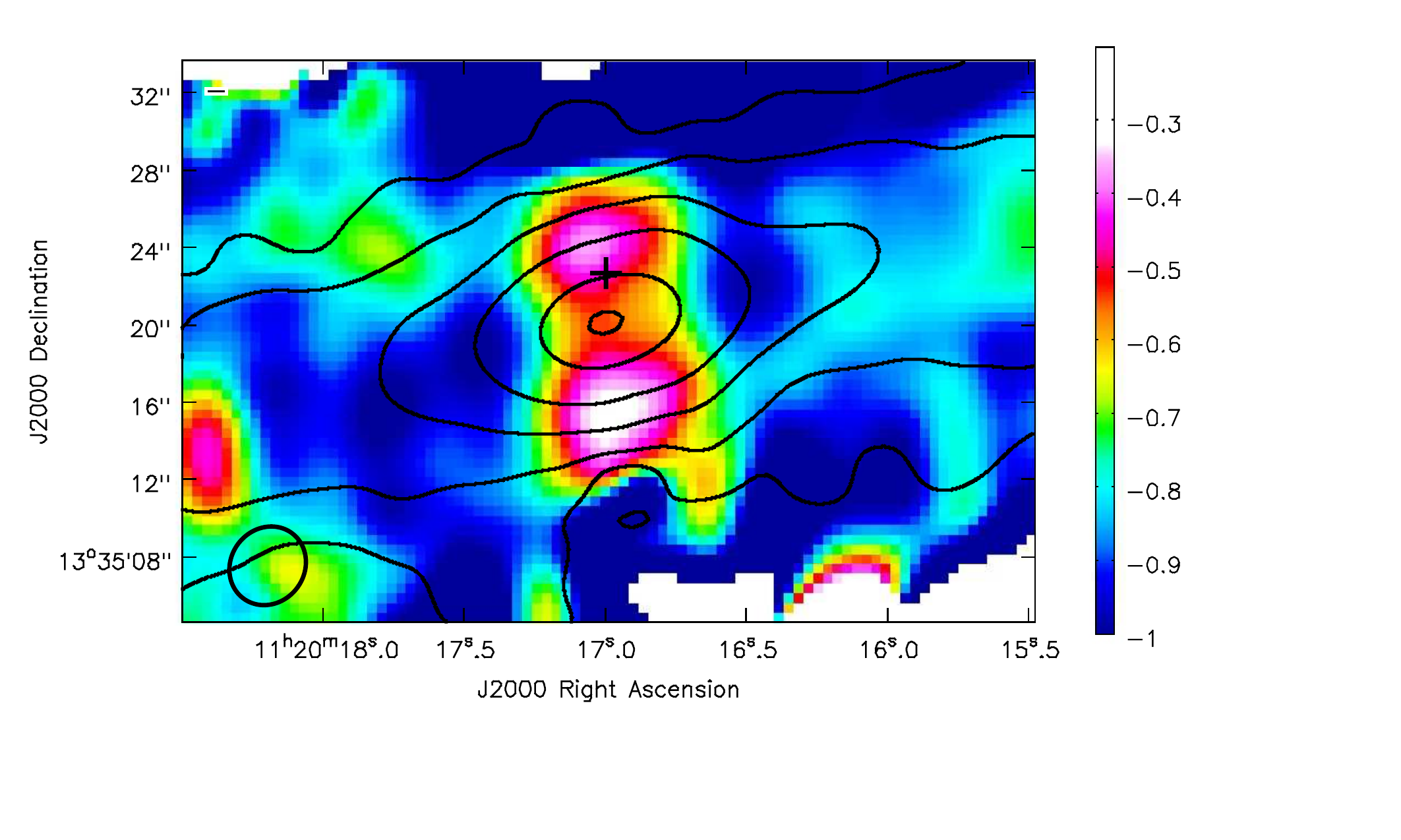}
  \end{minipage}
%\vspace{-2truein}
  \centering
  \hfill
  \begin{minipage}[b]{0.5\textwidth}
    \includegraphics[width=\textwidth,trim=0 12cm 0 0,clip]{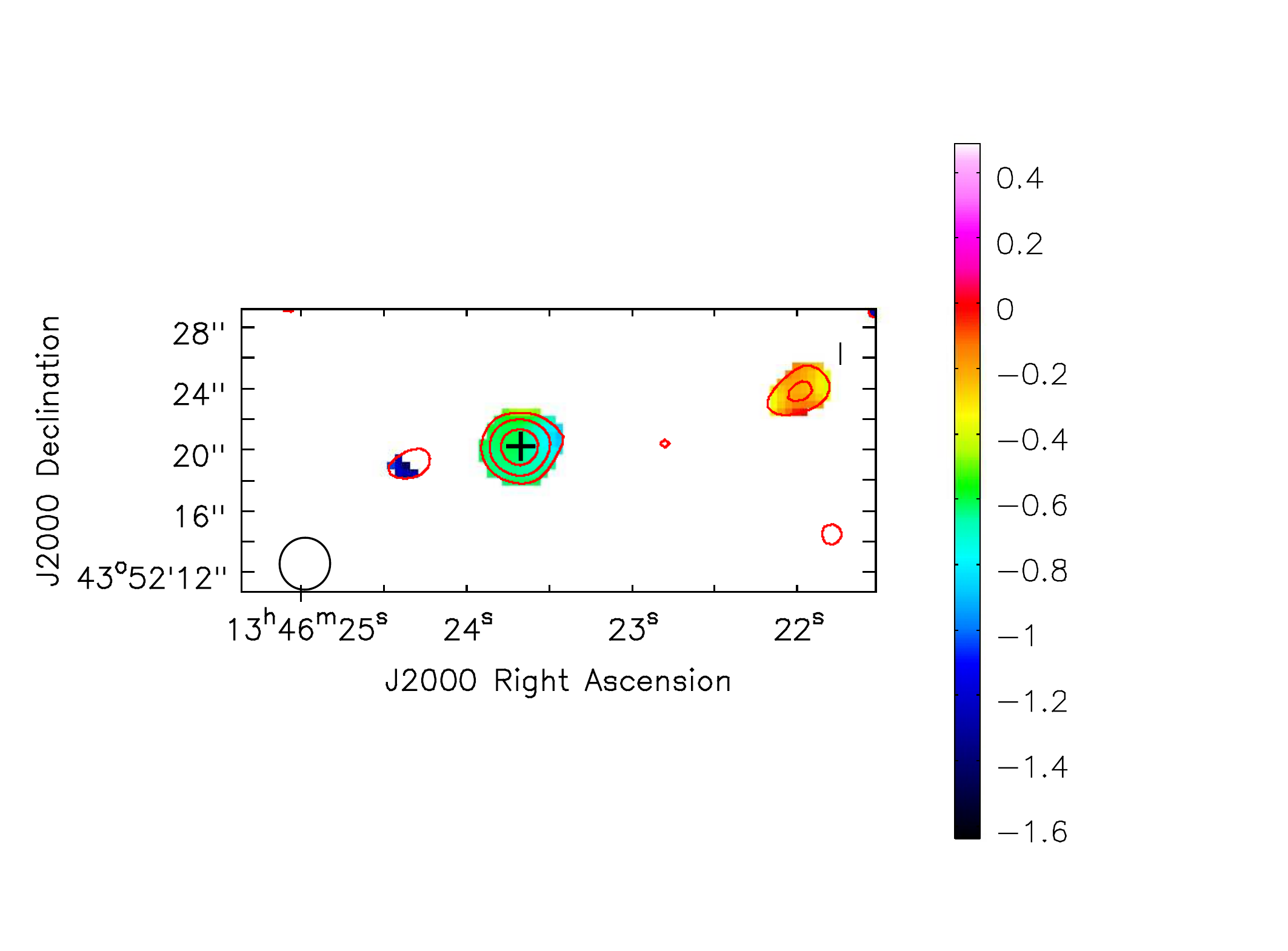}
  \end{minipage}
%  \vspace{-2truein}
  \caption{Total intensity contours for the rob 0 weighting images over spectral index image with colour scale shown at right. The beam of the spectral index image is shown at lower left and the NED center of the galaxy is marked with a 'plus' sign. 
   {\bf (Top)} NGC~4438. Contours are at 400 (10$\sigma_I$), 1000, 3000, 8000, and 25000  $\mu$Jy beam$^{-1}$. 
   {\bf (Center)}  NGC~3628. Contours are at 232 (10$\sigma_I$), 500, 1500, 5000, 30000, and 75000 $\mu$Jy beam$^{-1}$.
   {\bf (Bottom)}  NGC~5297. Contours are at 40.8 (3$\sigma_I$), 65, and 100 $\mu$Jy beam$^{-1}$.}
\label{fig:jets}
\end{figure}

A new and very interesting result, however, is that radio lobes could be revealed {\it only} in the spectral index (hidden lobes) and {\it not} in the total intensity image}; again, the weak lobes are masked by other sources of emission in total intensity {as previously described in Sect.~\ref{sec:images}}.  Presumably, the spectral index map {\it can} reveal the radio lobes because the energy spectral dependence of cosmic ray electrons will be different in the outflow than in the surrounding regions. 

A good example of this is in NGC~3628 (Fig.~\ref{fig:jets}, center).  Here we see the total intensity emission (contours) smoothly increasing to a peak at the center of the galaxy.  Above and below the nucleus are two regions of relatively flat spectral index,  $\approx -0.34$ for the south peak (Table~\ref{tab:alpha}) and $\approx -0.43$ for the north peak. The center of the galaxy must be located at the position of the total intensity radio continuum peak, i.e. RA = 11 20 17.01, DEC = 13 35 20.1.  This center is 2.6 arcsec below the NED peak which is marked with a '+' in the figure.  The NED position was adopted from CHANDRA X-ray data \citep{eva10}. % and it may be that the X-ray data is picking up emission from this northern lobe [is this true??].[check which side of the galaxy is the nearer one -- optical ned images look like the northern side is actually nearer so the northern side should have more dust]
Notice that the spectral index is flatter than that of the radio outflow for NGC~4438.  This suggests that the outflow in NGC~3628 could be young in comparison. %  AGN-related outflow that has not yet emerged from the galaxy disk.

In Table~\ref{tab:AGNs} we list galaxies that appear to show radio lobes.  The two galaxies with question marks are NGC~3556 which has three peaks along the disk in both total intensity and spectral index (Fig.~\ref{fig:alpha_shift} bottom) and NGC~5297 (Fig.~\ref{fig:jets} bottom) which has two very weak features on either side of the nucleus, the westmost `lobe' having a spectral index of -0.23.

In summary, we have searched through the BL data for evidence of new radio lobes (beyond what has been previously known) and find evidence for lobes in NGC~3628, and possibly NGC~3556 and NGC~5297; these features are revealed in the {\it spectral index} maps, rather than total intensity.

\subsubsection{`Flat' Spectral Indices in Nuclei}
\label{sec:flatspec}

With non-thermal emission dominating (Sect.~\ref{sec:alpha_thermal}) we expect flatter spectral indices for AGNs as opposed to a group of SN remnants (SNRs). It is well known, for example, that AGNs can have very flat and sometimes positive spectra 
%(where $I_\nu\,\propto\,\nu^\alpha$)
due to the fact that AGNS are compact, can consist of multiple components, and can show self-absorbed synchrotron spectra.

As pointed out in the previous section (e.g. NGC~4438), the core of an AGN could be blended with steeper spectrum outflows so this criterion will only select a subset of AGNs.  It is important, however, to ask whether a collection of SNRs (also non-thermal emission) at a galaxy's nucleus  could produce a flat spectral index and masquerade as an AGN.

For SNRs the theoretical value derived from diffusive shock acceleration (DSA) is $\alpha\,=\,-0.5$ and 20\% of all SNRs have measured values consistent with this value.  Observationally, even young SNRs have steep spectral indices, typically of about \mbox{-0.7} \citep{uro14} prior to significant cosmic ray electron (CRE) aging and spectral steepening.  The  youngest known SNR in our galaxy (about 100 years), for example,  has a spectral index of \mbox{-0.65} at 1 GHz \citep{rey08}. A sampling of Galactic and extragalactic SNRs by \cite{bel11} shows spectral indices ranging from \mbox{-0.4} to \mbox{-1.1}.

So-called `flat spectrum' SNRs have spectral indices in the range, $-0.2\,>\, \alpha\,>\,-0.5$ \citep{uro14}.
Theoretically, such flat spectral indices can occur intrinsically \citep[eg. see chapter 21 of][for second-order Fermi acceleration]{lon94} and have been observed for individual SNRs in the Galaxy \citep{uro14}. %Significant intrinsic flattening has also been observed at higher frequencies \citep[e.g. Cas A at 30 GHz]{oni15}.
A {\it collection of SNRs} in the nuclear region of a galaxy, however, is unlikely to show intrinsically flat spectral indices at L-band.

{\it Non-intrinsic} flattening of the spectral index could occur from the presence of a strong thermal Bremsstrahlung component.  Thermal absorption is unlikely for a collection of SNRs at L-band (Sect.~\ref{sec:flat_disks}), but flattening due to thermal emission  (for which $I_\nu\,\propto\,\nu^{-0.1}$) is sometimes observed for individual Galactic SNRs when they are embedded in molecular clouds \citep{oni12}. %A mixture of thermal gas could indeed flatten the spectral index of a collection of SNRs if the emission measure, EM, is very high throughout the region (see below and Sect.~\ref{sec:flat_disks}).  

A good example of emission that is primarily from a sample of supernovae in the nuclear region of a galaxy is, again, M~82.
Various authors find low frequency spectral turnovers in individual SNRs in M~82 due to thermal absorption, implying emission measures (EMs) of typically $10^6$ cm$^{-6}$ pc \citep{wil97,kro75,car91}, consistent with radio recombination line EM results \citep{sea85}, although a few $10^{7-8}$ cm$^{-6}$ pc have also been cited from very high resolution observations \citep[i.e. 1.6 pc,][]{mcd02}.  However, the mean spectral index from this collection of SNRs is still quite steep, on average: 
 $-0.6\,\pm\,0.2$ \citep{wil97} at 408 MHz $-0.4\,\pm\,0.3$ at 5 GHz \citep{kro92}, in a region
%of diameter, $\approx\,59$ arcsec %2.88e-4 radians 52 arcsec is 2.5e-4 radians=807 pc
%5ghz sample says 600 pc they both use the same distance
$\approx$ 600 pc in diameter which is similar in size (to order of magnitude) to the CHANG-ES galaxies (Tables~\ref{tab:pointcores} and \ref{tab:nopointcores}).  Between 74 MHz and 23 GHz, 19 out of 26 compact sources in M~82 have $\alpha\,<\,-0.49$.  For more up-to-date data on M~82, see, e.g. \citet{ade13} and \citet{var15}.

%\citet{var18} have also shown that non-thermal spectral indices as flat as -0.4 can occur in edge-on galaxies at lower spatial resolution ($\approx\,15$ arcsec or regions 500pc to 1.5 kpc in size).

In summary, we would not expect a collection of SNRs in the nuclear region of an external galaxy to show a mean spectral index flatter than $\alpha\,\approx\,-0.3$ unless thermal emission from gas with extremely high EMs ($>>10^6$ cm$^{-6}$ pc) is present throughout the region.  We therefore adopt $\alpha\,\ge\,-0.3$ at the nucleus as evidence for an AGN.

In Table~\ref{tab:alpha} we provide spectral index measurements near the nucleus of each CHANG-ES galaxy.
Since the NED nucleus does not always indicate the true center of the galaxy (e.g. NGC~3628, Fig.~\ref{fig:jets}, center), we make measurements at the location of the flattest spectral index that is closest to the NED center.  These locations (RA and DEC) are specified in the table.
Note also that not every galaxy shows a distinctly flatter spectral index anywhere near the nucleus; see notes to the table for further clarity. Measurements close to where the emission was cut off (blanked) were also avoided because of spurious spectral index values at the boundaries. 

{\renewcommand{\arraystretch}{1.1}
\begin{deluxetable}{lcccc}
\tabletypesize{\scriptsize}
%\rotate
\tablecaption{{Flattest Spectral Index Closest to the Galaxy's Center} \label{tab:alpha}}
\tablewidth{0pt}
\tablehead{
\colhead{Galaxy}  & \colhead{RA\tablenotemark{a}} & \colhead{DEC\tablenotemark{a}} &  \colhead{$\alpha_{BL-CC}$\tablenotemark{b}} &   \colhead{Avg. FWHM\tablenotemark{c}} \\
                    &(h m s) & ($^\circ$ $^\prime$ $^{\prime\prime}$ )&    &(pc)  
}
\startdata 
N 660       & 01 43 02.33 & +13 38 44.7  & +0.527$\pm$0.003         & 198     \\
N 891       & 02 22 33.29 & +42 20 57.9  & -0.446$\pm$0.002           & 138   \\
N 2613$^g$  & 08 33 22.76 & -22 58 25.0  & -0.102$\pm$0.140           &646     \\
N 2683      & 08 52 41.31 & +33 25 18.0  & -0.070$\pm$0.016          &90.2   \\
N 2820      & 09 21 46.04 & +64 15 27.6  & -0.450$\pm$0.015           &404    \\
N 2992      & 09 45 41.89 & -14 19 33.5  & -0.653$\pm$0.001           &664    \\
N 3003      & 09 48 35.62 & +33 25 17.3  & -0.363$\pm$0.030         &392    \\
N 3044           & 09 53 40.78 & +01 34 47.2  & -0.448$\pm$0.004          &349    \\
N 3079    & 10 01 57.86 & +55 40 47.3  & +0.116$\pm$0.005      &    341    \\
N 3432$^d$ & 10 52 31.13& +36 37 07.6  & -0.125$\pm$0.110      &   138    \\
N 3448              & 10 54 39.33 & +54 18 21.90 & -0.389$\pm$0.005         &379    \\
N 3556          & 11 11 31.50 & +55 40 29.3  & -0.079$\pm$0.008       &206    \\
N 3628    & 11 20 16.97 & +13 35 15.4  & -0.341$\pm$0.003      &    146    \\
N 3735           & 11 35 57.19 & +70 32 08.1  & -0.505$\pm$0.006          &641   \\
N 3877$^g$       & 11 46 07.69 & +47 29 39.6  & -0.342$\pm$0.007         &262   \\
N 4013     & 11 58 31.38 & +43 56 50.7  & -0.515$\pm$0.003      &    237    \\
N 4096$^d$      & 12 06 01.13 & +47 28 42.4  & -0.327$\pm$0.083         &149   \\
N 4157          & 12 11 04.57 & +50 29 03.3  & -0.477$\pm$0.031         &214   \\
N 4192$^d$  &12 13 48.29& +14 54 01.2& -0.663$\pm$0.004      &  198      \\
N 4217$^g$       & 12 15 50.90  & +47 05 29.3 & -0.614$\pm$0.006          &292   \\
N 4244$^e$       & ---    &  ---              & ---  & ---    \\
N 4302$^f$  & 12 21 41.99  & +14 35 44.2 & -0.345$\pm$0.130      &    317    \\
N 4388           & 12 25 46.68  & +12 39 46.0 & -0.158$\pm$0.005           &274    \\
N 4438        & 12 27 46.11  & +13 00 29.7 & -0.674$\pm$0.012      &     156    \\
N 4565    & 12 36 20.9   & +25 59 15.0 & -0.527$\pm$0.016      &   176    \\
N 4594$^g$ & 12 39 59.43 &-11 37 23.0& +0.432$\pm$0.005      &    230    \\
N 4631              & 12 42 07.80  & +32 32 34.9 & -0.388$\pm$0.020          &114    \\
N 4666$^d$  &12 45 08.59  & -00 27 42.8 & -0.672$\pm$0.004      &    466    \\
N 4845$^g$  & 12 58 01.19 & +01 34 32.5 & +0.376$\pm$0.005      &      270    \\
N 5084$^g$  & 13 20 16.83&-21 49 38.5 & +0.077$\pm$0.009      &     444    \\
N 5297$^g$  & 13 46 23.7 & +43 52 20.5& -0.638$\pm$0.100      &    581    \\
N 5775$^f$  & 14 53 57.57& +03 32 30.6&  -0.487$\pm$0.010     &     484   \\
N 5792$^d$  & 14 58 22.71& -01 05 27.9 & -0.606$\pm$0.003     &      544   \\
N 5907$^g$  & 15 15 53.50& +56 19 43.5&-0.624$\pm$0.033   &   248     \\
U 10288$^e$ & ---   & --- & ---  &---  \\
\enddata
%% Text for table notes should follow after the \enddata but before
%% the \end{deluxetable}. Make sure there is at least one \tablenotemark
%% in the table for each \tablenotetext.
\tablecomments{Measured from the rob 0 weighted maps.}
\tablenotetext{a}{RA and DEC of the location of a flattest spectral index feature near the nucleus unless otherwise indicated.  Note that this is not necessarily at the NED center of the galaxy or at the peak in total intensity (see Sect.~\ref{sec:flatspec}).}
\tablenotetext{b}{Mean spectral index within a FWHM (from Table~\ref{tab:smoothed_data}). The error is the mean value from the related error map {(Eqn.~\ref{eqn:error_maps})} in the same region, or the variation that results from altering the position of the FWHM by approximately 1 pixel, whichever is larger. {See Sect.~\ref{sec:alpha_errors} for further discussion of the errors.}}
%\tablenotetext{d}{Flux density in the same FWHM region as above.  Uncertainties are $\approx\,5$\%.} 
\tablenotetext{c}{Linear size corresponding to the average of the major and minor axes (bmaj and bmin of Table~\ref{tab:smoothed_data}) of the spectral index maps.}
%\tablenotetext{f}{Spectral power corresponding to the flux density measured in the FWHM.}
%\tablenotetext{g}{Number of average M~82 SNRs ($\bar L_\nu\,=\,3\,\times\,10^{18}$ W Hz$^{-1}$) corresponding to the measured spectral power, rounded to an integer.} 
\tablenotetext{d}{No distinct spectral index feature; measured at NED center.}
\tablenotetext{e}{Too faint to measure.}
\tablenotetext{f}{Peak of $\alpha_{BL-CC}$ is offset $\approx$ 10 arcsec from the nucleus.}
\tablenotetext{g}{Measured at the central total intensity peak.}% Spectral index shows a distinct feature but with a gradient.}%judith this is because there is a gradient in alpha -- for n5907, no distinct alpha peak but the ned center is clearly off.
\end{deluxetable}
}  %end array

In Table~\ref{tab:alpha}, we also provide the linear size corresponding to the average beam size of the synthesized spectral index maps as well as the number of SNRs that would have to be present in this region, should those SNRs have the average spectral power of the SNRs in the well-known starburst galaxy, M~82. % Recall that these values apply only to a single FWHM at the position specified.

Using the flat spectral index criterion, we find that eight galaxies show evidence for AGNs.  These are denoted in Table~\ref{tab:AGNs}.

\subsection{XMM Spectra}

\begin{figure*}[htp]
\centering
  %  \begin{center}
    \vspace{-2truein}
\resizebox{0.4\hsize}{!}{\includegraphics{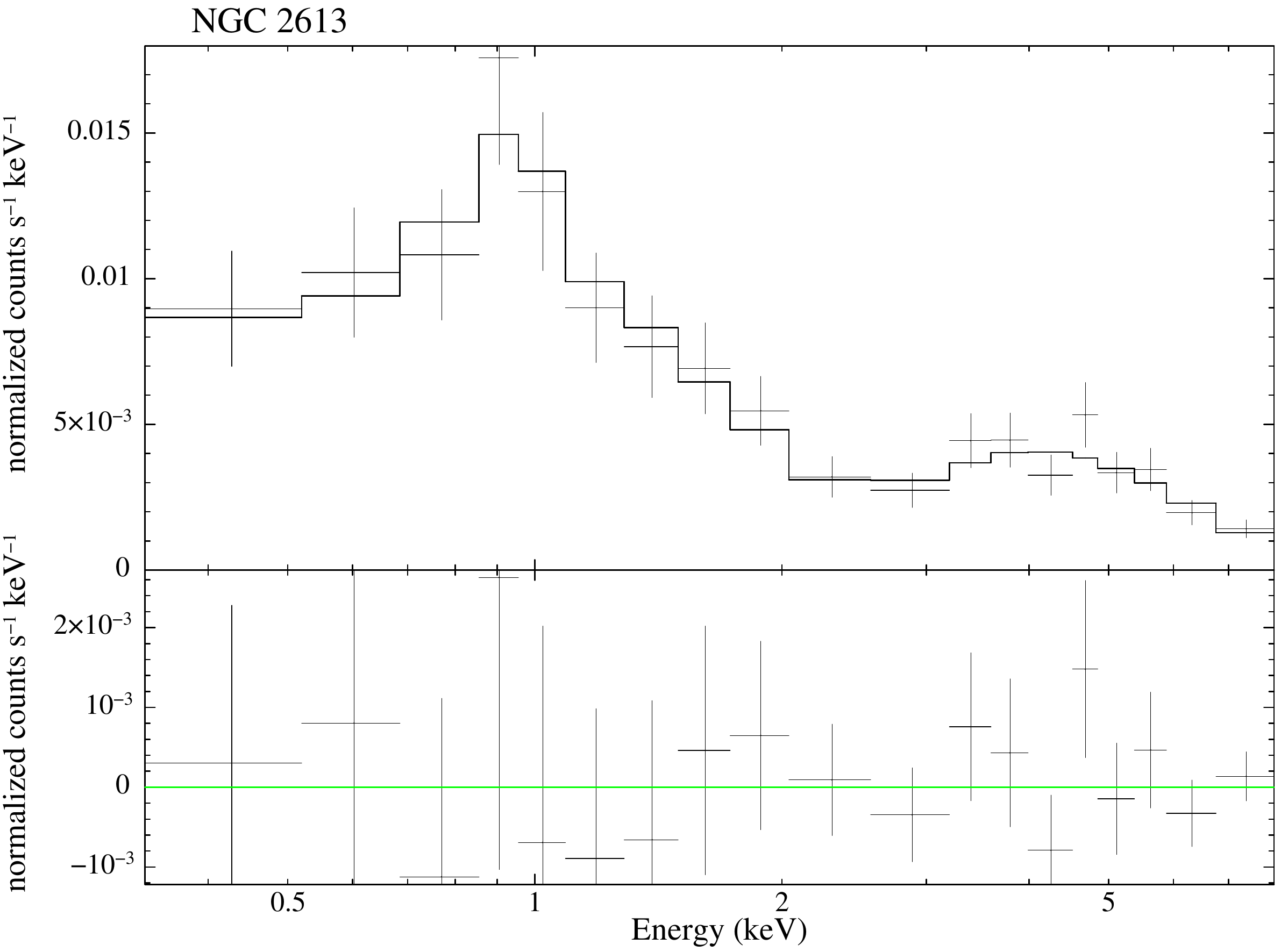}}
\resizebox{0.4\hsize}{!}{\includegraphics{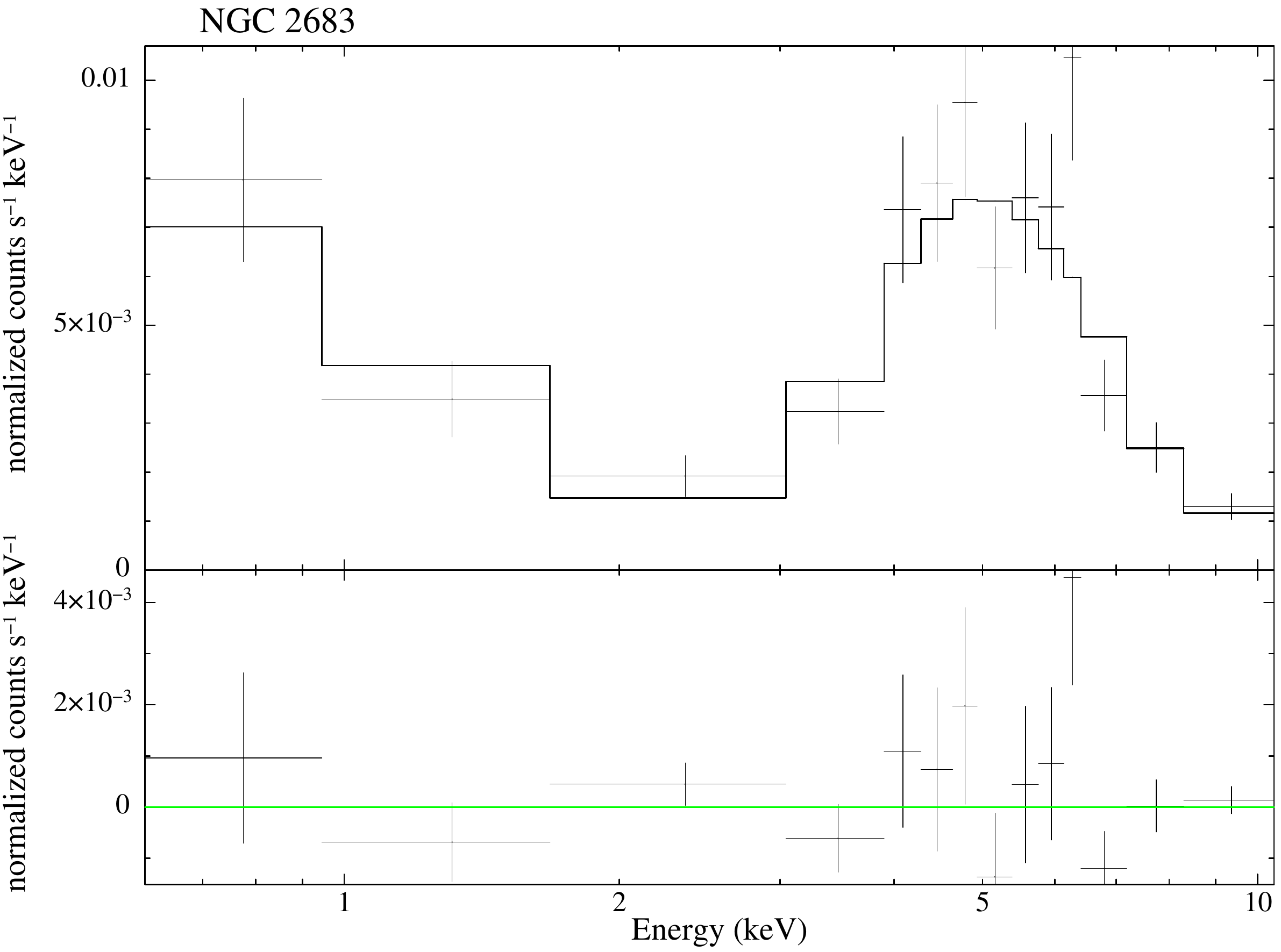}}
\resizebox{0.4\hsize}{!}{\includegraphics{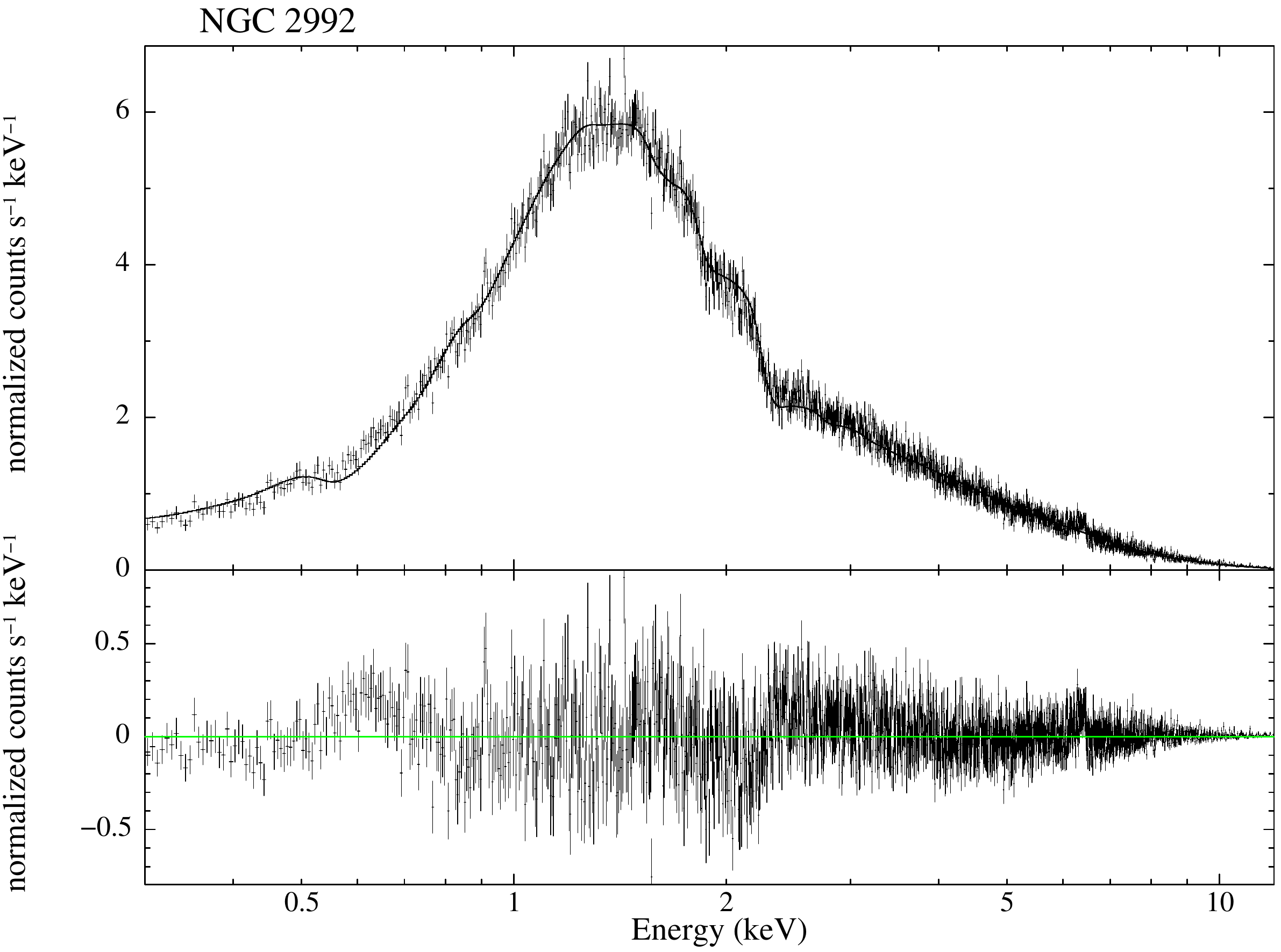}}
\resizebox{0.4\hsize}{!}{\includegraphics{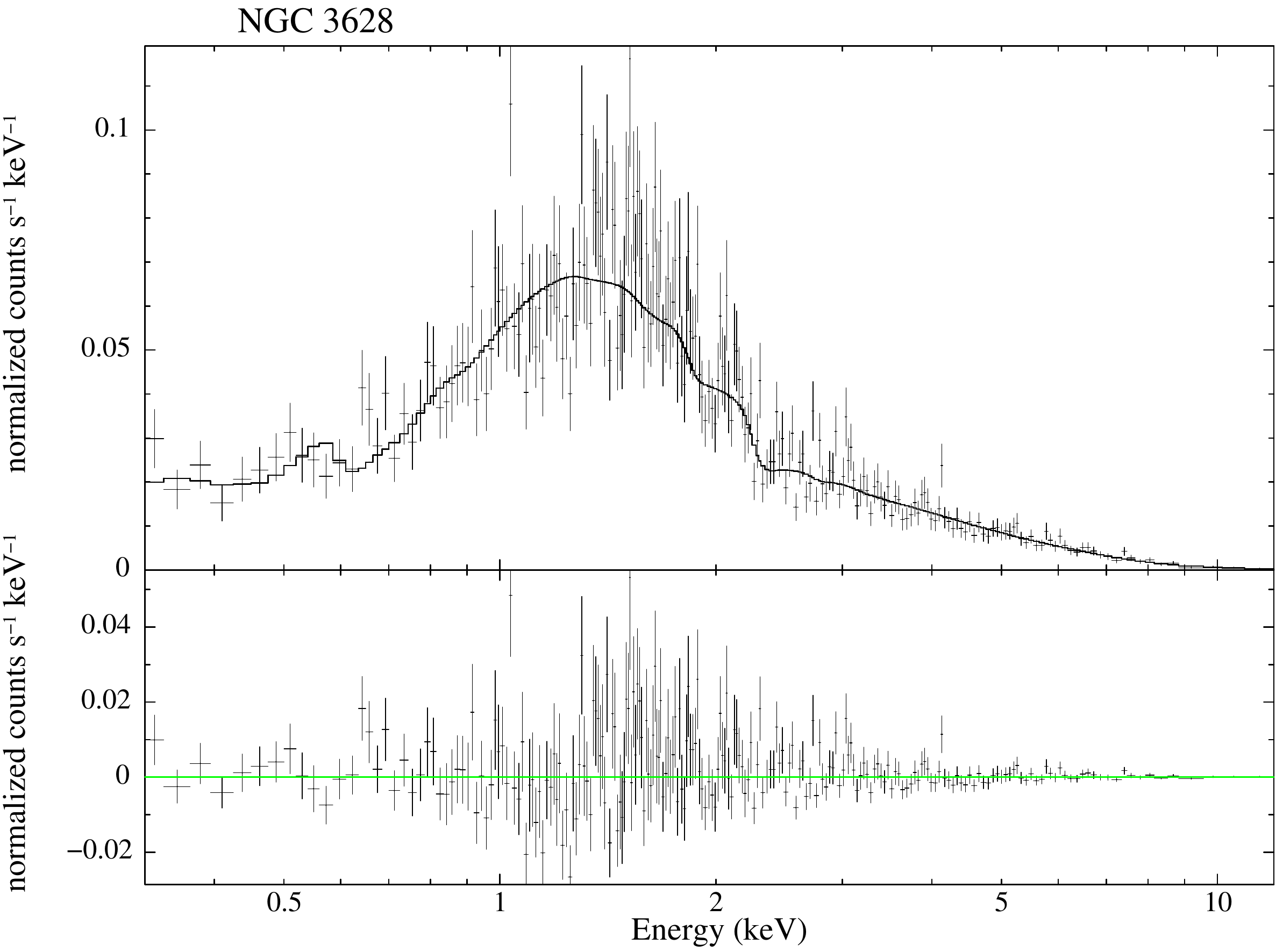}}
\resizebox{0.4\hsize}{!}{\includegraphics{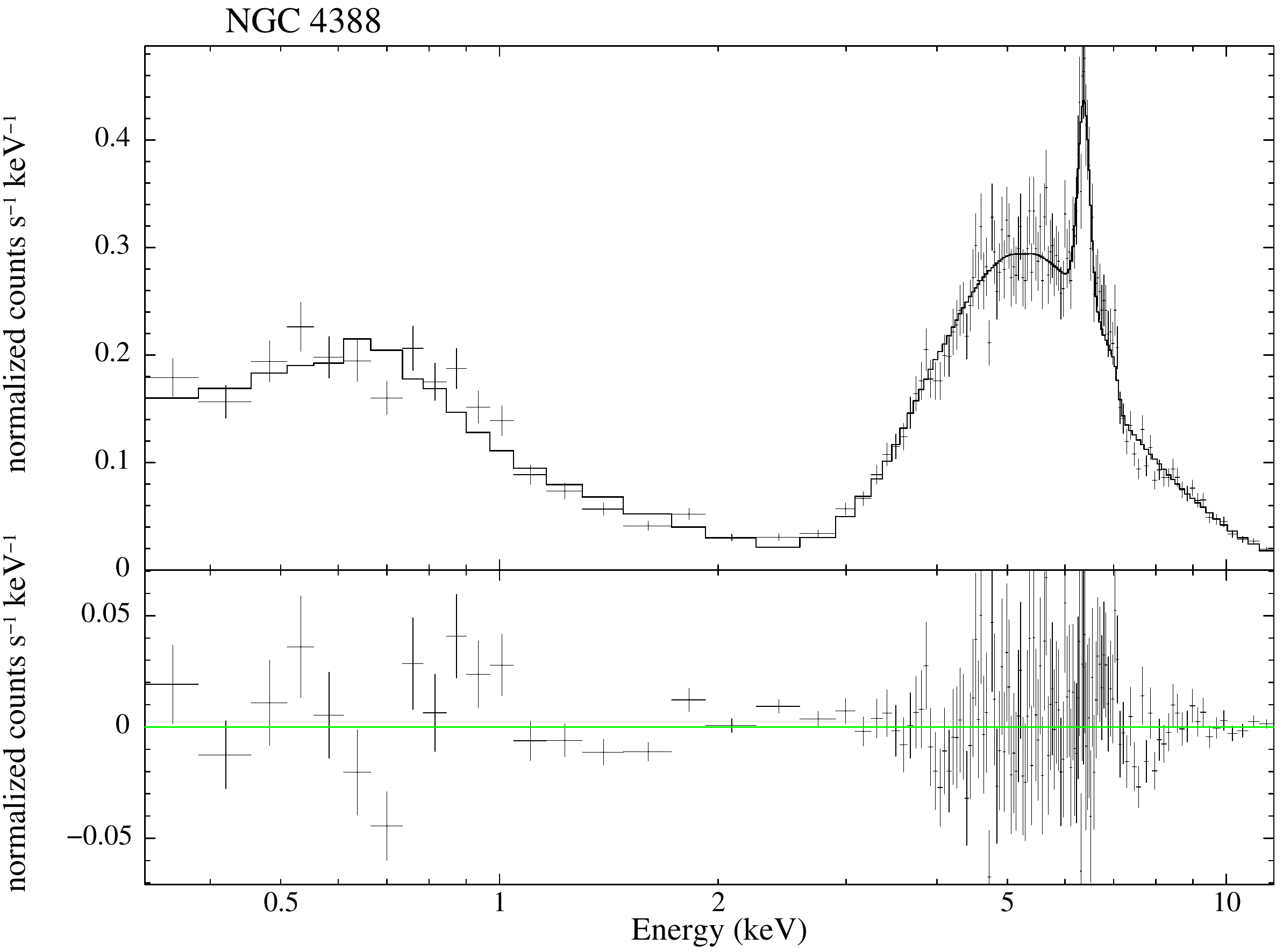}}
\resizebox{0.4\hsize}{!}{\includegraphics{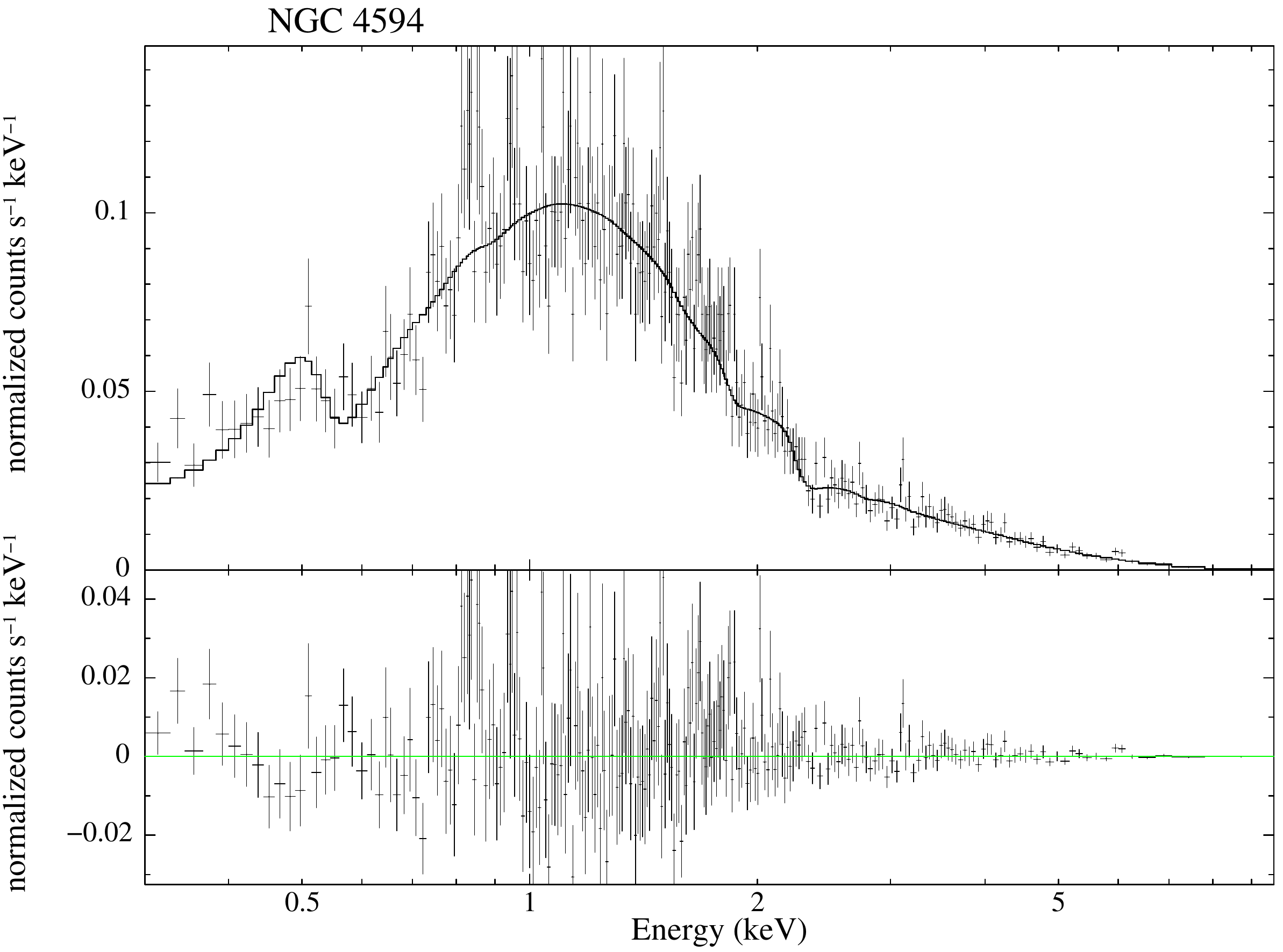}}
\resizebox{0.4\hsize}{!}{\includegraphics{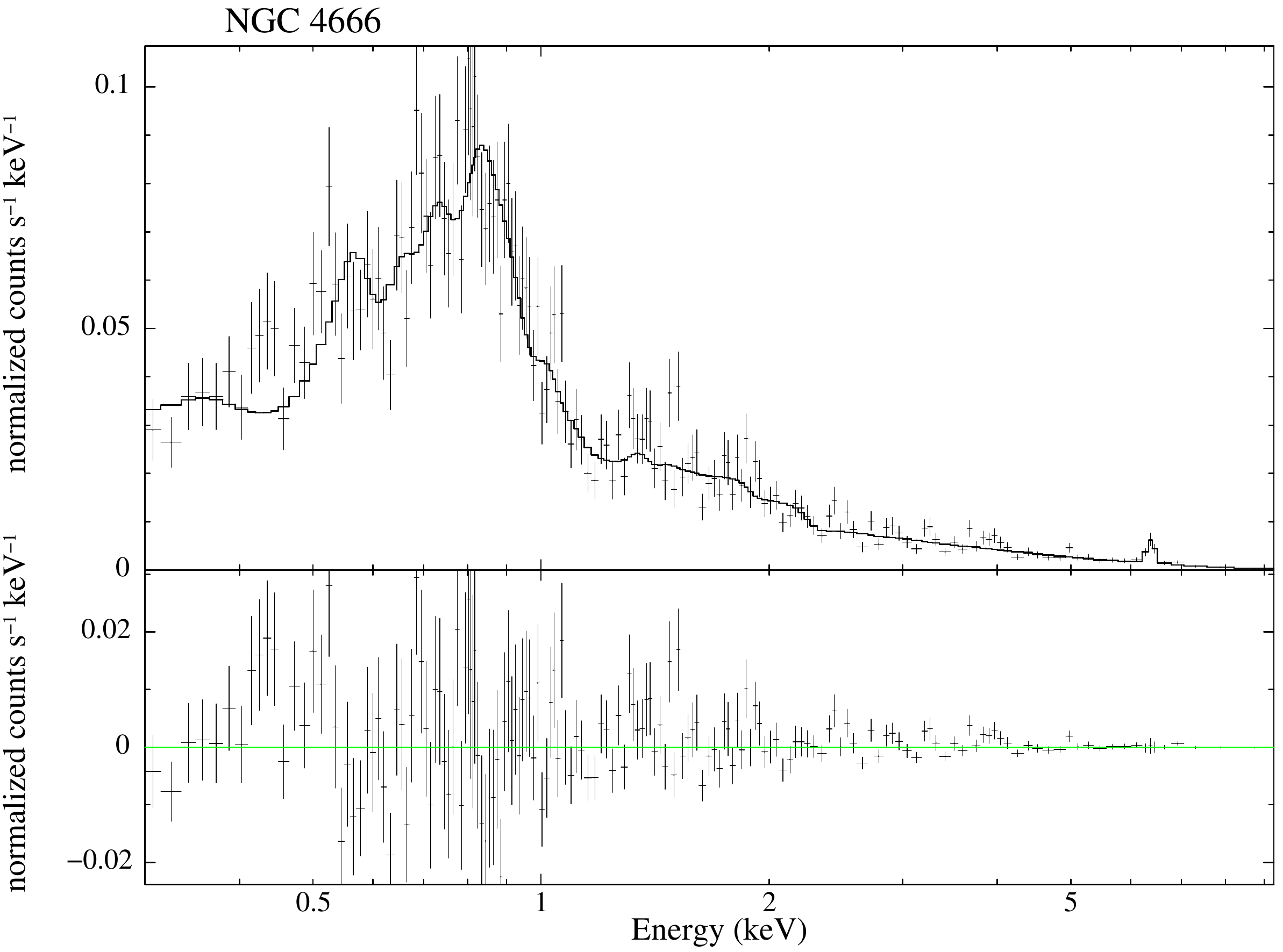}}
\resizebox{0.4\hsize}{!}{\includegraphics{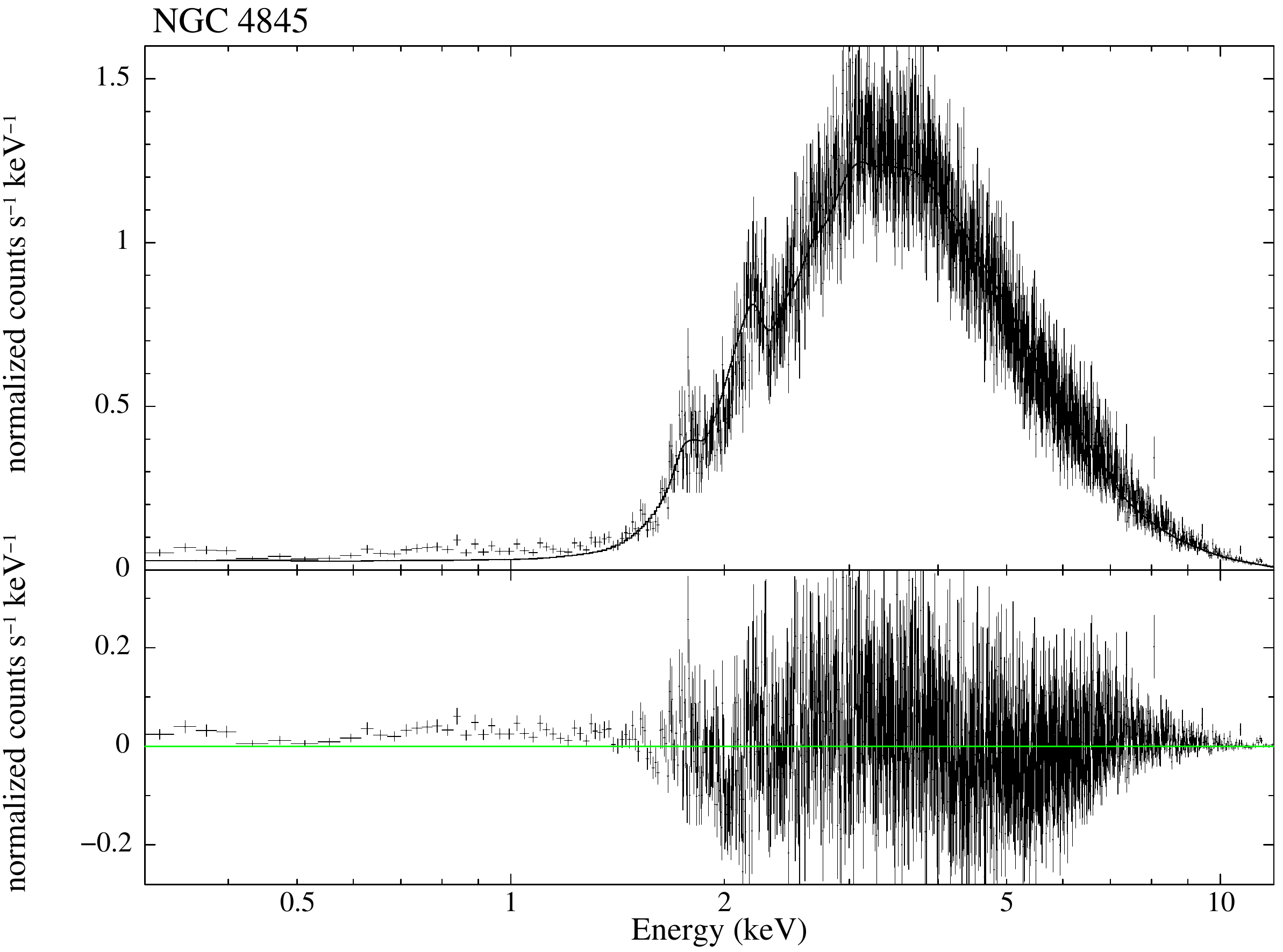}}
%\end{center}
\caption{Spectra of the X-ray selected AGNs. Data are points with error bars and curves are the best fit model.  Lower panels show the residuals (data - model).}
\label{spectraagns}
\end{figure*}

As noted in Sect.~\ref{sec:xmm_obs}, we have obtained new XMM data for 19 CHANG-ES galaxies.  Of these galaxies, eight show evidence from their X-ray emission as having an AGN (see Fig.~\ref{spectraagns}). The 19 galaxies are designated `Y' or `N' in Table~\ref{tab:AGNs}.

%(specified by either `Y' or `N' in Table~\ref{tab:AGNs}).  Of these galaxies, eight show evidence from their X-ray emission as having an AGN (see Fig.~\ref{spectraagns}) and we show the spectra of two more (Fig.~\ref{spectraradioagns}) which do not reveal an X-ray AGN but for which our radio criteria (see Sect.~\ref{sec:fraction_AGNS}) suggest that an AGN is present.

XMM data may simply not have the capability (sensitivity and resolution) to clearly detect the AGN in some `N' galaxies.  A scan through Table~\ref{tab:AGNs} illustrates that a number of galaxies which show evidence for a radio AGN don't always show an AGN from the XMM data.  For example, the `N' galaxy, NGC~660 clearly harbours an AGN \citep[][and CHANG-ES data]{arg15, sai18} as does NGC~3079 and others in this list.

Spectra of two galaxies, NGC~4388 and NGC~4666, as mentioned in Sect.~\ref{sec:xmm_obs}, show an iron Fe-K{$\alpha$} line around 6.4\,keV, which is a direct argument for the
existence of an AGN \citep[eg.][and the references therein]{guainazzi05}.

In two more galaxies, NGC~2683 and NGC~2992, a hint for such a line is visible in their spectra, although no reliable 
model fit of this line could be performed. Nevertheless, for these two galaxies, as for several other objects,
a significant level of hard emission (above 2\,keV) can be observed (see Fig.~\ref{spectraagns}), suggesting 
strong non-thermal source(s) of X-ray radiation in the galactic core. Although the resolution of the observations does not allow us to clearly determine that this emission comes from a single central 
source, we can still assume that it contributes significantly to the observed hard X-ray emission. This seems to be justified because typically we expect AGN luminosities 
of the order of 10$^{40}$\,{\rm erg\,s$^{-1}$} or more, while galactic X-ray binaries show only moderate hard emission and have luminosities of 10$^{38}$--10$^{39}$\,{\rm erg\,s$^{-1}$} 
\citep[see eg.][]{wezgowiec16}. Therefore, luminosities derived by us and presented in Table~\ref{lumins} are a reliable estimate for each galaxy, most likely agreeing  
with the real AGN luminosity within the presented uncertainty.

For the galaxies, NGC~2613, and NGC~2683, the poor quality of the extracted spectrum, resulting from short observations and lower apparent brightness, caused large uncertainties 
of the derived luminosity values (Table~\ref{lumins}). Nevertheless, the overall spectral distribution of the X-ray emission with the significant part in the hard range,
suggests that these might indeed be luminous AGN-type sources. For a clear confirmation, more sensitive data are certainly needed, especially since in the spectrum of 
NGC\,2683 a hint for an iron Fe-K{$\alpha$} line can be seen, as mentioned above. 

We note here, that the candidate AGNs presented in Fig.~\ref{spectraagns} have the highest X-ray luminosities (see Table~\ref{lumins}) of the studied galaxy sample.
%The two radio-selected AGNs of Fig.~\ref{spectraradioagns} are the two least luminous of the potential AGNs discussed here.

%For high resolution, however, we first search through the literature for information that may be relevent for determining whether or not AGNs are present.  For example, \cite{she17a}  have compiled a list of 719 galaxies within 50 Mpc and identified 314 of them to have an AGN, the majority being LLAGNs.  Twenty-one of the CHANG-ES galaxies were in their sample.
%see online table 2 from this paper
%N660: Y
%N891: Y
%2613: not obs
%N2683: Y
%N2820: not obs
%N2992: Y
%N3003:not obs
%N3044: not obs
%N3079: Y
%N3432: N
%N3448:not obs
%N3556: ditto
%N3628: N
%N3735: not obs
%N3877: Y
%N4013: Y
%N4096: N
%N4157: N
%N4192: not obs
%N4217: Y
%N4244: N
%N4302: not obs
%N4388: Y
%N4438: Y
%N4565: Y
%N4594: not obs
%N4631: N
%N4666: Y
%N4845: not obs
%N5084: Y
%N5297: not obs
%N5775: Y
%N5792: not obs
%N5907: N
%U10288: not obs

\subsection{The Incidence of AGNs in the CHANG-ES Sample}
\label{sec:fraction_AGNS}

Table~\ref{tab:AGNs} summarizes the evidence for AGNs in the CHANG-ES sample. A`Y' in the final column refers to galaxies for which the {\it radio} criteria point to an AGN.

Galaxies for which only a single radio criterion suggests AGN activity are NGC~3432 and NGC~4565 and therefore these could be considered to be the weaker cases. However, both of these galaxies have been  optically identified as either LINERS or Seyferts strengthening the case for an AGN for these galaxies.

The galaxies which have a single `Y' plus a question mark in the `Lobe(s)' column are NGC~3556 and NGC~5297.   For NGC~3556, \cite{sat08} find no optical evidence for an AGN, but \cite{wan03} have identified a possible AGN candidate from X-ray observations. As for NGC~5297, we know of no independent evidence for an AGN nor is there a NED nuclear classification for this galaxy. These two galaxies are new AGN candidates, i.e. galaxies that have previously been specified as having only an HII-type nuclear spectrum.

The remaining new detection is NGC~2613, which has a point-core with a flat spectrum.  Even though only a few SNRs could account for the luminosity of the emission (Table~\ref{tab:pointcores}), the very flat spectrum ($\alpha\,=\,-0.1$, Table~\ref{tab:alpha}) is unlikely to result from a collection of SNRs.  The spectrum is consistent with thermal emission alone, but we would then require a collection of HII regions right at the galaxy's nucleus without SNRs. Finally, the XMM data show a hard spectrum, confirming our conclusion from the radio criteria.

{\renewcommand{\arraystretch}{1.1}
\begin{deluxetable}{l|c|c|ccccc|c|}
\tabletypesize{\scriptsize}
%\rotate
\tablecaption{AGN Candidates \label{tab:AGNs}}
\tablewidth{0pt}
\tablehead{
  \colhead{Galaxy} &  \colhead{NED type$^a$} &\colhead{XMM$^b$} &\colhead{Flat $\alpha_{BL-CC}\,^{c}$} & \colhead{Point Core$^d$} &  \colhead{High $L_\nu\,^{e}$} & \colhead{Lobe(s)$^f$}
 &\colhead{CP$^g$} & \colhead{AGN$^h$}  
 }
\startdata 
N 660  & SB(s)a pec; HII,LINER   & N & Y &  Y  &  Y  & Y  & Y     & {\bf Y}   \\
N 891  & SA(s)b? sp; HII         & N &  &     &     &    &     & {\bf N}  \\
N 2613 & SA(s)b; HII             & Y & Y &   Y &     &    &     & {\bf Y}   \\
N 2683 & SA(rs)b; LINER,Sy2      & Y &  Y &   Y &     &    &     & {\bf Y}  \\
N 2820 &  SB(s)c pec sp          & --- &  &      &     &    &     & {\bf N}   \\
N 2992 & Sa pec; Sy1.9           & Y & &      &  Y  & Y  &     & {\bf Y}    \\
N 3003 &  SBbc                   & --- &  &     &     &    &     & {\bf Y}   \\
N 3044 & SB(s)c? sp; HII         & --- &   &     &     &    &     & {\bf N}   \\
N 3079 & SB(s)c; LINER,Sy2       & N & Y &   Y & Y   & Y  &  Y   & {\bf Y}    \\
N 3432 & SB(s)m; LINER,HII       & --- & Y&     &     &    &     & {\bf Y}  \\
N 3448 & I0                      & --- &  &     &     &    &     & {\bf N}   \\
N 3556 & SB(s)cd; HII            & --- & Y &     &     & Y? &     & {\bf Y}   \\
N 3628 & SAb pec sp; HII,LINER   & Y &  &     & Y   & Y  &  Y   & {\bf Y}   \\
N 3735 & SAc: sp;  Sy2           & --- &   &  Y  &  Y  &    &     & {\bf Y}   \\
N 3877 & Sc; HII                 & --- &   &    &     &    &     &  {\bf N}  \\
N 4013 & SAb; HII,LINER          & N &   &     &     &    &     & {\bf N}   \\
N 4096 &  SAB(rs)c; HII          & --- &   &    &     &    &     & {\bf N}  \\
N 4157 & SAB(s)b? sp; HII        & N &   &     &     &    &     & {\bf N}  \\
N 4192 & SAB(s)ab; HII,Sy,LINER  & --- & &    &     &    &     & {\bf N}   \\
N 4217 & SAb sp; HII             & --- & &        &     &    &     & {\bf N}  \\
N 4244$^i$ &SA(s)cd: sp; HII     & --- & ---     &  ---& ---& ---    &--- & ---   \\
N 4302 &Sc: sp; Sy,LINER         & N &  &  &  &          &     & {\bf N}  \\
N 4388 & SA(s)b: sp; Sy2,Sy1.9   & Y &Y &    Y  &  Y  & Y &  Y   & {\bf Y}  \\
N 4438 & SA(s)0/a pec:; LINER    & N & &    Y   &  Y  & Y &     & {\bf Y}  \\
N 4565 & SA(s)b? sp; Sy3,Sy1.9   & N &&   Y &   &         &     & {\bf Y}  \\
N 4594 &SA(s)a; LINER,Sy1.9      & Y & & Y      & Y   & &     & {\bf Y} \\
N 4631 &  SB(s)d;  HII           & N &  &    &    &      &     &  {\bf N}  \\
N 4666 &SABc:; HII,LINER         & Y &&  Y     &  Y   &  &     & {\bf Y}  \\
N 4845 &SA(s)ab sp; HII,LINER    & Y & & Y      & Y  &  &  Y   & {\bf Y}  \\
N 5084 & S0; poss. LINER         & --- &Y & Y       &Y   & Y    &    & {\bf Y}   \\
N 5297 & SAB(s)c: sp             & --- & &  Y   &    & Y?&     & {\bf Y}  \\
N 5775 & Sb(f)                   & N & & &  &          &     & {\bf N}  \\
N 5792 &SB(rs)b; HII             & --- &  &      &     & &     & {\bf N}  \\
N 5907 &SA(s)c: sp; HII          & N & &  &  &     &     & {\bf N} \\
U 10288$^j$&   Sc                & --- & ---   & --- & ---  &--- & ---& ---\\
\enddata
%% Text for table notes should follow after the \enddata but before
%% the \end{deluxetable}. Make sure there is at least one \tablenotemark
%% in the table for each \tablenotetext.
\tablecomments{Radio data were measured from the rob 0 weighted maps.}
\tablenotetext{a}{Galaxy classification and `activity type' from NED using the classic interface. Some of these entries have been updated since \cite{irwI}.}
\tablenotetext{b}{Galaxies for which XMM spectra suggest an AGN are designated `Y' and shown in Fig.~\ref{spectraagns}.
Galaxies for which XMM data do not clearly detect an AGN are designated `N' and --- indicates that no data are available.  See also \citet{ste19} for details on NGC~4666.}
\tablenotetext{c}{Galaxies that have a flat spectral index near the nucleus (Sect.~\ref{sec:flatspec} and Table~\ref{tab:alpha}).}
\tablenotetext{d}{Galaxies with nuclear point-like cores (Sect.~\ref{sec:point-like} and Table~\ref{tab:pointcores}).} 
\tablenotetext{e}{Galaxies that have a nuclear luminosity that exceeds what is normally expected for SNRs (Sect.~\ref{sec:radio_lums} and Tables~\ref{tab:pointcores} and \ref{tab:nopointcores}). }
\tablenotetext{f}{Galaxies showing evidence for radio lobes (Sect.~\ref{sec:jets}).}
\tablenotetext{g}{Galaxies showing Circular Polarization \citep{irwXI}.} 
\tablenotetext{h}{Final summary of AGN candidates based on radio criteria.}
\tablenotetext{i}{Emission too weak for measurements near the core.}
\tablenotetext{j}{Emission confused with background source \citep{irwIII}.}
\end{deluxetable}
}  %end arraystretch

%N3556:Hi Judith,
%Wang, Chaves and Irwin 2003, where you found a source with a typical AGN
%x-ray spectrum.
%Also Satyapal 2018 found that the (optical?) emission of 3556 could have 10\%
%coming from the AGN.
%But judging from the coordinates that are given in wang et. al, and the
%B Array L Band image (3" resolution), this is certainly not in the
%center of the galaxy, but offset to north-east. Maybe a background
%AGN..., but yes, with little contribution.

It is interesting that all galaxies for which the X-ray data suggest an AGN (Fig.~\ref{spectraagns}) have also been identified as having an AGN via our radio criteria.  The converse, however, is not true.  %Two galaxies which show strong radio evidence for an AGN, NGC~3079 and NGC~660 (Fig.~\ref{spectraradioagns}) show no evidence for an X-ray selected AGN.
This could be because of the resolution of the XMM data, but given the fact that these data are capable of identifying AGNs in many of our other sources, the mismatch could be because of outburst timescales; for example, an X-ray decline could occur in advance of a radio decline.

Several galaxies with known or suspected nuclear activity have been missed using our criteria, namely
NGC~4013 and NGC~4192, and NGC~4302.  All three galaxies have very strong centrally concentrated radio emission (see Appendix~\ref{app:B}) which, nevertheless, is somewhat resolved \citep[see also][for details of N~4013]{ste19b}.  They also have steep nuclear spectral indices.  The same is true for the corresponding CC data at similar resolution.  These three galaxies may have been missed because of our strict AGN requirement for an unresolved nuclear radio source.  It is likely that a radio AGN is indeed embedded in other emission and the steep spectral indices are the result of masking by the other emission, similar to what was found for NGC~4438 (Sect.~\ref{sec:jets} and Fig.~\ref{fig:jets}).  Higher resolution radio observations are recommended for these galaxies.

Our detection rate (excluding the three galaxies noted in the above paragraph) is 18/33 or 55\%.  We take this as a lower limit to the AGN detection rate in our sample based on radio criteria alone.  Note that the selection criteria for CHANG-ES galaxies \citep{irwI} did not include any reference to galaxies with AGNs nor did it favour the inclusion of AGNs.  If we include the three galaxies for which independent evidence exists for an AGN, then the detection rate of AGNs amongst `normal' nearby spiral galaxies is 64\%.

Chandra observations by \cite{she17a} and \cite{she17b} suggest that the incidence of AGNs in nearby galaxies ranges from 60\% in elliptical galaxies to 20\% in Sc and later types, and they find 314 AGN candidates that had previously only been designated as HII region-type nuclear spectra. Although we have many fewer galaxies, our detection rate is roughly consistent with \cite{she17b} (see their Table~1) in which they find about 50\% in S0 to 39\% in Sc and later types for nearby galaxies.

She et al. identify AGN candidates by finding those galaxies that have an X-ray point source at the location of the galaxy center.  Their AGN candidates include the CHANG-ES galaxies  NGC~891, NGC~3877, NGC~4217, and NGC~5775, for which we find no evidence for a radio AGN.  It is possible that their list may have been overestimated \citep[for example, see][for NGC~891]{hod18}. However, an intriguing notion is that a radio AGN with sufficiently high luminosity for detection has yet to emerge after an X-ray detection.  A detailed investigation of the radio/X-ray connection in AGNs for the CHANG-ES sample is beyond the scope of this paper, but the 
association between radio and X-ray emission in AGN \citep[e.g.][]{pan14} as well as low luminosity AGNs (LLAGNs) \citep[][]{su17} is well established.

%In addition, we supplement the VLA data with X-ray spectral information from a subset of 19 CHANG-ES galaxies obtained from the X-ray Multi-Mirror Mission, or XMM-Newton (hereafter XMM) satellite.
%The association between radio and X-ray emission in AGN \citep[e.g.][]{pan14} as well as in low luminosity AGNs (LLAGNs) \citep[][]{su17} is well established.
%{ She 2017b find that 31\%
%51/163
%of optically classified HII
%nuclei have an X-ray core. ... The luminosity of low mass x-ray binaries cuts off at 
%a few 10$^{38}$ erg/s and there are few sources above $10^{39}$ (Gilfanov 2004) so higher values suggest an AGN %-- implications for low-mass BHs and also BHs in bulge-less galaxies}
%{ See Satyapal 2018 -- for late type galaxies}

\section{High Resolution In-Disk Emission in CHANG-ES Galaxies}
\label{sec:disks}

An abundance of information is available from the high resolution BL data and to help in understanding the origin of the discrete radio emission, we provide overlays of radio emission on H$\alpha$ maps obtained by \cite{var19}  (Appendix~\ref{app:B}, Frame g).  The H$\alpha$ maps, in standard FITS format have been presented in Data Release 2. In this section, we examine only two issues of relevance to high resolution in-disk emission.

\subsection{Flat Spectral Indices in the Disks of CHANG-ES Galaxies}
\label{sec:flat_disks}

In Sect.~\ref{sec:alpha_thermal}, we noted that the thermal component is typicaly about 8\% of the total emission at L-band.  However, since the resolution of our BL data is of order a few hundred parsecs (Tables~\ref{tab:pointcores}, and \ref{tab:nopointcores}), specific HII regions or HII region complexes in the disk could be resolved as discrete objects.  In such regions, non-thermal emission from SNRs can still be important, but the thermal contribution could also be significant. In fact, various CHANG-ES galaxies show many regions of flat spectral index in the disk far from the nucleus.  Such regions in the disk obviously cannot be attributed to AGNs. 

\begin{figure}[!tbp]
%   \vspace{-1.5truein}
  \centering
  \begin{minipage}[b]{0.50\textwidth} 
  \includegraphics[width=0.85\textwidth,trim=0 12cm 0 0,clip]{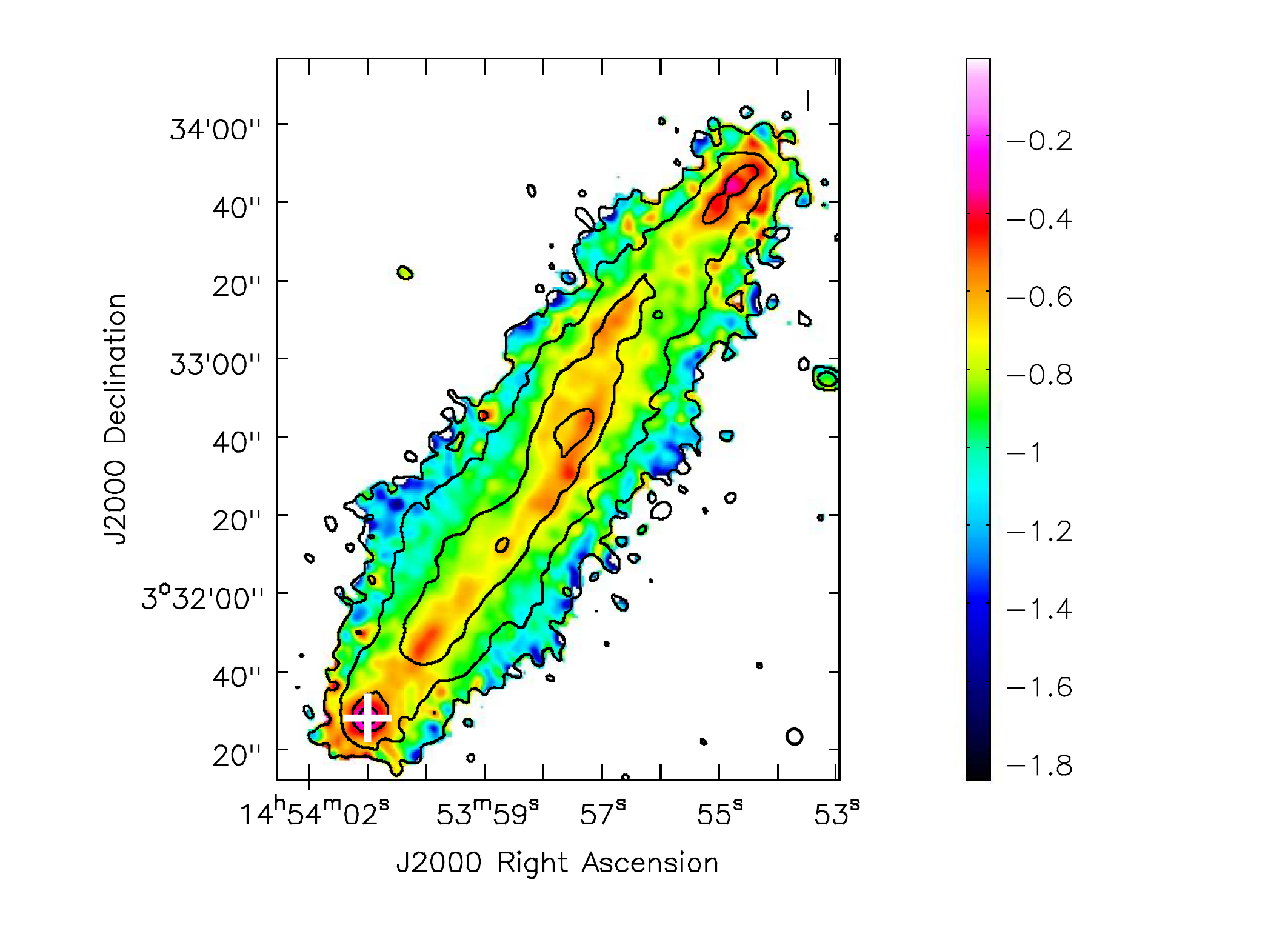}
  \end{minipage}
%    \vspace{-2truein}
  \centering
  \hfill
  \vspace{0.4truein}
  \begin{minipage}[b]{0.7\textwidth} 
  \includegraphics[width=0.6\textwidth,trim=0 12cm 0 0,clip]{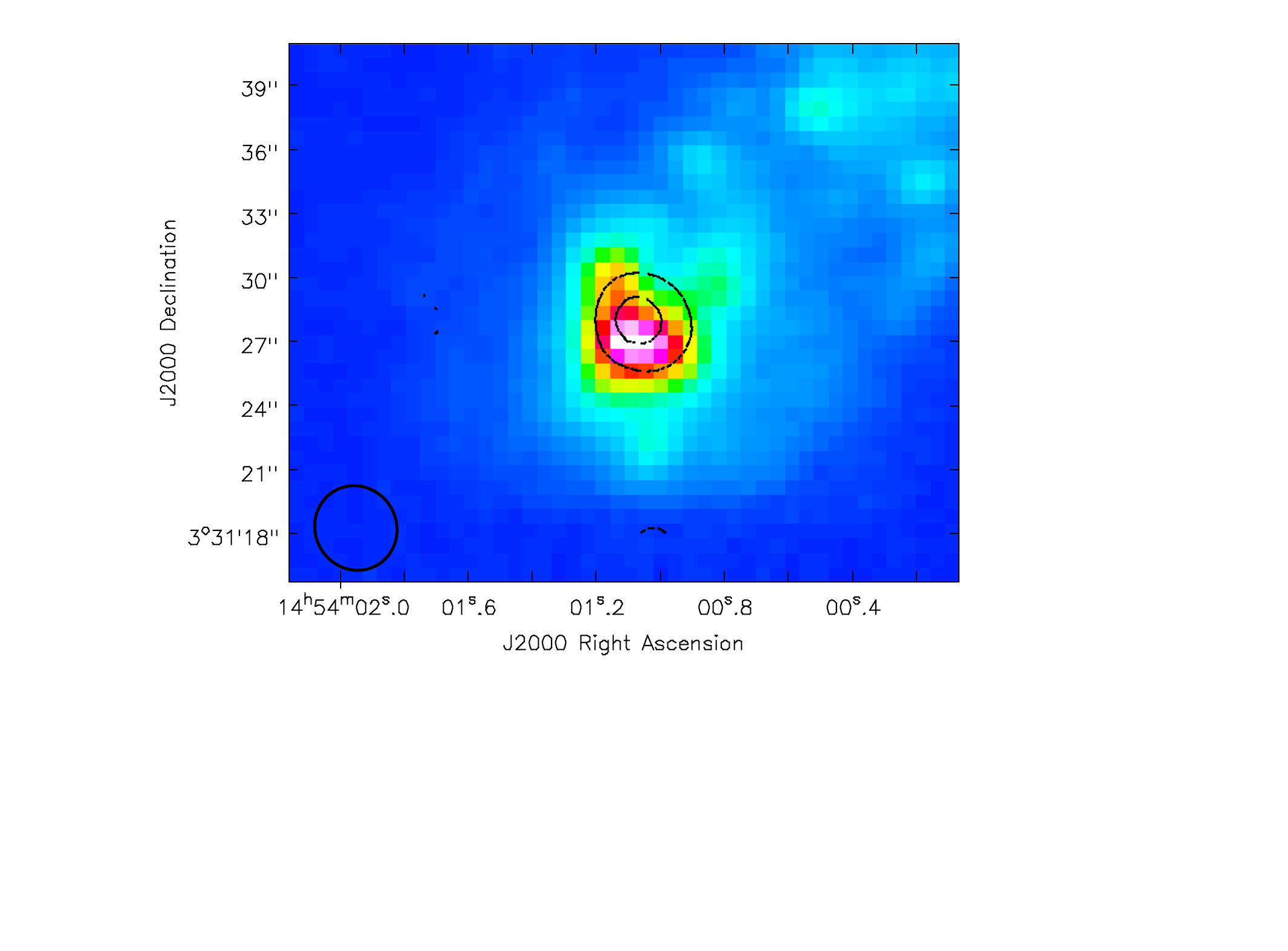}
  \end{minipage}
%    \vspace{-2truein}
  \caption{NGC~5775: {\bf (Top)} Total intensity BL rob0 contours over spectral index image with colour scale shown at right. The beam of the spectral index image is shown at lower right. Contours are at 42 (3$\sigma_I$), 150, 500, and 1400 $\mu$Jy beam$^{-1}$. The HII region complex discussed in Sect.~\ref{sec:flat_disks} is at the far SE of the disk and denoted with a `plus' sign.
  {\bf (Bottom)} Blow-up of the SE HII region complex showing two dashed spectral index contours ($\alpha\,=\,-0.3$ and $-0.25$, the latter is smaller) over the H$\alpha$ image from \cite{col00} in colour. The H$\alpha$ pixel size is 0.68 arcsec.  The spectral index beam size is shown at lower left.}
\label{fig:n5775}
\end{figure}

We consider only one example, the galaxy, NGC~5775, and examine a region of relatively flat spectral index far out in its disk, marked with a `plus' sign in Fig.~\ref{fig:n5775} (top: total intensity BL contours over $\alpha_{BL-CC}$ in colour). A blow-up of this region is displayed in the same figure (bottom:  $\alpha_{BL-CC}$ contours over an H$\alpha$ image in colour).  Here, the mean spectral index in a region of $\approx\,546$ pc in diameter (FWHM beam size) is $\alpha_{BL-CC}=-0.26\,\pm\,0.01$, where the error is also an average over the beam applying the (in this case negligible) corrections of  Appendix~\ref{app:A}. %using regions =-0.25926 in viewer mean of error map is 0.00629 but increased from maple script
%gives factor of 1.02 at a distance of 1.49 arcmin from the center so no real change
Emission from purely thermal gas should give a flatter spectral index ($I_\nu\,\propto\,\nu^{-0.1}$) and emission from purely non-thermal gas (e.g. a collection of SNRs) should give steeper spectral indices, on average ($\alpha\,\approx\,-0.7$, see Sect.~\ref{sec:flatspec} for examples).  We consider possible explanations below. In the following, we designate the observed $\alpha_{BL-CC}$ as simply $\alpha$.
%judith  BL-CC spectral index map average fwhm beam size is 546.4 pc.

\subsubsection{Flattening of the Spectral Index due to Foreground Thermal Absorption}
\label{sec:flattening_1}

Here we consider non-thermal emission whose spectral index is flattened by thermal absorption from a foreground screen of thermal gas.

Free-free absorption requires an EM of $5\,\times\,10^7$ pc cm$^{-6}$ for a thermal optical depth, $\tau_{\nu_{TH}}\,=\,1$, at 3.8 GHz with electron temperature, $T_e\,=\,10^4$ K (Eqn.~\ref{eqn:tauth} below). The corresponding electron density would have to be $n_e\,\approx\,300$ cm$^{-3}$ throughout a line-of-sight equal to the diameter of the region (546 pc). However, we require thermal absorption from a foreground in which case the line-of-sight should be reduced and the density correspondingly increased (e.g. a factor of 2 decrease in distance corresponds to a factor of $\sqrt{2}$ increase in density).  These are extreme values for the far `edge' of a disk although other exceptional extragalactic HII regions do exist.  For example, 30 Doradus shows a wide range of densities in complex structures \citep[e.g.][]{sco98, tsa05} with typical values of $\approx\,100$ cm$^{-3}$ and much higher densities in smaller filaments \citep{pel10}.

A lower but significant optical depth could also flatten the spectral index.  For example, for otherwise the same parameters but  $n_e\,=\,150$ cm$^{-3}$ ($\tau\,=\,0.25$), if the non-thermal spectral index were $\alpha_{NT}\,=\,-0.76$, the observed spectral index would be flattened to the observed value of $\alpha\,=\,-0.26$
\citep[see][their Eqn.~56]{irwV}.

However, we have stronger constraints available on the spectral index from the observed {\it in-band} spectral indices at L-band (over a 512 MHz bandwidth) and at C-band (2 GHz bandwidth).  Although these in-band spectral indices have a lower S/N, they are still sufficiently accurate for comparison (e.g. the BL signal-to-noise at this location is $>100/1$). We find that the observed $\alpha$ at {\it both} L-band and C-band agree with the band-to-band value within errors.  That is, the observed spectral index is constant from L-band through C-band.

We can now ask what the non-thermal spectral index would be at each frequency if  $n_e\,=\,150$ cm$^{-3}$ so that the observed spectral index remains constant at -0.26.   We find that we require $\alpha_{NT}\,=-3.4$ within L-band,  $\alpha_{NT}\,=-0.75$ band-to-band (as calculated above) and $\alpha_{NT}\,=-0.45$ within C-band.  In other words,  $\alpha_{NT}$ would have to be strongly varying with frequency in order for the observed value to remain constant.  This scenario seems highly unlikely and we therefore rule out thermal absorption as the cause of a relatively flat spectral index at the location of the SE HII region in NGC~5775.

\subsubsection{Flattening of the Spectral Index due to a Thermal Contribution}
\label{sec:flattening_2}

Here, we consider that $\alpha$ has been flattened due to an optically thin thermal emission fraction that is higher than the estimated global values of 8\% at L-band and 20\% at C-band (Sect.~\ref{sec:alpha_thermal}).
%JUDITH equation 56 of the paper in appendix D.  alpha_eff=alpha+2kappa_nu*s.

 Then, with $NT$, and $TH$ referring to  non-thermal and thermal quantities, respectively, and $I$ being the specific intensity, the standard equations are,
%more extreme values that smooth out at lower res.  e.g. N5775 average alpha are
%bl-cc 3sigma:  -0.823
%cl-dc 5sigma:  -0.892
%bl-cc 3sigma smoothed to cl-dc resolution: -0.806   all measured over same region
%In the following, we consider contributions from both non-thermal emission, $I_\nu(NT)$, as well as thermal emission, $I_\nu(TH)$, allowing for the possibility of thermal absorption of a non-thermal background and thermal self-absorption.  
\begin{eqnarray}
  I_{\nu_{obs}}&=& K\nu^{\alpha}\,=\,\,I_{\nu_{NT}}\,+\,I_{\nu_{TH}}\label{eqn:Itot}\\
              &=& C\nu^{\alpha_{NT}}+{\left(\frac{2\nu^2k}{c^2}\right)}{T_e\tau_{\nu_{TH}}}\label{eqn:Itot2}
\end{eqnarray}
where $K$ and $C$ are constants, and $k$ and $c$ are Boltzmann's constant and the speed of light, respectively, and we have employed the Rayleigh-Jeans relation in Eqn.~\ref{eqn:Itot2}. We have two sets of equations, one for each of BL and CC data at matching resolutions.
The thermal optical depth is well known, i.e.
\begin{equation}
  \label{eqn:tauth}
 \tau_{\nu_{TH}}\,=\,8.24\,\times\,10^{-2}\left[\frac{T_e}{\rm K}\right]^{-1.35}\left[\frac{\nu}{\rm GHz}\right]^{-2.1}\left[\frac{\rm EM}{{\rm pc \,cm}^{-6}}\right]
\end{equation}

For an adopted value of EM (assuming $T_e\,=\,10^4$ K),  $I_{\nu_{TH}}$ is easily calculated, and $I_{\nu_{NT}}$ is obtained from Eqn.~\ref{eqn:Itot}, from which the thermal fraction,  $f_{\nu_{TH}}=I_{\nu_{TH}}/I_{\nu_{obs}}$ follows for each band.  With $I_{\nu_{NT}}$ at both bands, the non-thermal spectral index, $\alpha_{NT}$ is found along with the difference, $\Delta\,\alpha\,=\,\alpha - \alpha_{{NT}}$.  This latter quantity indicates how much the spectrum has been flattened by the addition of thermal emission. 

\begin{figure}[!tbp]
  \centering
  \begin{minipage}[b]{0.40\textwidth}
    \includegraphics[width=\textwidth]{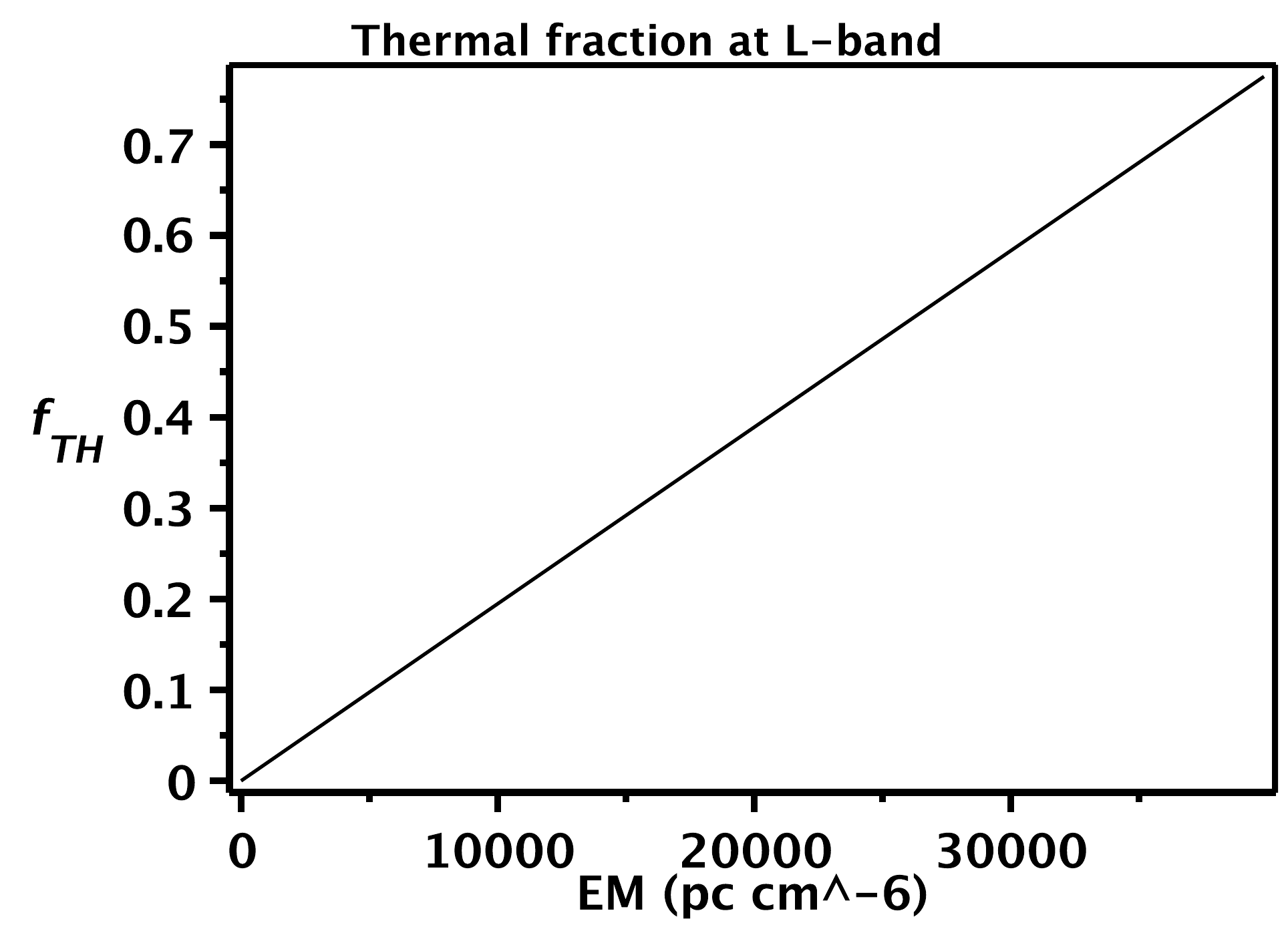}
  \end{minipage}
  \hfill
  \begin{minipage}[b]{0.4\textwidth}
    \includegraphics[width=\textwidth]{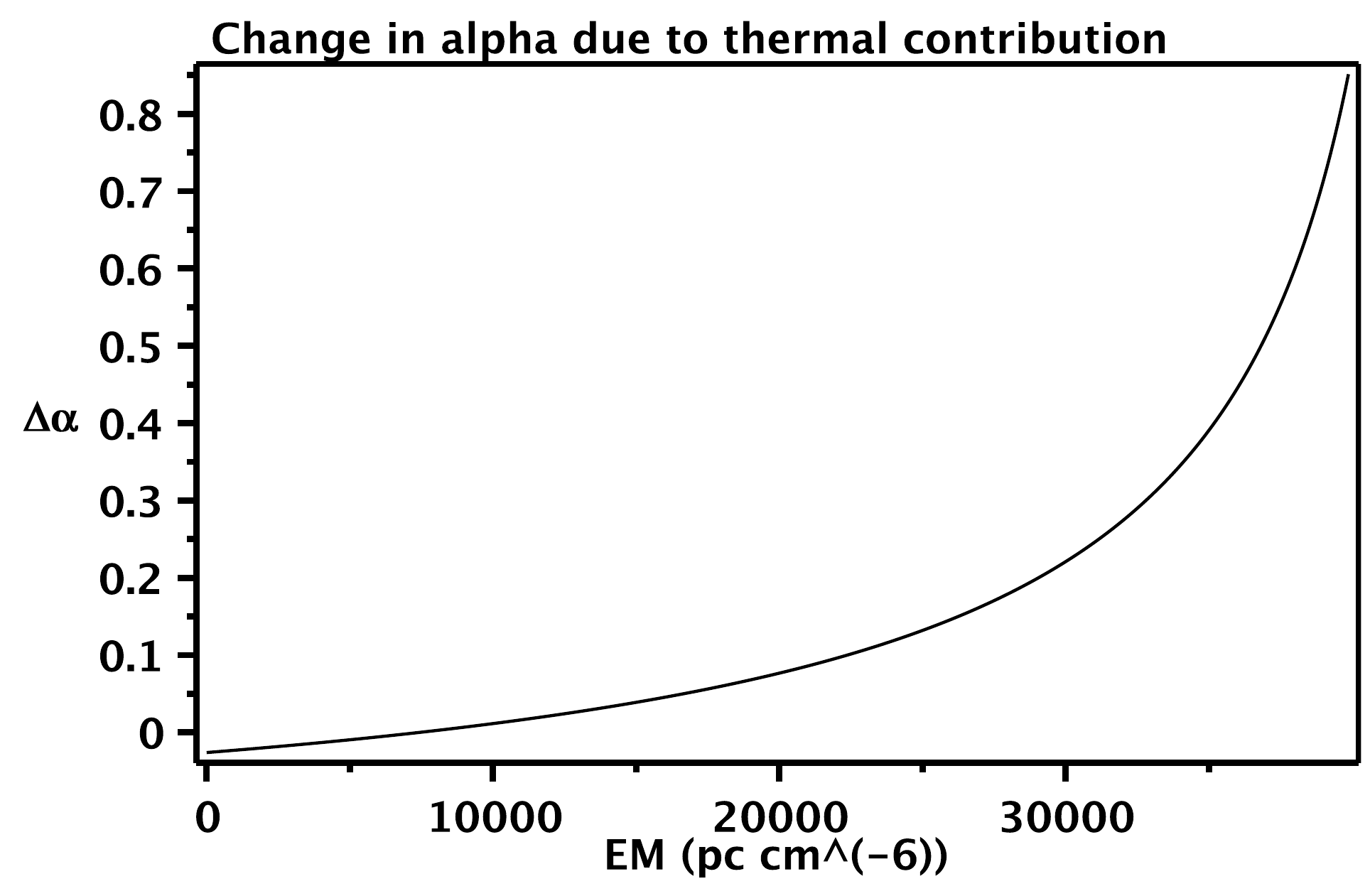}
  \end{minipage}
  \caption{Thermal fraction, $f_{\nu_{TH}}$ at L-band {\bf (top)} and change in spectral index, $\Delta\,\alpha\,=\,\alpha\,-\,\alpha_{NT}$ {\bf (bottom)} at 3.8 GHz that results from a range of emission measures, EM (see Sect.~\ref{sec:flat_disks}).
   }
\label{fig:mapleout}
\end{figure}

Fig.~\ref{fig:mapleout} shows the thermal fraction at L-band and $\Delta\,\alpha$ at 3.8 GHz (corresponding to the mid-point between the L-band and C-band observations) for a variety of values of EM. % It is clear that if EM = 0, then all emission must be non-thermal and there is no change in the spectral index; then $\alpha\,=\,\alpha_{NT}$.
As the contribution of thermal gas increases, so does $\Delta\alpha$. For example, if $\alpha_{NT}\,=\,-0.72$ at 3.8 GHz, then
$\Delta\,\alpha\,=\,0.46$, $EM\,=\,3.62\,\times\,10^4$ pc cm$^{-6}$,  $f_{th}\,=\,0.7$ at L-band and  $f_{th}\,=\,0.8$ at C-band. The thermal gas is highly optically thin at both frequencies.  The average electron density over the line of sight would be $n_e\,=\,8$ cm$^{-3}$.  Unlike the case for thermal absorption (Sect.~\ref{sec:flattening_1}) the non-thermal spectral index does not vary with frequency and we can clearly see that the required electron density is a more moderate value.

We finally can compare the emission measure with the observed value of 2958 cm$^{-6}$ pc obtained from the H$\alpha$ image \citep{col00}. Fig.~\ref{fig:mapleout} shows that such a low value of EM would produce only a negligible amount of spectral flattening.  Thus, either $\alpha_{NT}$ is quite flat to begin with, or the optically measured EM is heavily extincted by dust.  We suspect the latter and suggest that further analysis of this region to determine the dust contribution would provide such verification. 

In summary, flat spectral indices in discrete regions in the disks of the CHANG-ES galaxies are most likely due to a higher-than-average contribution from thermal emission.

%feature at far SE disk -- cl/dc alpha is -0.46   bl/cc -0.25 so 0.2 flatter at the higher resolution  -- why?
%avg beam is 484 pc  how many snrs??

%Bl flux measurement (imfit) gives 3.6 p/m 0.15 mJy which at distance of N5775 is 3.6e20 W/Hz which is 120 mean M82 SNe.
%how much thermal absorption would flatten alpha by 0.2?

\subsection{Background Sources}
\label{sec:background_sources}

At L-band, there are many background sources in the various fields. % The detailed analysis of such sources is beyond the scope of this paper but
We note that, since 35 fields at each of two different UV weightings are available, the BL images can provide an independent measurement set for source count studies.
With rms values typically of order 20 $\mu$Jy beam$^{-1}$ (Table\,~\ref{tab:imagingparameters}), the CHANG-ES data can provide source count estimates to very low levels (though subject to PB weighting).  In addition, our matching resolution CC data set \citep{wal19} has rms values that are typically much less than 10 $\mu$Jy beam$^{-1}$. 
%{ before uploading these fields to the data release website, I want to correct the L-band fields for the new primary beams.  Such correction can be done after this paper is submitted.}
In this section we consider only the possibility that background sources may be seen through the disks of our galaxies.

Numerous studies have examined radio source counts historically, and more recently to increasingly lower flux levels, especially in anticipation of the full power of the SKA\footnote{Square Kilometre Array}.  Unfortunately, considerable scatter is seen in the differential source counts at the lowest flux levels.  For example, the 1.4 GHz normalized differential source counts vary by a factor of $\approx\,6$ at a level of 100 $\mu$Jy \citep{dez10,pra18}, underscoring the need for good deep surveys.

More relevant to an analysis of the edge-on galaxies is how many background sources might be shining through the galaxy disks.  Such sources may need to be removed before detailed analysis of the galaxies can be carried out.

We have several clear examples of known background sources seen through the disk.  An example is NGC~5907 which has a very strong double-lobed radio source behind the far SE tip of the major axis \citep{dum00}  as can clearly be seen in Appendix~\ref{app:B}).
In this case, the extragalactic nature of the source is obvious and its spectral index is also discordant compared to the galaxy's disk.  Another example is UGC~10288 (Appendix~\ref{app:B}) whose radio emission is actually dominated by the background double-lobed radio source as described in detail in \cite{irwIII}.

However, sources such as the point source at RA = 12 13 47.1, DEC = 14 54 49.2 apparently in the disk of NGC~4192 (Appendix~\ref{app:B}) with a peak specific intensity of 0.60 mJy beam$^{-1}$ may or may not be a background source.

Bearing in mind the uncertainties noted above, one can make a rough estimate of the number of sources that might be seen through the disk of such a galaxy.  Following \cite{rah16}, who employed the simulations of \cite{wil08}, we would expect $\approx$ 108 background sources to be present in any 1000 arcsec x 1000 arcsec field above a typical $3\sigma$ threshold of $60~\mu$Jy beam$^{-1}$.  For the specific NGC~4192 field (rob0), taking an ellipse for the galaxy area, of major x minor axis size, 250 arcsec x 45 arcsec, and adopting a more stringent $5\sigma$ limit of $87~\mu$Jy beam$^{-1}$ for that field, then we expect 0.6 background sources to be shining through the disk.  For NGC~4192, then, the point source noted above could indeed be a background source.  Higher resolution observations are really required to confirm this.
%JUDITH see maple sheet called source_counts_worksheet.mw

\section{Summary and Conclusions}
\label{sec:conclusions}

We have now presented the third data release of the CHANG-ES galaxy sample of 35 edge-on nearby galaxies.  This data release consists of FITS images at 1.58 GHz taken in the B-configuration (BL data) of the VLA with a spatial resolution of $\approx$ 3 arcsec.  For most galaxies, uvtapered images with resolution of $\approx$ 6 arcsec were also made and are in the data release. In addition to these images, we include band-to-band spectral index maps made from BL and C-configuration 6.00 GHz data (CC data) at matching spatial resolutions.

{\it In-band spectral indices} (over 512 MHz L-band bandwidth and 2 GHz bandwidht at C-band) have been made but are not released, given the relatively low S/N for many galaxies.   Polarization at L-band is not detected except for two galaxies: NGC~3079 and NGC~4388 (Fig.~\ref{fig:lin_pol}) because of Faraday depolarization at L-band; consequently, polarization images are also not included in the data release.  Non-release images for this configuration can be requested from the first author, if required. Otherwise, release data can be downloaded from https://www.queensu.ca/changes.

Panels displaying the data products are provide in Appendix~\ref{app:B} for all galaxies.  The C configuration data from which the spectral index maps were made will be released separately \citep{wal19} and the H$\alpha$ images that are displayed in the panels can already been found at our data-release website \citep[see][]{var19}.

Several new and unique results have been identified.  We see structures in the spectral index maps that are not observed in total intensity, presumably because the structure is swamped by other emission in total intensity.  An example is apparent spiral arms in NGC~3448 and two radio lobes on either side of the nucleus in NGC~3628.  Thus, high resolution spectral index maps are essential for revealing otherwise hidden features and show us how the energy spectral index of cosmic ray electrons varies with position.

We have looked for radio-only criteria for identifying AGNs in these galaxies, including unresolved point-like cores, high radio luminosities, radio lobes, and relatively flat spectra in the cores.  Other criteria, such as time variability, circular polarization and high brightness temperatures have been examined elsewhere or are not applicable to this data set (Sect.~\ref{sec:criteria}).  We find a lower limit to the incidence of radio AGNs of 55\% in nearby spiral galaxies.  This incidence increases to 64\% if we include galaxies that have strongly peaked emission in the core but that emission is still slightly resolved.

All galaxies for which available X-ray data suggest the presence of an AGN, are also radio AGN.  However, some other galaxies (e.g. NGC~660 and NGC~3079) clearly have radio AGN but show only little evidence for an X-ray AGN.
This could be because of timescales, with the X-ray emission fading before the radio, or could be related to the resolution and sensitivity of the XMM data.

An examination of discrete emission in galaxy disks reveals some regions of quite flat spectral index which can reasonably be explained by a contribution from free-free thermal emission that is higher than the global average.   We suggest that one such HII region in the outer disk of NGC~5775 could have a thermal fraction of as much as 70\% at L-band.  A careful comparison between the H$\alpha$ images and the radio continuum will likely prove to be of importance for other discrete regions, provided dust obscuration is also considered carefully.

The L-band fields contain many background sources, some of which may be shining through the disks of the nearby galaxies.

\acknowledgments

The first author would like to thank the Natural Sciences and Engineering Research Council of Canada for a Discovery Grant.
The National Radio Astronomy Observatory is a facility of the National Science Foundation operated under cooperative agreement by Associated Universities, Inc.
This research has made use of the NASA/IPAC Extragalactic Database (NED) which is operated by the Jet Propulsion Laboratory, California Institute of Technology, under contract with the National Aeronautics and Space Administration. 

%% To help institutions obtain information on the effectiveness of their
%% telescopes, the AAS Journals has created a group of keywords for telescope
%% facilities. A common set of keywords will make these types of searches
%% significantly easier and more accurate. In addition, they will also be
%% useful in linking papers together which utilize the same telescopes
%% within the framework of the National Virtual Observatory.
%% See the AASTeX Web site at http://aastex.aas.org/
%% for information on obtaining the facility keywords.

%% After the acknowledgments section, use the following syntax and the
%% \facility{} macro to list the keywords of facilities used in the research
%% for the paper.  Each keyword will be checked against the master list during
%% copy editing.  Individual instruments or configurations can be provided 
%% in parentheses, after the keyword, but they will not be verified.

{\it Facilities:} \facility{VLA}.

%% Appendix material should be preceded with a single \appendix command.
%% There should be a \section command for each appendix. Mark appendix
%% subsections with the same markup you use in the main body of the paper.

%% Each Appendix (indicated with \section) will be lettered A, B, C, etc.
%% The equation counter will reset when it encounters the \appendix
%% command and will number appendix equations (A1), (A2), etc.

\appendix

\section{Corrections to the Spectral Index Error Maps}
\label{app:A}

For highest accuracy, the spectral index error maps
should take into account the fact that the rms noise increases with distance from the field (pointing) center.
%as computed from Eqn.~\ref{eqn:error_maps} and shown in Fig.xx should be modified to take into account the fact that the rms noise increases with distance from the field (pointing) center.
The corrected uncertainty is
\begin{equation}
  \label{eqn:error_maps_corr}
  {\sigma_\alpha}_{corr}\,=\,\frac{1}{ln\left(\frac{\nu_L}{\nu_C}\right)}
  \sqrt{\left(   \frac{\sigma_L}{PB_L(r) I_L}   \right)^2\,+\,
  \left(\frac{\sigma_C}{PB_C(r) I_C}\right)^2}
\end{equation}
where $PB_L(r)$ and $PB_C(r)$ are the primary beam responses at L-band and C-band, respectively, and $r$ is the distance from the field center. Rearranging this equation and using
Eqn.~\ref{eqn:alpha_defn} to eliminate $I_L$ and $I_C$ we find
\begin{equation}
  \label{eqn:error_maps_corr2}
 {\sigma_\alpha}_{corr} \,=\,\frac{\sigma_L}{I_L\,ln\left(\frac{\nu_L}{\nu_C}\right)}\sqrt{
          \frac{1}{\left(PB_L(r)\right)^2}\,+\,\frac{1}{\left(PB_C(r)\right)^2}\left(\frac{\sigma_C}{\sigma_L}\right)^2
          \left(\frac{\nu_L}{\nu_C}\right)^{2\alpha}
  }%end sqrt
\end{equation}

The primary beam shape, $PB(r)$ is given in \cite{per16} as a 6th order polynomial.  However, for the regions shown in  our  spectral index maps, the polynomial is well fitted by a normalized Gaussian, where the half-width at half maximum, HWHM = 12.76 arcmin at 1.57 GHz and HWHM = 3.51 arcmin at 6.00 GHz, i.e.
%HWHM = 20.03 arcmin$\cdot$GHz at L-band and HWHM = 21.03 arcmin$\cdot$GHz at C-band. %to within $\approx\,1$\% at both frequencies. judith see file alpha_rms_error_estimate.mw
%% L-band HPHM=20.03arcminGHz so FWHM=2*20.03/2.35482 to turn fwhm into sigma sigma=17.01arcminGHz
%NOW turn it into arcmin so =10.84
%% C-band HPHM=21.07arcminGHz so FWHM=2*21.07/2.35482 to turn fwhm into sigma sigma= 17.90 arcminGHz
%%NOW turn it into arcmin s = 2.98
\begin{equation}\label{eqn:PBs}
  PB_L(r)\,=\,e^{\left(-\frac{1}{2}\left(\frac{r}{10.8}\right)^2\right)}~~~~~
  PB_C(r)\,=\,e^{\left(-\frac{1}{2}\left(\frac{r}{2.98}\right)^2\right)}
  \end{equation}
where $r$ is in units of arcmin and the standard deviation, $\sigma\,=\,10.8$ for a HWHM of 12.76, and $\sigma\,=\,2.98$ for a HWHM of 3.51 in the same units.

%The user can correct the error maps by multiplying by a factor
%5\begin{equation}\label{eqn:alphacor1}
 % \label{eqn:error_maps_corr}
 % f\,=\,\frac{{\sigma_\alpha}_{corr}}{\sigma_\alpha}\,=\,\sqrt{
 % \frac
 % {{\left(   \frac{\sigma_L}{PB_L(r) I_L}   \right)^2\,+\,
 %     \left(\frac{\sigma_C}{PB_C(r) I_C}\right)^2}}
 % {{\left(   \frac{\sigma_L}{ I_L}   \right)^2\,+\,
 %     \left(\frac{\sigma_C}{I_C}\right)^2}}}
%\end{equation}

Thus, given an input spectral index map, $\alpha$, Eqn.~\ref{eqn:error_maps_corr2} together with Eqns.~\ref{eqn:PBs} can be used to determine a corrected value of the error in the spectral index.  
The quantities, $\sigma_L$ and $\sigma_C$ are taken as constant for a given map and are read from Table~\ref{tab:smoothed_data}, whereas $\alpha$ varies with position and the PBs are functions of $r$.

Alternatively, the given error maps can be corrected by multiplying by a factor
\begin{equation}\label{eqn:errorfactor}
  f(r)\,=\,\frac{{\sigma_\alpha}_{corr}}{\sigma_\alpha}\,=\,
\sqrt{
    \frac{
          \frac{1}{\left(PB_L(r)\right)^2}\,+\,\frac{1}{\left(PB_C(r)\right)^2}\left(\frac{\sigma_C}{\sigma_L}\right)^2
          \left(0.26\right)^{2\alpha}
        }
         {
           {1}\,+\,\left(\frac{\sigma_C}{\sigma_L}\right)^2
          \left(0.26\right)^{2\alpha}
         }     
  }%end sqrt
\end{equation}
where $\sigma_\alpha$ is given by Eqn.~\ref{eqn:error_maps} and we have explicitly evaluated the frequency ratio for CHANG-ES.
%hwhm=20.03 arcminGHz at lband and 21.07 arcminGHz at Cband

%Then, for any value of $\alpha$ at some position, $r$, from the field center, $\sigma_\alpha$ can be corrected by multiplying the value given in the error map by the factor, $f(r)$.
For example, let $r\,=\,3$ arcmin, $\alpha\,=\,-0.8$, $\sigma_L\,=\,16$ $\mu$Jy beam$^{-1}$ and $\sigma_C\,=\,3$ $\mu$Jy beam$^{-1}$. Then $PB_L(r)\,=\,0.96$, $PB_C(r)\,=\,0.60$, and $f(r)\,=\,1.2$.  At the adopted position, the error map value would be increased by this factor.

\vfill\eject

\section{Galaxy Panels}
\label{app:B}

The figures in this Appendix display images of the data products in Data Release 3 of the CHANG-ES project, described below. Each field of view is the same for any given galaxy.  {Missing panels may occur if images were not satisfactorily made (e.g. uvtapered).}

{For the large galaxies, NGC~891, NGC~4565, and NGC~5907, in addition to the complete panels, we include blow-ups of panels (a), (b), (g) and (h) as described below.}

%Rather than using an arbitrary midpoint for our colour tables, we use conventional spectral index thresholds to specify divergent points. (A different hue appears on each side of a divergent point.)  The “flat” (thermal) region of the spectrum has alpha values greater than -0.1. On the increasing value side of this point, appears  dark blue-purple. The purple aspect appears stronger as alpha increases until a pure purple is associated with the maximum alpha value.   More negative than -0.1 the colour diverges to dark cyan.  The “steep” (synchrotron) region has alpha less than -0.8. This boundary is  represented by saturated orange. This colour appears more yellow with more negative spectral index values until yellow represents the minimum alpha.  Between the alpha thresholds of -0.8 and -0.1 orange and dark cyan blend with each other (creating another divergent point around alpha = -0.6). Although binning the colour map is a redundancy, it should aid in visually selecting alpha values in  cases where perception illusions hinder associating a map colour with the equivalent one in the colour bar.

%In summary, flat regions, such as those containing AGN,  are clearly distinguished by their purple colours while orange through yellow demark steep regions, indicating non-thermal emission is dominant. 

\noindent{\bf First row:}

\noindent{\bf a}.  Contours and a colour scale (logarithmically stretched from background (2$\sigma$) to peak emission) for total intensity images with a Briggs robust=0 (rob 0) uv weighting.
 The I contours are at 3, 6, 12, 24, 48, 96 and 192 times the rms value.  The rms values are taken from Table~\ref{tab:imagingparameters} where the higher value near the galaxy has been used in cases where there is strong variation.
 
\noindent{\bf b}. Same as (a) except that these maps are total intensity images with a uv taper applied onto the rob 0 weighting (Sect.~\ref{sec:obs_red} and Table~\ref{tab:imagingparameters}).

\noindent{\bf Second row:}

\noindent{\bf c}. Band-to-band (BL to CC) spectral index maps for rob 0 weighting.  A cut-off of 3 times the rms value has been applied to each map (both BL and CC) as described in Sect.~\ref{sec:spectral_indices}.  In this case, the rms values for each band are listed in Table~\ref{tab:smoothed_data}. The colour scheme is displayed to the right of Panel d. These maps tend to look noisier than the others because the PB correction for each band results in an increasing rms with distance from the map center (Sect.~\ref{sec:spectral_indices}). Note that no correction for thermal emission has been made (Sect.~\ref{sec:alpha_thermal}).

Note that care has been taken for the colour scheme. We show flatter spectral indices as dark blue-purple and steeper spectral indices appear orange-yellow.  Thus flatter regions, such as those containing AGNs (synchrotron-dominated) or discrete HII regions (thermal dominated), may be identified by their blue-purple colour.  Regions in which diffuse synchrotron emission dominates will appear orange or even yellow if the spectral index is very steep.

\noindent{\bf d}. Same as Panel c but for the rob 0 maps with a uv taper.

\noindent{\bf Third row:}

\noindent{\bf e}. Spectral index uncertainty map corresponding to Panel c.  The colour scheme is displayed to the right of Panel f.  See Sect.~\ref{sec:alpha_errors} for a discussion of the uncertainties and Appendix~\ref{app:A} for corrections with distance from the center.

\noindent{\bf f}. Same as Panel e but for the rob 0 maps with a uv taper.

\noindent{\bf Fourth row:}

\noindent{\bf g}. Same contours as in (a) over a colour H$\alpha$ image from \cite{var19} (log stretch from background to peak emission).   H$\alpha$ FITS images are available on our data release web site (see Sect.~\ref{sec:disks}).

\noindent{\bf h}. The 3 $\sigma$ contour from (b) over an optical image of the galaxy.  {When panel (b) is unavailable, we use the 3$\sigma$ contour from panel (a) instead.} The optical images were created using a combination of Sloan Digital Sky Survey (SDSS) g, r, and i bands {\it or} Digitized Sky Survey 2 (DSS2) blue, red, and infrared bands, for the galaxies not available in SDSS.

%% thebibliography produces citations in the text using \bibitem-\cite
%% cross-referencing. Each reference is preceded by a
%% \bibitem command that defines in curly braces the KEY that corresponds
%% to the KEY in the \cite commands (see the first section above).
%% Make sure that you provide a unique KEY for every \bibitem or else the
%% paper will not LaTeX. The square brackets should contain
%% the citation text that LaTeX will insert in
%% place of the \cite commands.

%% We have used macros to produce journal name abbreviations.
%% AASTeX provides a number of these for the more frequently-cited journals.
%% See the Author Guide for a list of them.

%% Note that the style of the \bibitem labels (in []) is slightly
%% different from previous examples.  T
%he natbib system solves a host
%% of citation expression problems, but it is necessary to clearly
%% delimit the year from the author name used in the citation.
%% See the natbib documentation for more details and options.

\clearpage
%% Use the figure environment and \plotone or \plottwo to include
%% figures and captions in your electronic submission.
%% To embed the sample graphics in
%% the file, uncomment the \plotone, \plottwo, and
%% \includegraphics commands
%%
%% If you need a layout that cannot be achieved with \plotone or
%% \plottwo, you can invoke the graphicx package directly with the
%% \includegraphics command or use \plotfiddle. For more information,
%% please see the tutorial on "Using Electronic Art with AASTeX" in the
%% documentation section at the AASTeX Web site, http://aastex.aas.org/
%%
%% The examples below also include sample markup for submission of
%% supplemental electronic materials. As always, be sure to check
%% the instructions to authors for the journal you are submitting to
%% for specific submissions guidelines as they vary from
%% journal to journal.

%% This table also includes a table comment indicating that the full
%% version will be available in machine-readable format in the electronic
%% edition.
\clearpage

%[start, end, duration, phasecal distance to galaxy, flux of phasecal]
%[REF TO BIG TABLES HERE - observation dates, start, end, duration, SB number, cals, phasecal distance to galaxy, flux of phasecal]

%% If you use the table environment, please indicate horizontal rules using
%% \tableline, not \hline.
%% Do not put multiple tabular environments within a single table.
%% The optional \label should appear inside the \caption command.

%% If the table is more than one page long, the width of the table can vary
%% from page to page when the default \tablewidth is used, as below.  The
%% individual table widths for each page will be written to the log file; a
%% maximum tablewidth for the table can be computed from these values.
%% The \tablewidth argument can then be reset and the file reprocessed, so
%% that the table is of uniform width throughout. Try getting the widths
%% from the log file and changing the \tablewidth parameter to see how
%% adjusting this value affects table formatting.

%% The \dataset{} macro has also been applied to a few of the objects to
%% show how many observations can be tagged in a table.

\clearpage

\clearpage

%[start, end, duration, phasecal distance to galaxy, flux of phasecal]
%[REF TO BIG TABLES HERE - observation dates, start, end, duration, SB number, cals, phasecal distance to galaxy, flux of phasecal]

\clearpage

\clearpage

\clearpage

\end{document}